\documentclass{elsart}
\usepackage{amssymb}
\usepackage{epsfig}
\usepackage{pifont}
\usepackage{ulem}

\newcommand{\D}{\mathrm{d}}

\begin{document}

\begin{frontmatter}

\title{A high performance likelihood reconstruction of gamma-rays for Imaging Atmospheric Cherenkov Telescopes}

\author{Mathieu de Naurois}
\address{LPNHE, IN2P3 - CNRS - Universit\'es Paris VI et Paris VII, Paris, France}
\ead{denauroi@in2p3.fr}
\author{Lo\"ic Rolland}
\address{LAPP, Universit\'e de Savoie, CNRS/IN2P3, Annecy-le-Vieux, France}
\ead{loic.rolland@lapp.in2p3.fr}
\begin{abstract}

We present a sophisticated gamma-ray likelihood reconstruction technique for
Imaging Atmospheric Cerenkov Telescopes. The technique is based on the
comparison of the raw Cherenkov camera pixel images of a photon induced atmospheric particle shower with the predictions from a
semi-analytical model. The approach was initiated by the CAT
experiment in the 1990's, and has been further developed by a new fit
algorithm based on a log-likelihood minimisation using all pixels in
the camera, a precise treatment of night sky background noise,
the use of stereoscopy and the introduction of first interaction depth
as parameter of the model.

The reconstruction technique provides a more precise direction and
energy reconstruction of the photon induced shower compared to other
techniques in use, together with a better gamma efficiency, especially
at low energies, as well as an improved background rejection. For data
taken with the H.E.S.S.\ experiment, the reconstruction technique yielded a
factor of $\sim$2 better sensitivity compared to the
H.E.S.S.\ standard reconstruction techniques based on second moments of the
camera images (Hillas Parameter technique).

\end{abstract}

\begin{keyword}
Cherenkov \sep IACT \sep analysis techniques \sep VHE Gamma-ray Astronomy


\end{keyword}

\end{frontmatter}

\section*{Introduction}

The last decade saw the emergence of very high energy (VHE; $E > 100$
GeV) gamma-ray astronomy as a new observational discipline, with the
number of VHE gamma-ray sources now approaching 100. This success was
driven by the third generation of ground based Imaging Atmospheric
Cherenkov Telescope Systems (IACT), such as H.E.S.S., with an order of
magnitude better sensitivity compared to the previous generation
instruments (see e.g. \cite{Buckley,VoelkBernloehr} for a recent
review).The improvement was made possible by the use of stereoscopic systems 
of large telescopes equipped with finely pixelated fast cameras.

To reconstruct the direction and energy of the primary gamma-ray and
to discriminate them from charged cosmic rays most of the current
experiments use reconstruction techniques based on the second moments of the
pixel amplitudes in the camera (Hillas parameters)
\cite{HillasParam1,HillasParam2}. These techniques are very robust and
efficient. However, a more sophisticated albeit more computing time
intensive reconstruction technique was pioneered by the CAT experiment
\cite{Bohec}, taking advantage of its very finely pixelized camera. The
technique is based on a $\chi^2$ comparison of the recorded Cherenkov
light distributions of a photon induced electromagnetic shower in the
camera, i.e. the shower images, with calculated shower images from a
model of the Cherenkov light distribution in electromagnetic
showers. The reconstruction technique presented in this paper ({\it Model Analysis})
is a further development and improvement of this early work.


The calculated shower images are derived from the Cherenkov light
distribution of charged particles in electromagnetic showers taking
into account light collection efficiencies, atmospheric absorption
etc. The Cherenkov light distribution of a shower is determined by the
longitudinal, lateral, and angular distributions of charged particles
in the shower. These distributions are derived from Monte Carlo
simulations and parametrised to yield an analytical description of the
shower, i.e.\ the {\it shower model}, including the depth of the first
interaction as a new parameter in the parametrisation. Additionally,
the contribution of the night sky background noise in the camera in
every pixel is modelled on the basis of a detailed statistical analysis. 
Thus the fit procedure does not require a dedicated image cleaning
procedure to extract the pixels illuminated by the shower. The
parameters of the calculated shower that best fits the measured shower
image are determined in a minimisation procedure which yields a
selection criteria to discriminate the gamma-ray induced shower from
the hadronic background.

The different parts of the model, i.e. the semi-analytical description
of the shower development are described in section \ref{sec:model},
and are used in section \ref{sec:generation} to generate the shower image
templates. In section \ref{sec:reconst} the fit algorithm is
described.  Finally, section \ref{sec:results} is devoted to the
description of the performance of source analysis using this model reconstruction, with detailed
comparison with alternate reconstruction techniques.

\section{Charged particle distributions in an electromagnetic shower}
\label{sec:model}

The model of the pixel illumination by Cherenkov light in the camera
is based on analytical descriptions of the energy-dependent
longitudinal, lateral, and angular distributions of charged particles
in electromagnetic showers. The electromagnetic showers were generated
with the KASKADE \cite{Kaskade} shower simulation program. The KASKADE
program has been improved to include among others a more precise
treatment of Bhabha and Moller scattering
\cite{TheseJulien,NoteJulien}. A detailed comparison of several
generators in the H.E.S.S.-collaboration, including CORSIKA
\cite{Corsika}, showed no noticeable difference between the different
generators.

In order to cover the dynamical range of the H.E.S.S. instruments, showers were generated at 
energies of 10~GeV, 50~GeV, 100~GeV, 500~GeV, 1~TeV, 5~TeV, 10~TeV and 20~TeV.
Typical number of showers required for the generation of a smooth model range from $\sim 200 000$ at low
energy to a few hundred at the highest energies. The varying number of showers ensure that 
different primary energies have similar weights in the model construction.

The model for atmospheric electromagnetic showers presented here is an
extension of the model proposed by Hillas in \cite{Hillas}. The
nomenclature used here follows closely the nomenclature of the model by
Hillas.

As the dominant source of shower-to-shower fluctuations, the depth of the first interaction 
is included as a new parameter in the parametrisation. This helps reducing the discrepancy between
actual shower images and model predictions, and thus improves the reconstruction and discrimination efficiency.

\subsection{Longitudinal distribution of charged particles}

\begin{figure}[ht]
\begin{center}
\epsfig{file=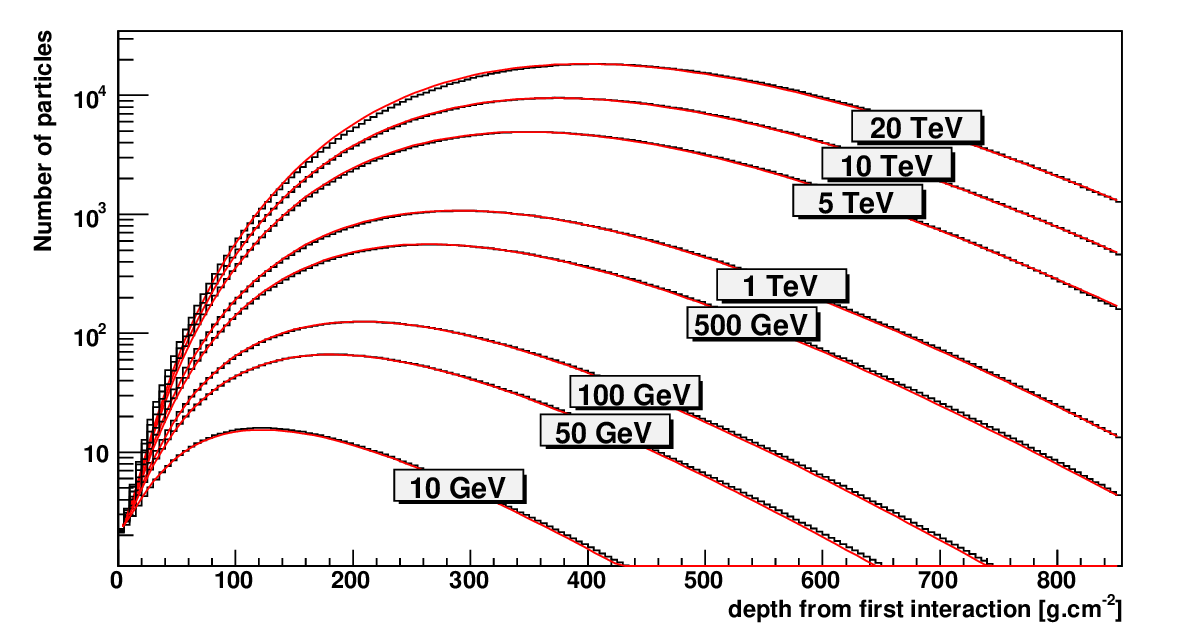,width=0.95\textwidth}
\epsfig{file=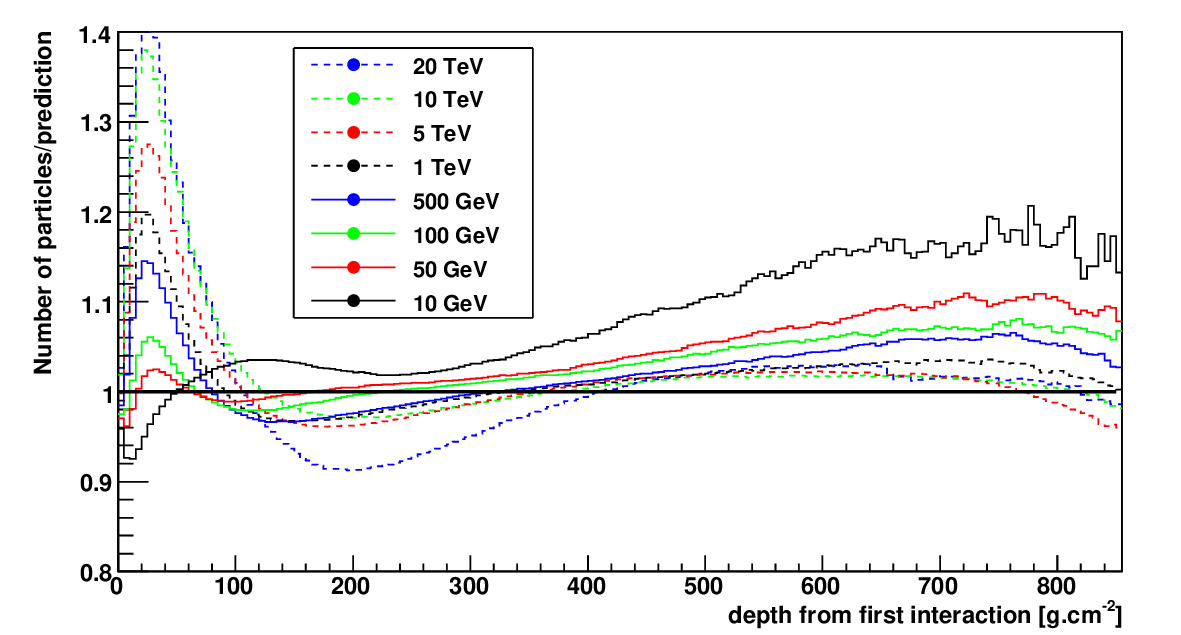,width=0.95\textwidth}
\end{center}
\caption{\label{fig:Dist}Top: Longitudinal shower development measured from
  the first interaction point: Number of charged particles in the
  shower as a function of the distance to the first interaction point
  for several primary energies ranging from $10~\mathrm{GeV}$ to
  $20~\mathrm{TeV}$. The black histograms show the results of a
  simulation compared to the analytical expression of
  eq. \ref{eq:NeeDeconvolved} (solid red curve).
   Bottom: Ratio of simulation to prediction from eq. \ref{eq:NeeDeconvolved} 
  for various primary energies, showing that eq. \ref{eq:NeeDeconvolved} is a good description
  at the level of  $\sim 5\%$ in the central depth range, degrading to  $\sim 10\%$ at the lowest energies.}
\end{figure}

The average number of charged particles ${\mathcal
  N}_e(y,t)$ in simulated electromagnetic showers as a function of the
distance to the first interaction point $t$ for different primary
photon energies are shown in fig.\,\ref{fig:Dist}. The longitudinal
distributions are modelled by a modified Greisen formula\cite{Greisen}:

\begin{equation}
{\mathcal N}_e(y,t) = \frac{a}{\sqrt{y}} \times \exp
\left[t \times \left(1 - \frac{b}{b-1}\times \ln(s)\right)\right]
\nonumber + \left(2 - \frac{a}{\sqrt{y}} \right) \times
\exp(-t) \label{eq:NeeDeconvolved}
\end{equation}

where $y=\ln(E_{prim}/E_{crit})$ is the scaled primary photon energy
with $E_{prim}$ being the primary photon energy and $E_{crit}$ being
the critical energy ($E_{crit}\approx 83~\mathrm{MeV}$ in air). 
The first part of the expression corresponds to
the standard Greisen formula, with parameters $a$ and $b$ left free. The second part 
describes the decay of the two charged particles created at the first interaction point
$t=0$ and ensures ${\mathcal N}_e(y,t=0) =2$.
The shower depth $t$ is given in units of radiation lengths. The shower
age $s$ measured from the point of the first interaction is given by

\begin{equation}
s = \frac{b}{1 + c \times (b-1)/t} \label{eq:AgeDeconvolved}
\end{equation}

This expression is constructed so that the shower age being $s=0$ at the first interaction and $s=1$ at the shower
maximum. The parameter $c$ is the depth of shower maximum measured
from the first interaction ($s=1$ for $t=c$) and $b$ is a scaling
factor related to speed of the shower development. The parameters $a$
and $c$ are found to be linearly dependent on the scaled energy $y$.
A fit of the analytical description on the simulated distributions
yielded for the parameters $a$, $b$ and $c$:

\begin{equation}
a  =  1.05 + 0.033 \times y, \quad
b  =  2.66, \quad
c  =  0.97 \times y - 1.32  \label{eq:NeeDeconvolvedParameters}
\end{equation}

As can be seen in fig.\,\ref{fig:Dist} the analytical description is
in good agreement (at a level of $\sim 5\%$ in the central depth range,
degrading to $\sim 20\%$ in the shower tail  and  $\sim 40\%$ in the shower head) 
with the simulation for primary energies ranging from $10~\mathrm{GeV}$ to more
than $20~\mathrm{TeV}$,with slightly worse performances at the very high
and very low energies.

The Cherenkov light distribution of an electromagnetic shower depends
on the energy dependent longitudinal distributions of charged
particles in the shower ${\D{\mathcal N}_e}/{\D E}$ with

\begin{equation}
{\mathcal N}_e = \int \frac{\D{\mathcal N}_e}{\D E} \D E., ~~~~
{\mathcal N}_e(E \geq E_0) = \int_{E_0}^\infty \frac{\D{\mathcal N}_e}{\D E} \D E.
\label{eq:dNeedE}
\end{equation}

\begin{figure}[p!]
\begin{center}
\begin{tabular}{cc}
\epsfig{file=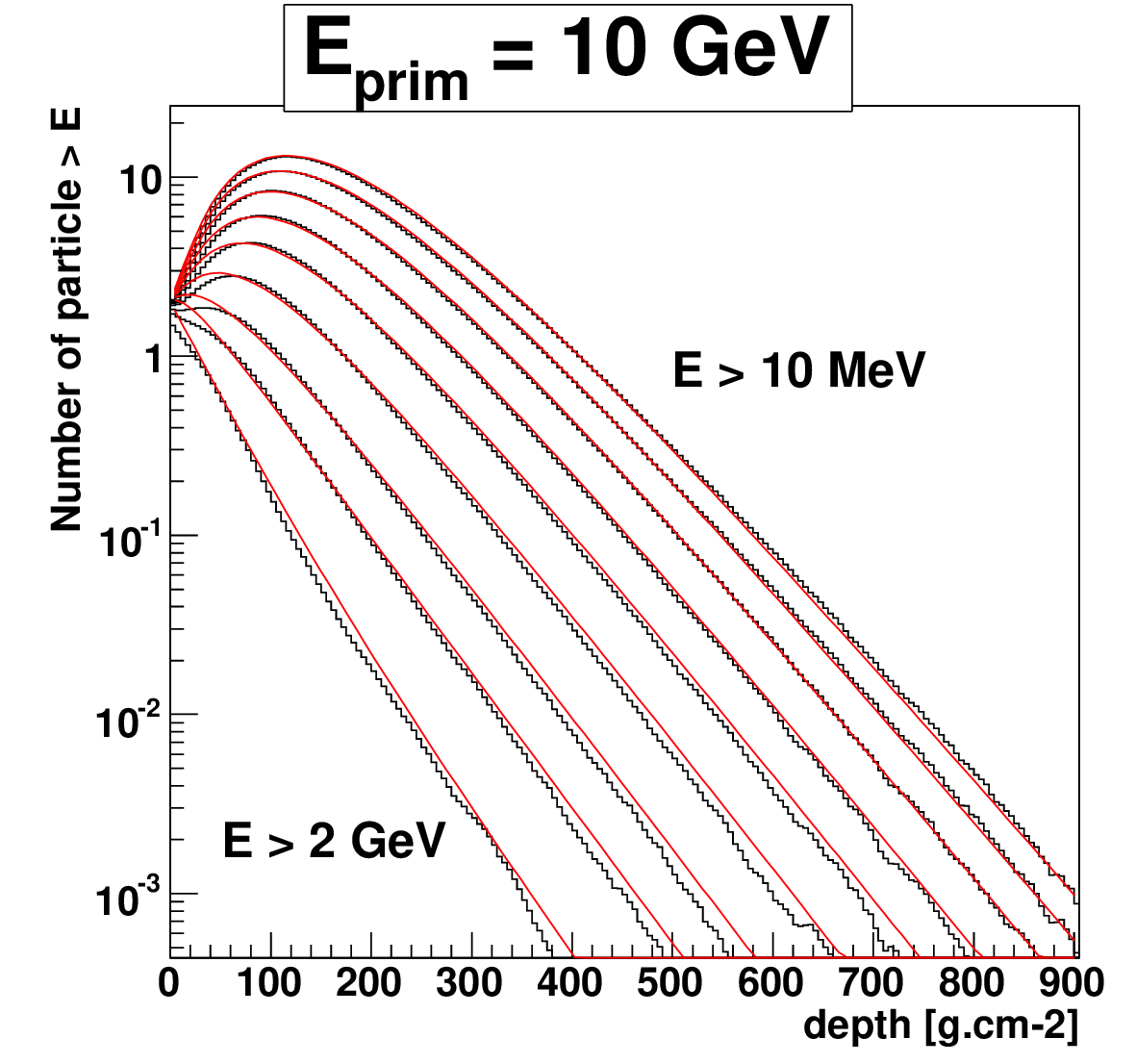,width=0.48\textwidth} & \epsfig{file=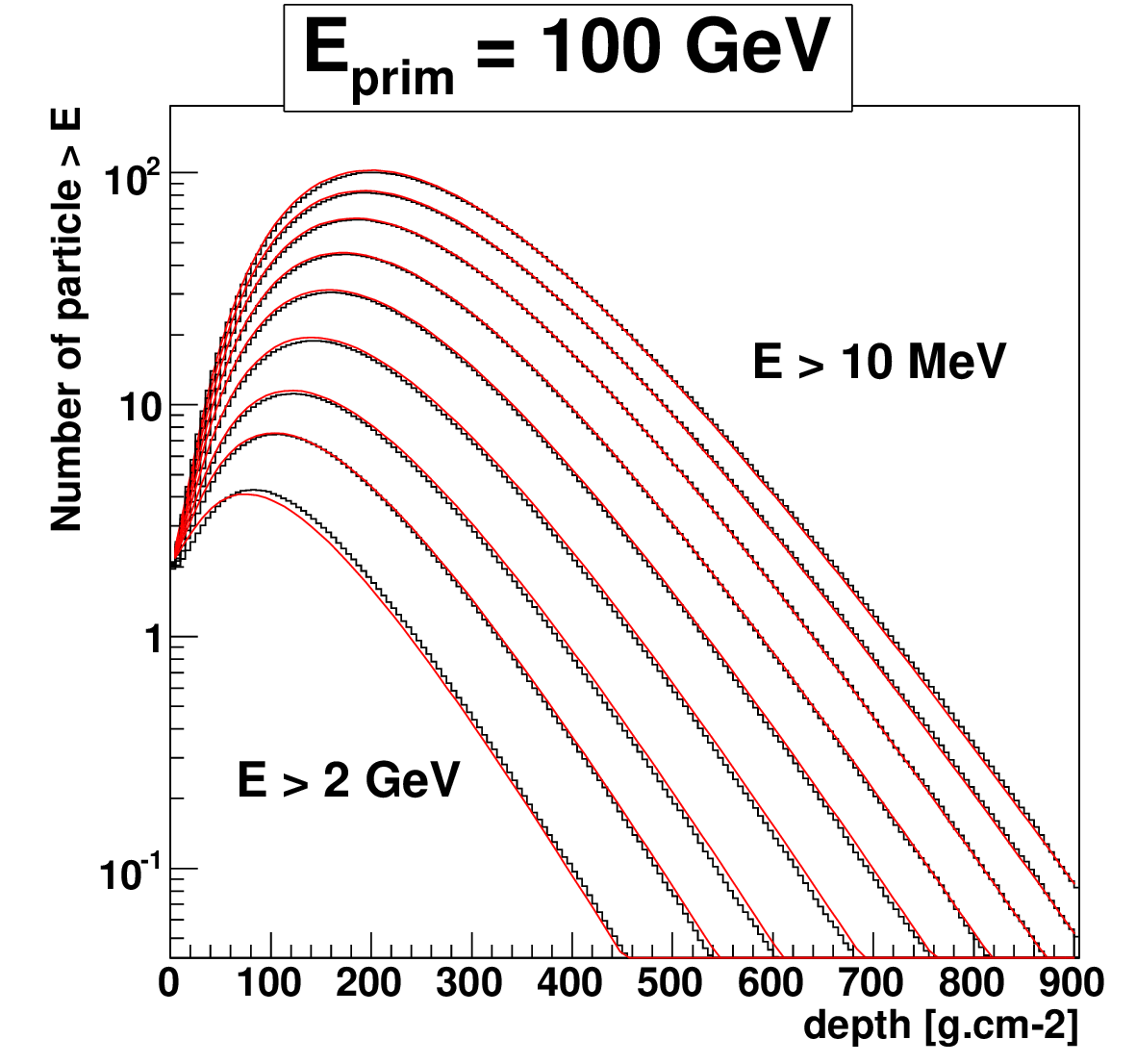,width=0.48\textwidth} \\
\epsfig{file=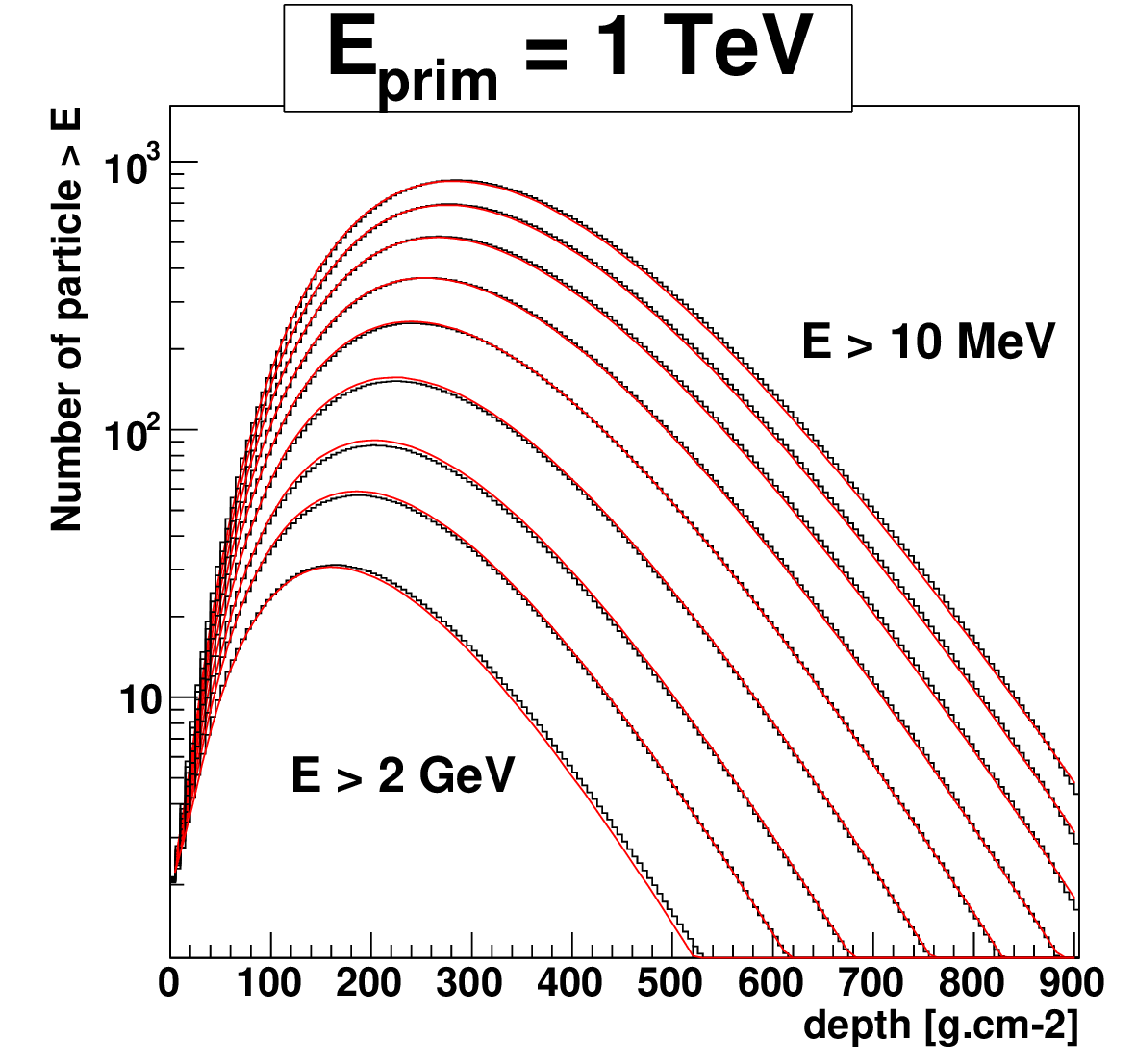,width=0.48\textwidth} & \epsfig{file=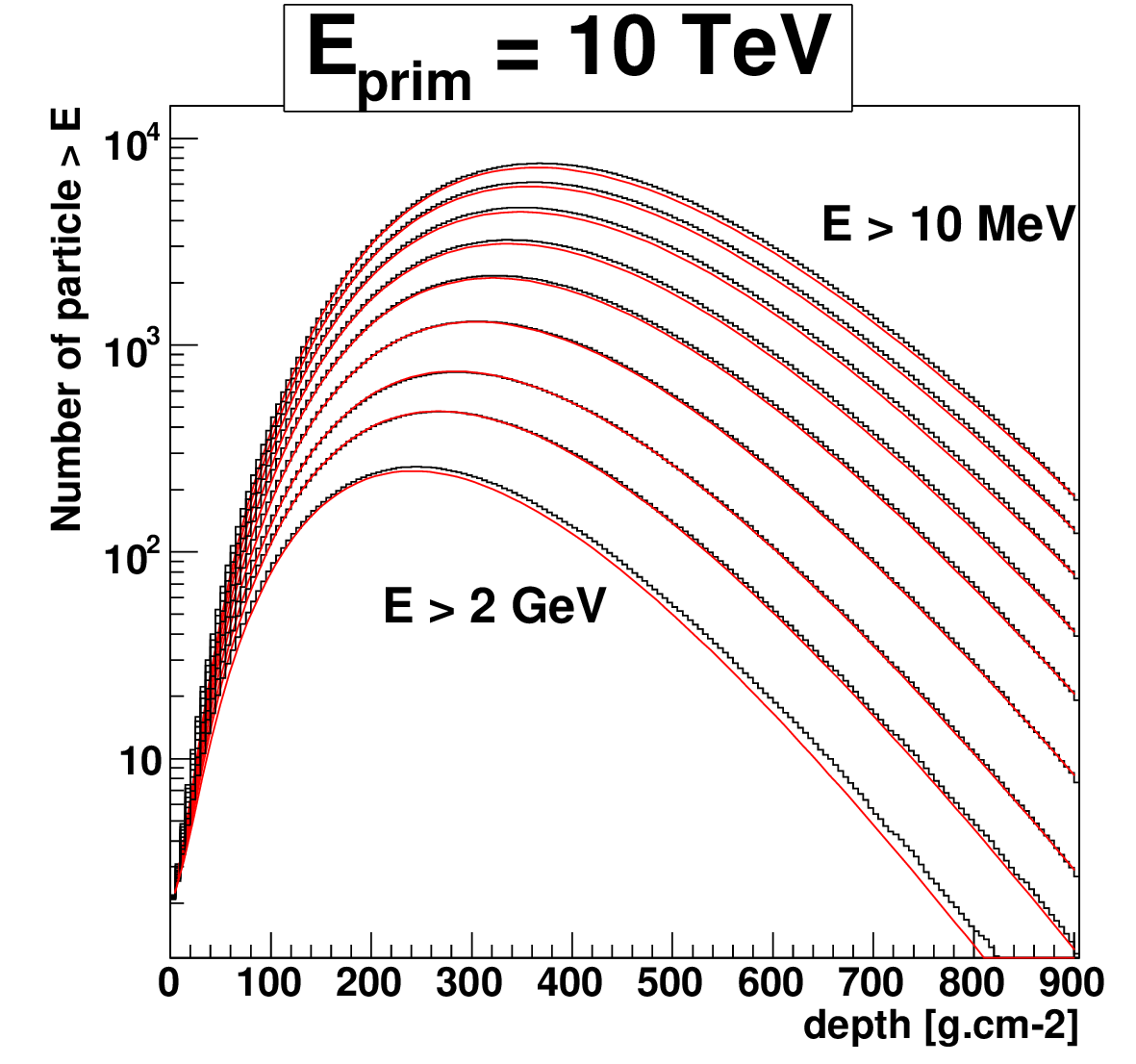,width=0.48\textwidth} \\
\epsfig{file=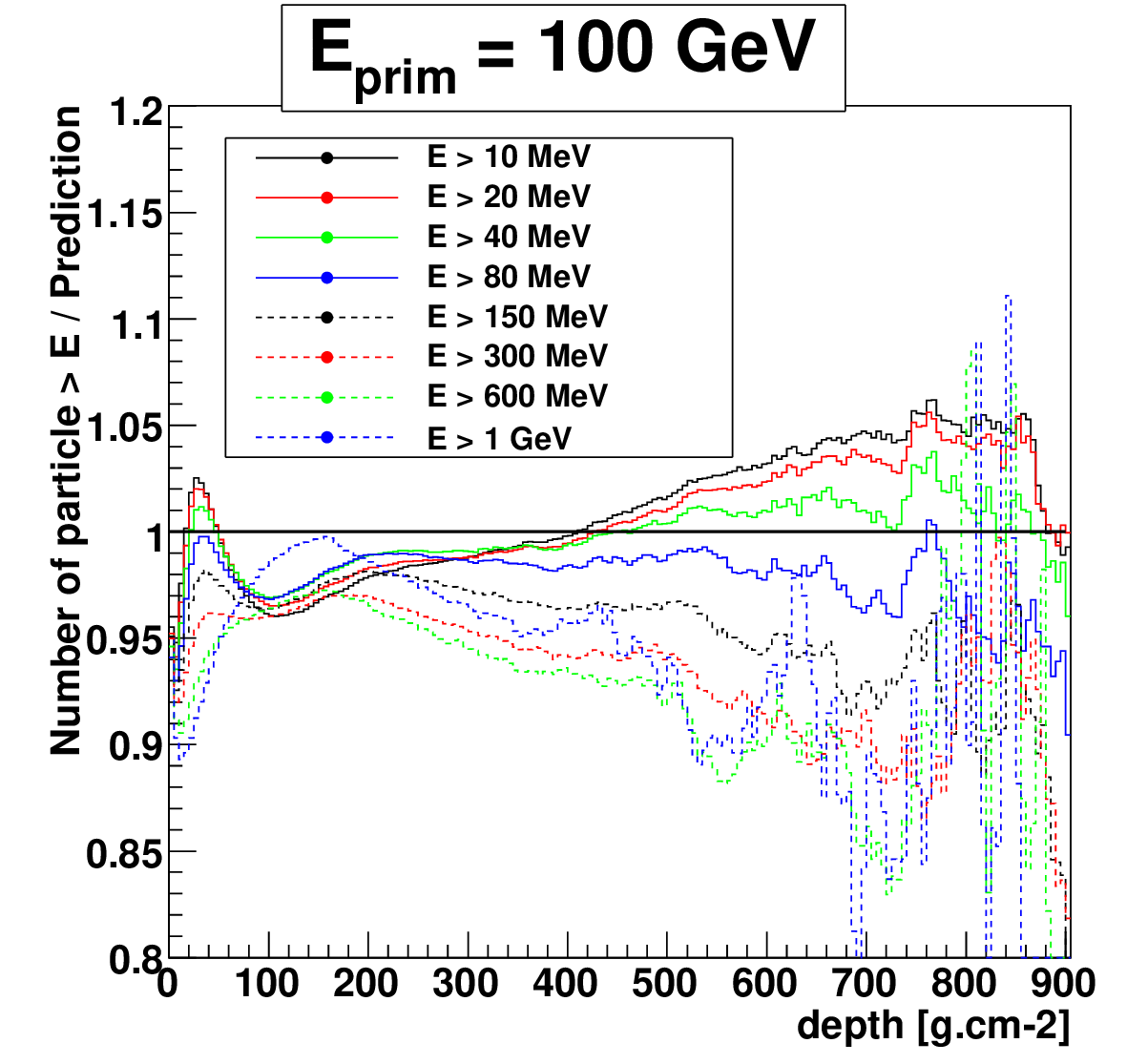,width=0.48\textwidth} & \epsfig{file=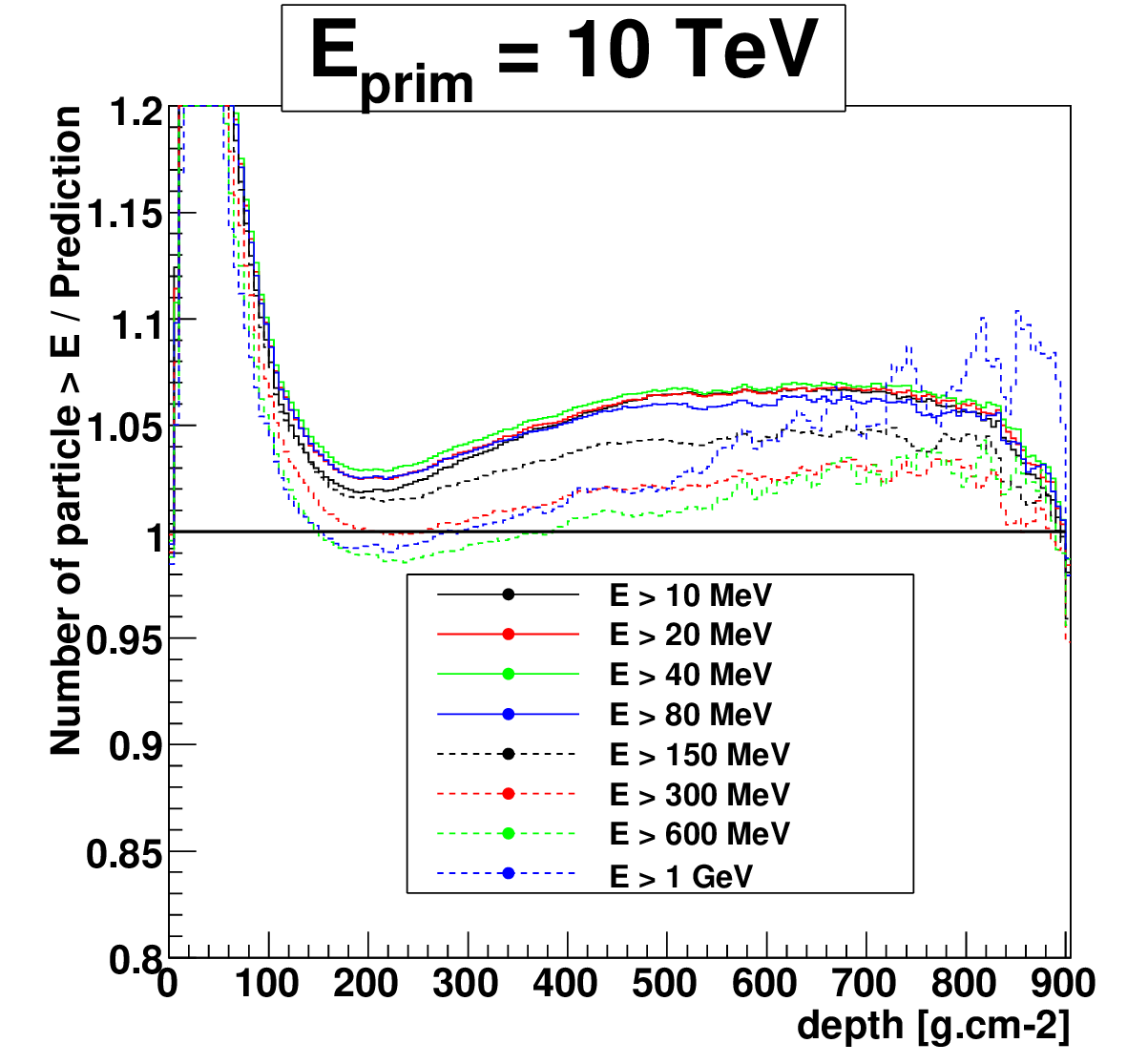,width=0.48\textwidth} 
\end{tabular}
\caption{\label{fig:EDist}Top and Middle: Energy dependent longitudinal shower development for primary energies of
(top to bottom and left to right) 10~GeV, 100~GeV, 1~TeV and 10~TeV, compared to analytical approximation 
from eq. \ref{eq:NeeEDeconvolvedParameters}(solid red lines). Each line gives the number of charged particles above
some energy (10~MeV, 20~MeV, 40~MeV, 80~MeV, 150~MeV, 300~MeV, 600~MeV, 1~GeV and 2~GeV from top to bottom) as function of atmospheric depth.
Bottom: Ratio of simulation to analytical prediction from eq. \ref{eq:NeeEDeconvolvedParameters}
for primary energies of 100~GeV and 10~TeV.}
\end{center}
\end{figure}

Examples of the longitudinal distributions of charged particles ${\D{\mathcal N}_e}/{\D t}(E \geq E_0)$  in the
shower integrated over charge particle energy $E$ above some threshold $E_0$
(10~MeV, 20~MeV, 40~MeV, 80~MeV, 150~MeV, 300~MeV, 600~MeV, 1~GeV and 2~GeV) and
for different primary photon energies is shown in
Fig.\,\ref{fig:EDist} . The distributions are modelled using the same
analytical expression as in Eq.\,\ref{eq:NeeDeconvolved} but with a
different set of parameters:


\begin{eqnarray}
a & = & (1.058 + 0.014 \times y) + 1.6\times 10^{-2} \times \left|\ln E_\mathrm{MeV} - 6\right|^{1.5}  \nonumber \\
b & = & 2.55 + 0.067 \times \ln E_\mathrm{MeV}\label{eq:NeeEDeconvolvedParameters}\\
c & = & 0.97 \times y - 1.43  + 0.137 \times \ln E_\mathrm{MeV} - (0.0712 + 0.0005 \times y) \ln^2 E_\mathrm{MeV}  \nonumber 
\end{eqnarray}

where $E_\mathrm{MeV}$ is the charged particle energy, expressed in MeV.

The comparison of the analytical function with the distributions from
simulations with different primary energies is shown in
Fig. \ref{fig:EDist}. The agreement is very good (at the level of
$\sim 5\%$) up to particle energies of a few GeV above which the
particles contribute very little to the overall Cherenkov light
distribution.

\subsection{Angular distribution of particles in a shower}


The angular distribution of charged particles in the shower together
with the velocity dependent Cherenkov-angle determine the angular
distribution of the Cherenkov-photons. As described in Moliere theory,
the relevant angle of the charged particles is the reduced angle $w$
given as

\begin{equation}
w = 2(1-\cos\theta) \times \left(\frac{E}{21\,\mathrm{MeV}}\right)^2 \approx \left(\frac{\theta E}{21\,\mathrm{MeV}}\right)^2 
\end{equation}

where $\theta$ is the angle between the direction of the primary
particle and the direction of the charged particle of energy $E$ in
the shower. The angular distribution of the charged particles in the
shower is decomposed into the mean reduced angle $\left\langle w
\right\rangle$ as a function of the particle energy of the charged
particle and the distribution of reduced angles around the mean
reduced angle.

The mean reduced angle is found to depend only very weakly on the
primary particle energy and is parametrised by

\begin{equation}
\left\langle w\right\rangle = \frac{0.435}{1 + \displaystyle\left(\frac{111}{E_\mathrm{MeV}}\right)^{0.92}}\label{eq:MeanW}
\end{equation}

\begin{figure}[ht]
\begin{center}
\epsfig{file=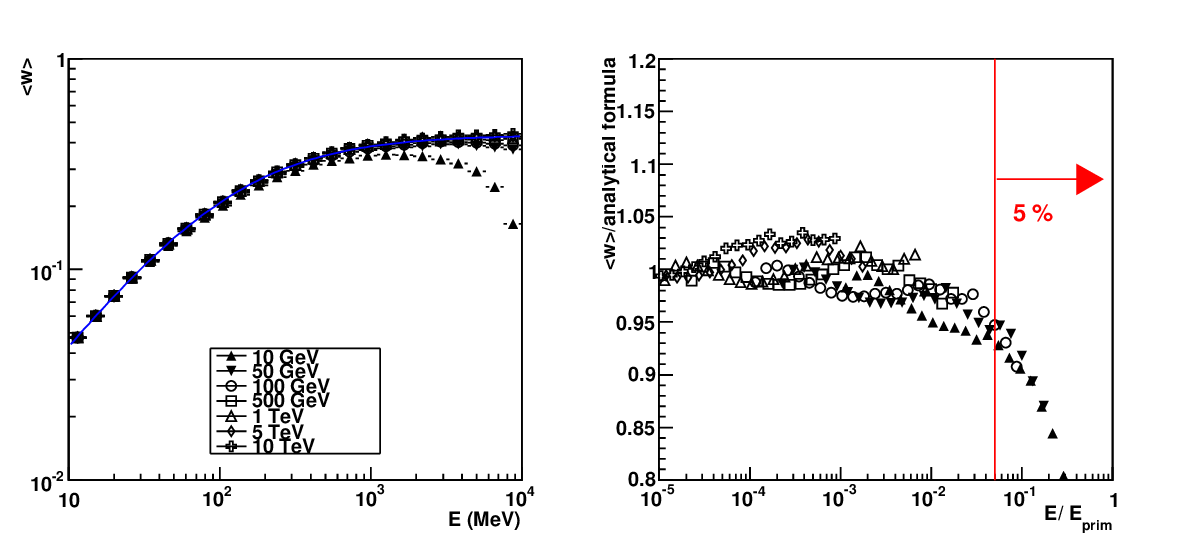,width=0.95\textwidth}
\end{center}
\caption{\label{fig:MeanW}Left: average value of the reduced angle
  cosine $\left\langle w\right\rangle$ as a function of charged
  particles energy. The solid line is the model parametrisation from
  equation \ref{eq:MeanW}.  Right: Ratio between simulation
  distribution and model prediction as function of charged particles
  kinetic energy (scaled to primary particle energy).}
\end{figure}

The mean reduced angle for different primary particle energies as a
function of the charged particle energy in units of the primary
particle energy together with the parametrisation is shown in
Fig.\,\ref{fig:MeanW}. There is a good overall agreement within a few
percent between the mean reduced angle from the simulation and the
parametrisation, in the range of small kinetic energies compared to
the primary particle energy. Fig.\,\ref{fig:MeanW} right shows the
ratio between the simulation and the parametrisation. A deviation is
seen when the charged particle kinetic energy exceeds about $5\%$ of
the primary particle energy (corresponding to a red vertical line in
fig. \ref{fig:MeanW}, right), affecting only a very small fraction of
the particles in the shower.

The detail modelling of the Cherenkov light distribution in a shower
requires the dependency of $\left\langle w\right\rangle$ on the shower
age $s$ to be taken into account.  The parametrisation of
eq. \ref{eq:MeanW} is kept, but with parameters varying with shower
age:

\begin{equation}
\left< w\right>(s) = \frac{p_0(s)}{1 + \displaystyle\left(\frac{p_1(s)}{E_\mathrm{MeV}}\right)^{p_2(s)}}\label{eq:MeanWS}
\end{equation}

\begin{figure}[ht]
\begin{center}
\begin{tabular}{ccc}
\epsfig{file=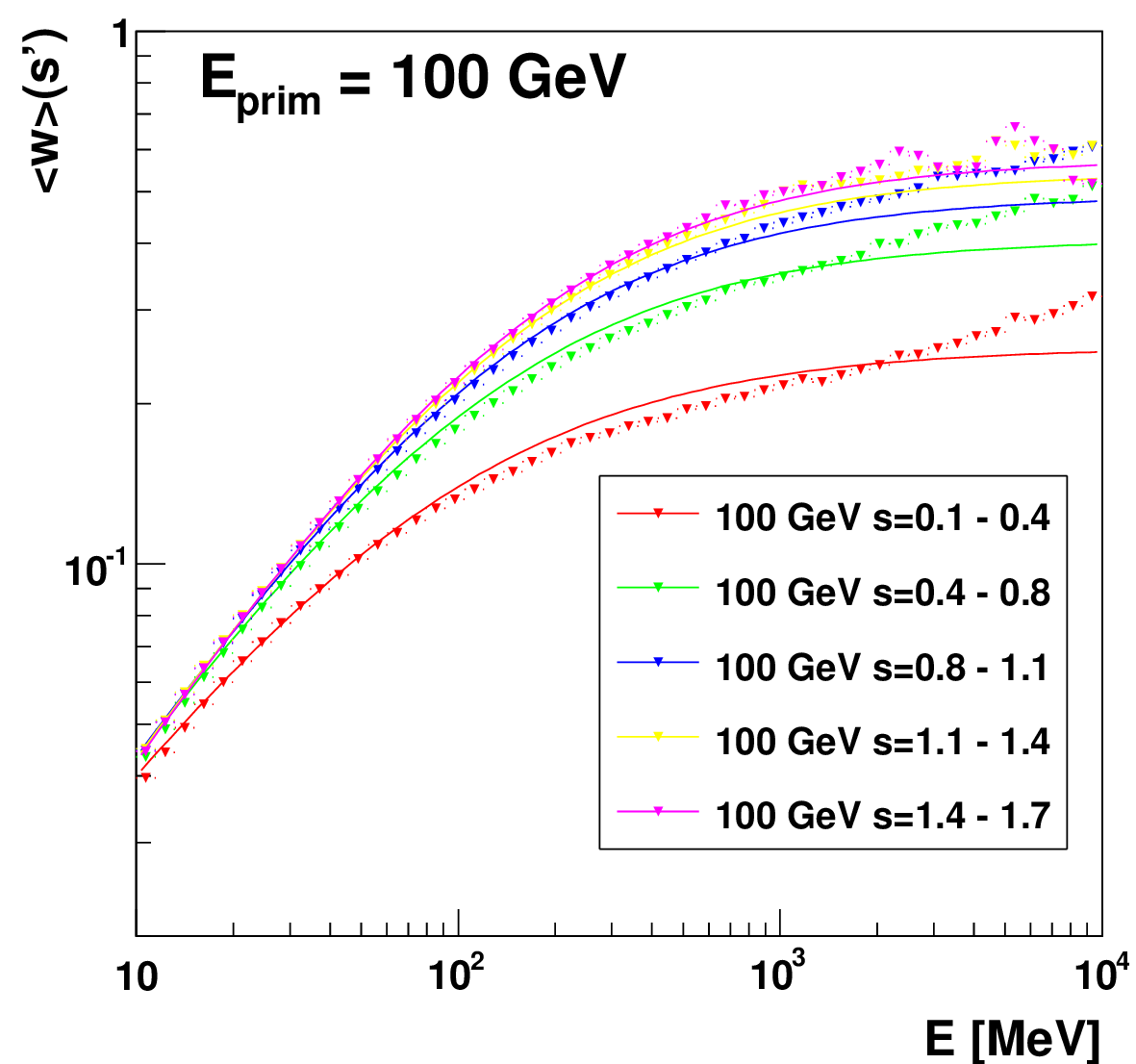,width=0.32\textwidth} & \epsfig{file=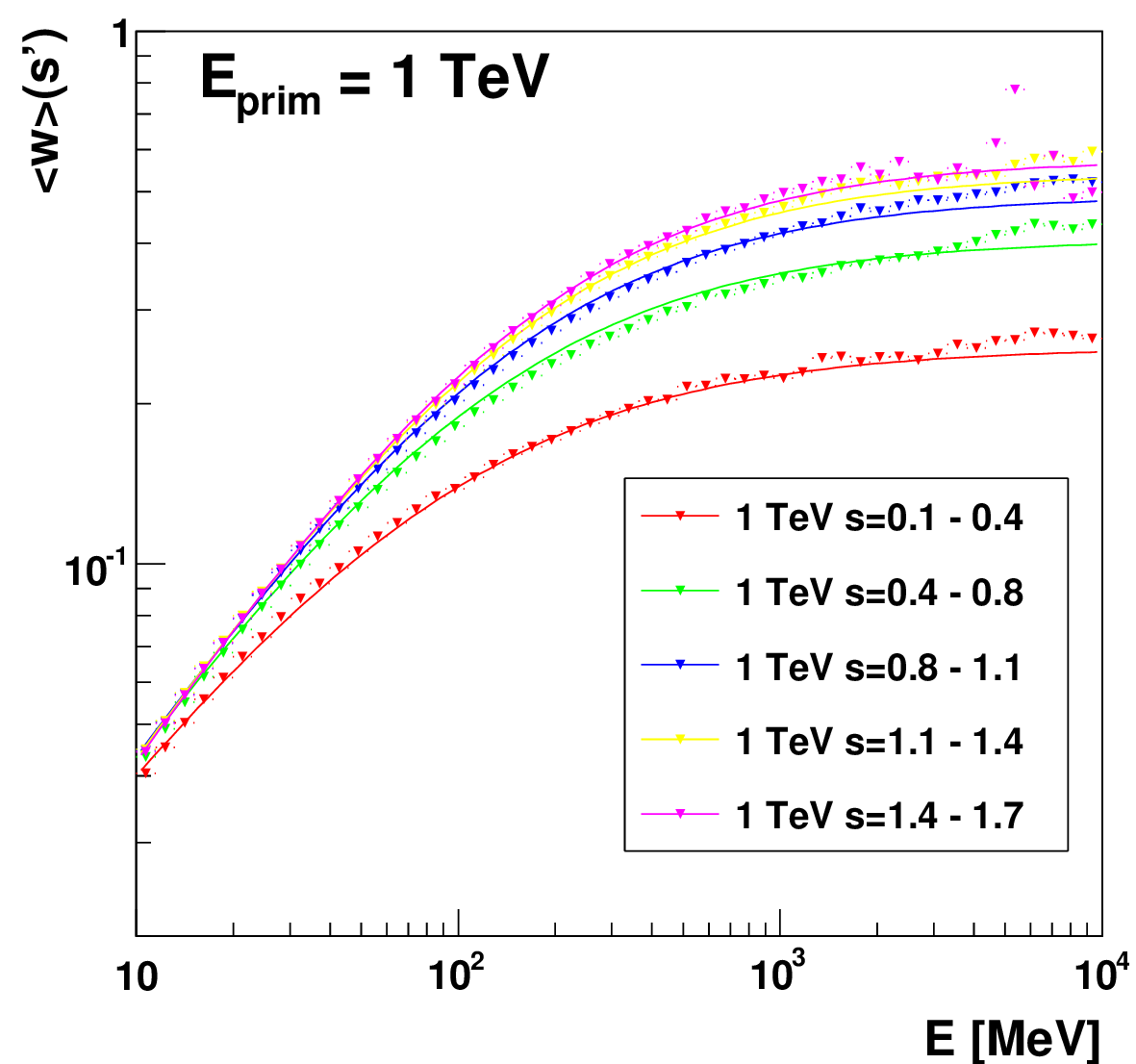,width=0.32\textwidth} & \epsfig{file=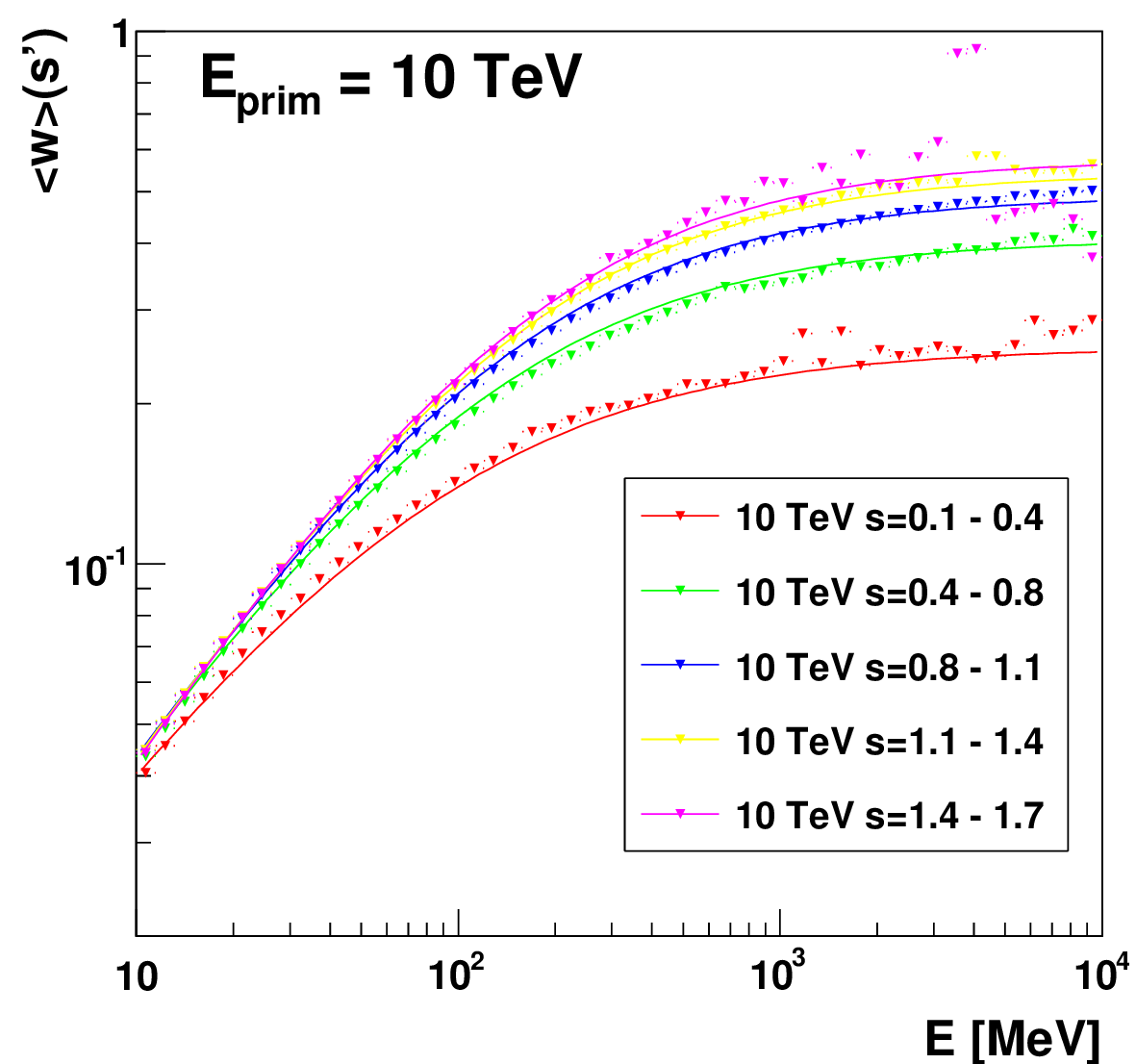,width=0.32\textwidth} \\
\end{tabular}
\end{center}
\caption{\label{fig:MeanWS}Mean value of scaled angle $\left\langle w\right\rangle(s)$ as function of particle energy and shower age, 
for different primary particle energies ($100~\mathrm{GeV}$ to $10~\mathrm{TeV}$ from left to right). 
The solid lines correspond to the parametrisation of eq. \ref{eq:MeanWS}. The average scaled particle
angle is found to be mostly independent of primary particle energy.}
\end{figure}

The parameters $p_0$, $p_1$ and $p_2$ are given by
(fig. \ref{fig:MeanWS}):
\begin{eqnarray}
p_0(s) &=& 0.506 \times \exp(0.351 \times \ln s - 0.147 \times \ln^2 s) \nonumber \\
p_1(s) &=& 144 \times \exp(0.33 \times \ln s - 0.11 \times \ln^2 s) \\
p_2(s) &=& 0.887 \times \exp(0.0507 \times \ln s - 0.014  \times \ln^2 s) \nonumber 
\end{eqnarray}

The distribution of angles around the mean is found to be quite
independent of shower age or primary particle energy. The {\it
  rescaled} angle $u$ is used here:

\begin{equation}
u = \frac{w}{\left\langle w\right\rangle}
\end{equation}

\begin{figure}[ht]
\begin{center}
\begin{tabular}{cc}
\epsfig{file=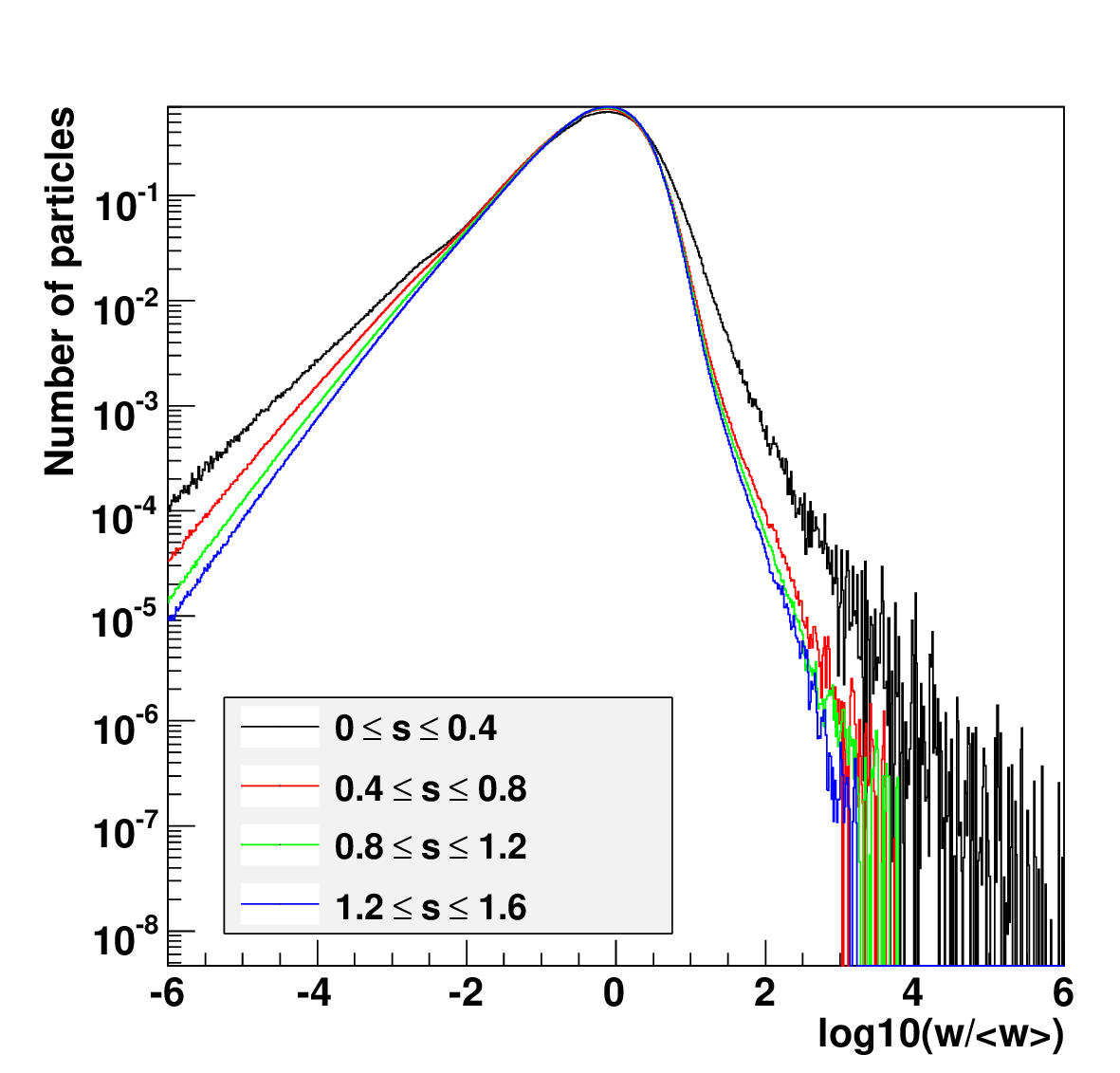,width=0.48\textwidth} &
\epsfig{file=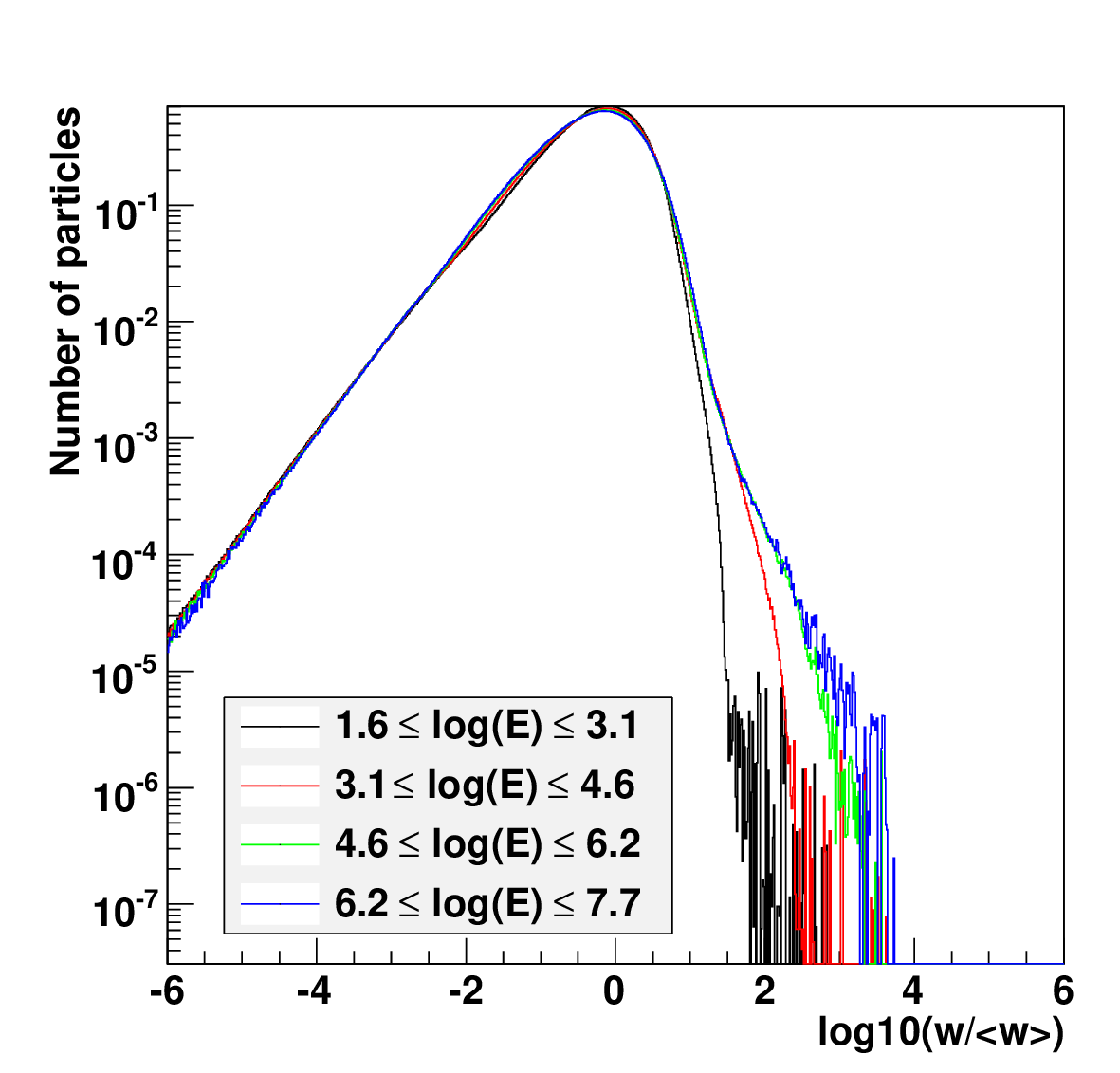,width=0.48\textwidth}
\end{tabular}
\end{center}
\caption{\label{fig:U_vs_SE}Left: Variation of the distribution of $\log_{10} u$ with shower age (for 1~TeV primaries). 
Right: Variation of the distribution of $\log_{10} u$ with particle energy.
The distribution is slightly broader at small shower ages and at large energies, however the discrepancy is not
very significant. Moreover, these regions of the parameters space correspond to a small number of charged particulars.}
\end{figure}

The distribution of $d\log_{10} (u)$ for varying shower age and particle energy is shown
in fig.~\ref{fig:U_vs_SE}, and is found to be almost independent of these parameters.
There is a slight broadening of the distribution at small shower ages,
but this corresponds to the beginning of the shower when the number of charge particles is small. 

\begin{figure}[ht]
\begin{center}
\epsfig{file=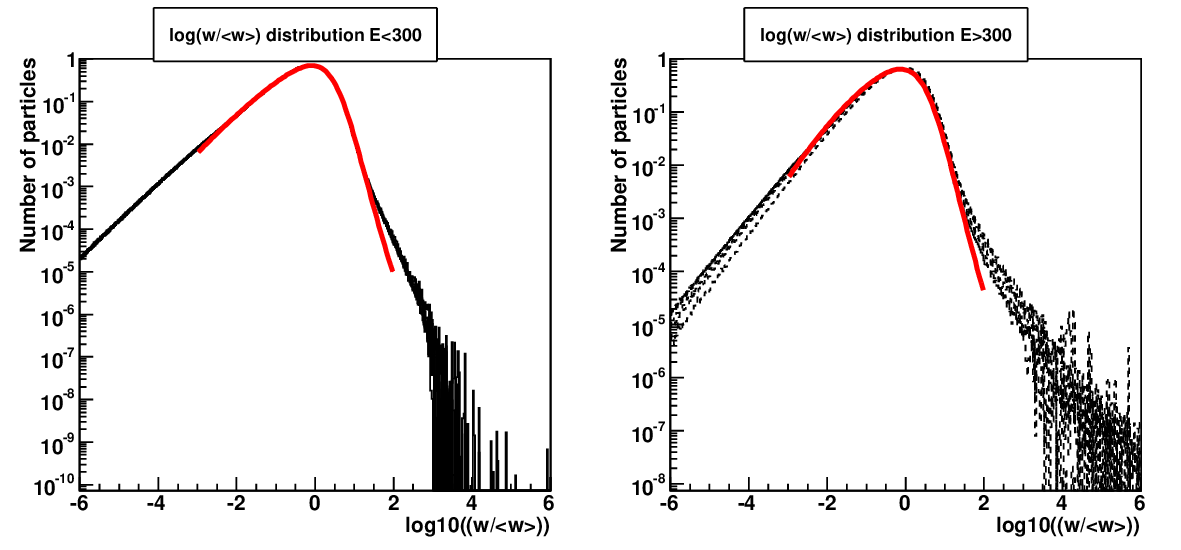,width=0.95\textwidth}
\end{center}
\caption{\label{fig:WDistribution}Distribution of rescaled angle $u$ for different primary energies
ranging from $10~\mathrm{GeV}$ to $10~\mathrm{TeV}$, and for charged particle energies of
$E \leq 300~\mathrm{MeV}$ (left) and $E > 300~\mathrm{MeV}$ (right). }
\end{figure}

The distribution of $d\log_{10} (u)$, assumed to be independent of shower age or primary particle energy, 
is shown in fig.~\ref{fig:WDistribution} and can be modelled by the expression\footnote{there was a typo in published paper}

\begin{equation}
\frac{\D N}{\D \log_{10} u} = A \times \exp\left[ - \frac{1}{2}\times \left(p_0 + p_1 \times \log_{10} u  + p_2\times \arctan(\log_{10} u - p_3) \right)^2 \right] \label{eq:dNdlogu}
\end{equation}

with:

\begin{eqnarray}
p_0 = 1.55, \quad p_1 = 0.29, \quad p_2 = 2.5, \quad p_3 = 0.73 \quad &\hbox{for}&\quad E\leq 300\,\mathrm{MeV} \nonumber \\
p_0 = 1.16, \quad p_1 = 0.50, \quad p_2 = 1.8, \quad p_3 = 0.57 \quad &\hbox{for}&\quad E\geq 300\,\mathrm{MeV} 
\end{eqnarray}

The distribution $dN/d\log_{10} (u)$ for different primary energies and for charged particles energies $E\leq 300\,\mathrm{MeV}$ (left) and
$E\geq 300\,\mathrm{MeV}$ (right) is shown in fig.~\ref{fig:WDistribution} with the analytical formula from eq. \ref{eq:dNdlogu} 
superimposed (solid red line). A good agreement up to a few $\%$ is observed in the central part of the distribution,
encompassing the majority of particles in the shower.

\subsection{Lateral distribution of particles in a shower}

\begin{figure}[ht]
\begin{center}
\begin{tabular}{cc}
\epsfig{file=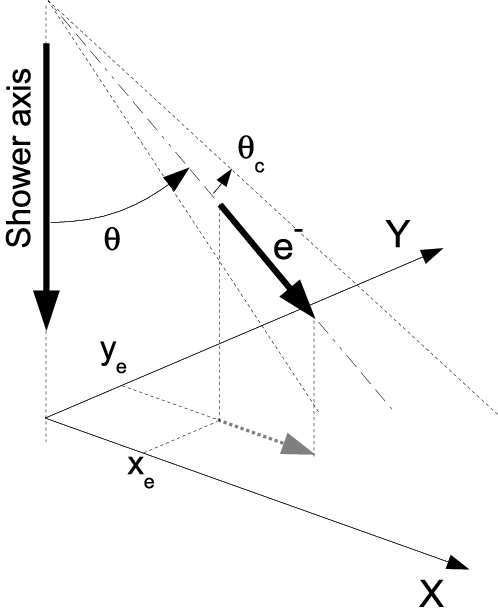,width=0.4\textwidth} &
\end{tabular}
\end{center}
\caption{\label{fig:Coordinate}Definition of coordinate system attached to each electron/positron in the shower.}
\end{figure}

The lateral distribution of the charged particles (electron or positron) in a shower are
given in a system of coordinates attached to each charged particle in
the shower as introduced in \cite{Hillas}. The $X$-direction is defined as the projection of the particle flight 
direction on the plane perpendicular to the shower axis, and the $Y$ direction is perpendicular to it (and therefore
perpendicular to the particle velocity) (see
fig. \ref{fig:Coordinate}).  The mean particle displacement $\langle
x_e \rangle$ with respect to the shower axis along the $X$ axis is
non-zero, whereas the average displacement along the $Y$ axis $\langle
y_e \rangle$ is zero for symmetry reasons.


Assuming a factorisation of the expressions of mean displacement and spread in terms of
energy, particle angle and shower age, the following expressions are found to properly describe the particle spread 
(expressed in units of $\mathrm{g}\times\mathrm{cm}^{-2}$)\footnote{In the publised paper, eq \ref{eq:SigmaX} had a typo.}:
 
\begin{eqnarray}
\langle x_e \rangle &=& 24.3 \times \exp(0.63 \times \log w + 0.025 \times \log^2 w ) \nonumber \\
&& \times (1 + 0.2 \times s^{4.6}) \times \Big(1 - \exp(- 0.11 \times s^{0.7} ) \Big) \label{eq:MeanX}\\
&& \times \frac{21}{E_\mathrm{MeV}} \times \exp(0.47 \times \log E_\mathrm{MeV} - 0.023 \times \log E_\mathrm{MeV}^2) \nonumber \\
\noalign{\vspace{0.5em}}
\sigma_{x_e} &=& 28.7 \times \exp(0.085 \times \log w) \times (1 + 0.9 \times w^{0.6}) \nonumber \\
&& \times (1 + 0.37 \times s^{3.9}) \times \Big(1 - \exp(- 0.03 \times s^{0.8} ) \Big) \label{eq:SigmaX}\\
&& \times \frac{21}{E_\mathrm{MeV}} \times \exp(0.55 \times \log E_\mathrm{MeV} - 0.028 \times \log E_\mathrm{MeV}^2) \nonumber \\
\noalign{\vspace{0.5em}}
\langle y_e \rangle &=& 0 \label{eq:MeanY}\\
\noalign{\vspace{0.5em}}
\sigma_{y_e} &=& 2.65 \times \exp(0.03 \times \log w) \times (1 + 0.51 \times w^{0.5}) \nonumber \\
&& \times (1 + 0.2 \times s^{4.5}) \times \Big(1 - \exp(- 0.2 \times s ) \Big) \label{eq:SigmaY}\\
&& \times \frac{21}{E_\mathrm{MeV}} \times \exp(0.675 \times \log E_\mathrm{MeV} - 0.035 \times \log E_\mathrm{MeV}^2) \nonumber 
\end{eqnarray}

The above expression is a good description of the mean lateral
displacement and spread at the level of 10\%.

\begin{figure}[ht]
\begin{center}
\epsfig{file=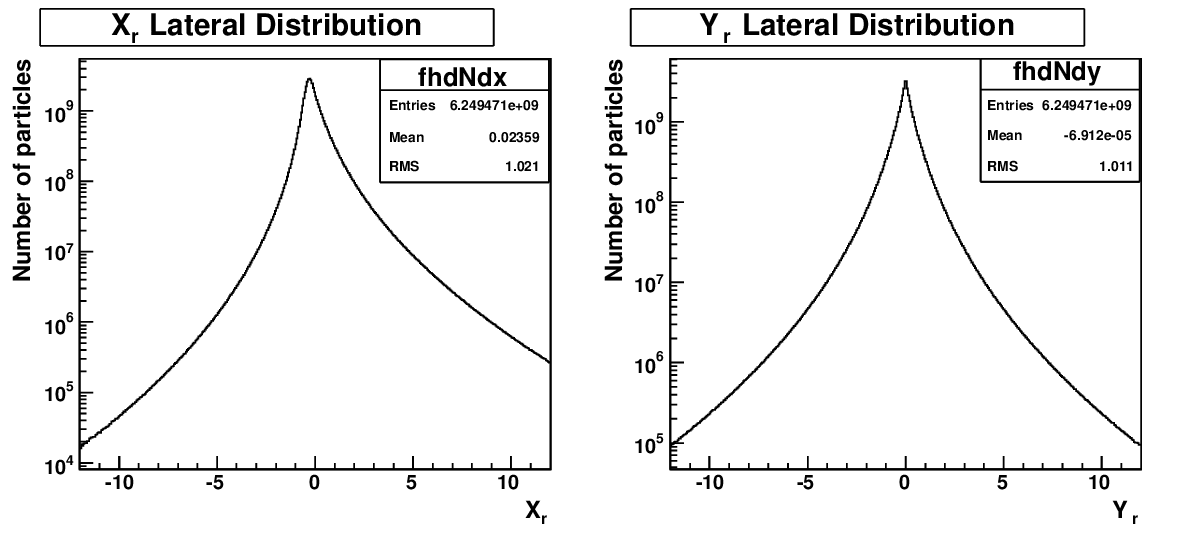,width=0.95\textwidth}
\end{center}
\caption{\label{fig:ReducedLateralDistri}Distributions of reduced lateral displacement
$X_r$ and $Y_r$ (eq. \ref{eq:ReducedLateralDistri}) for 1~TeV showers, along X and Y axis. As expected,
the distribution along Y axis is symmetrical while the distribution along the X axis is not.}
\end{figure}

The reduced variables $X_r$ and $Y_r$ are used for simplicity:

\begin{equation}
X_r = \frac{x_e - \langle x_e \rangle}{\sigma_{x_e}}, \quad Y_r = \frac{y_e}{\sigma_{y_e}} \label{eq:ReducedLateralDistri}
\end{equation}

were $\langle x_e \rangle$,  $\sigma_{x_e}$ and $\sigma_{y_e}$ are obtained from eq. \ref{eq:MeanX} to \ref{eq:SigmaY}.
The distributions of $X_r$ and $Y_r$ for 1~TeV gamma-ray showers are shown in fig.~\ref{fig:ReducedLateralDistri}.
The average values and RMS of $X_r$ and $Y_r$ are respectively $0$ and $1$, as expected if equations
\ref{eq:MeanX}-\ref{eq:SigmaY} are correct. As expected, the $Y_r$ distribution is symmetric while the $X_r$ distribution is not.

The $Y_r$ distributions can be modelled by the following expression:

\begin{equation}
\frac{\D N}{\D Y_r} = A \times \exp\left(-2.5 \frac{Y_r^{2}}{0.2 + \left|Y_r\right|^{1.4}}\right) \label{eq:LateralDistriY}
\end{equation}

A similar expression can be derived for $X_r$ taking into account the non-centred
position of the distribution maximum, and with different coefficients on both sides.
For the sake of simplicity, a tabulated version of this distribution is used in the model production
instead of a complicated analytical expression.


\begin{figure}[ht]
\begin{center}
\begin{tabular}{cc}
\epsfig{file=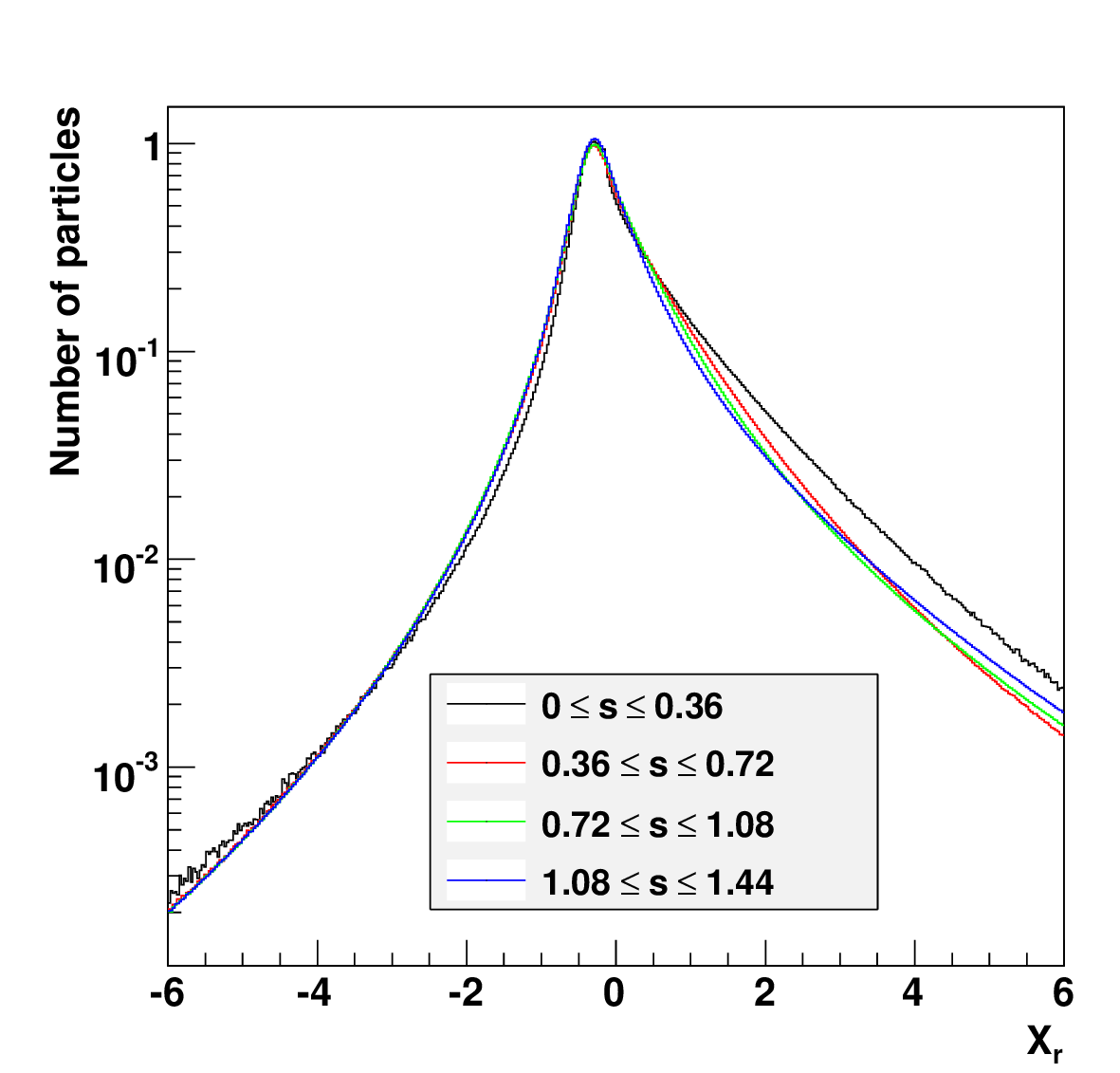,width=0.48\textwidth} & \epsfig{file=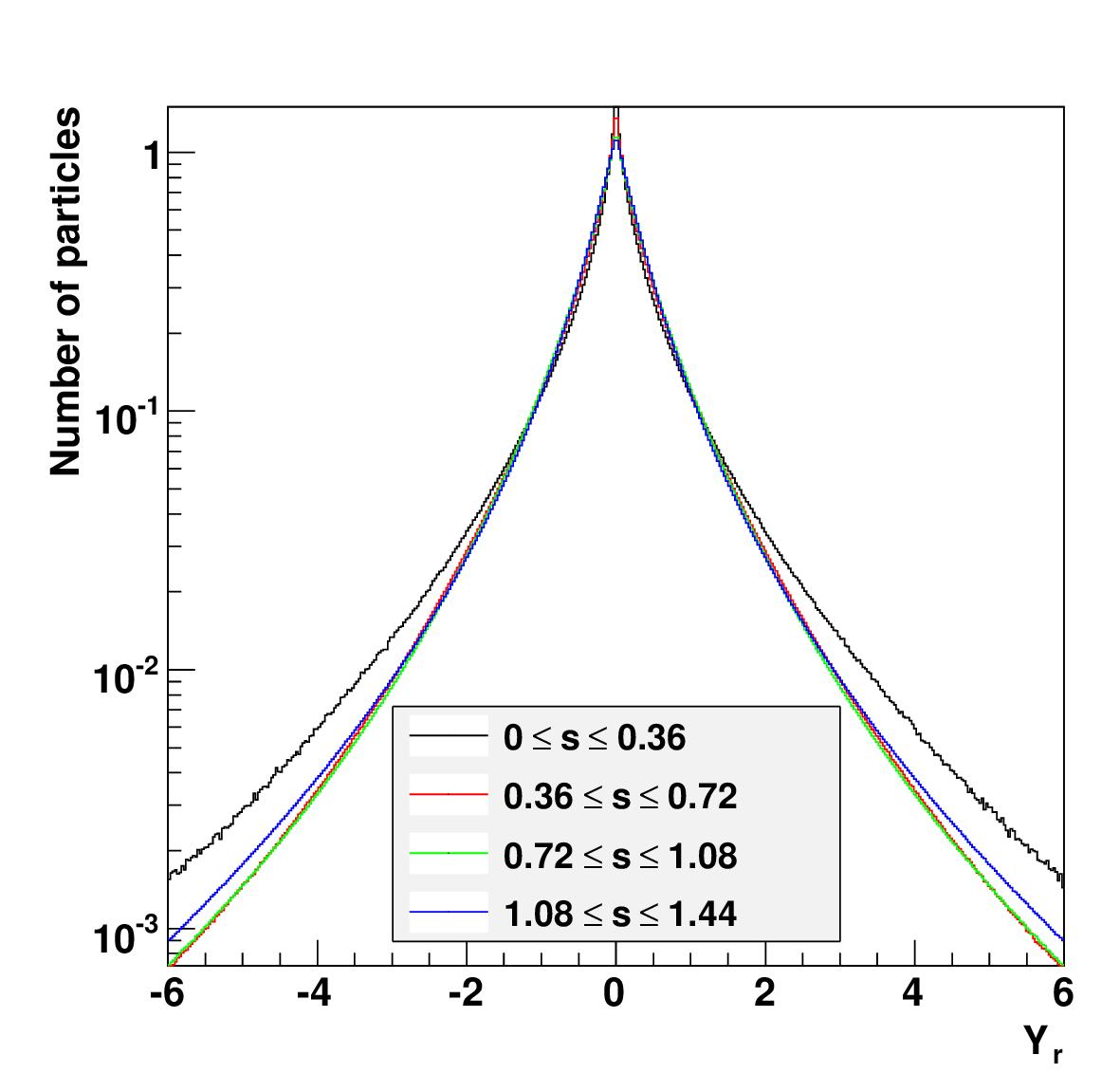,width=0.48\textwidth}\\
\epsfig{file=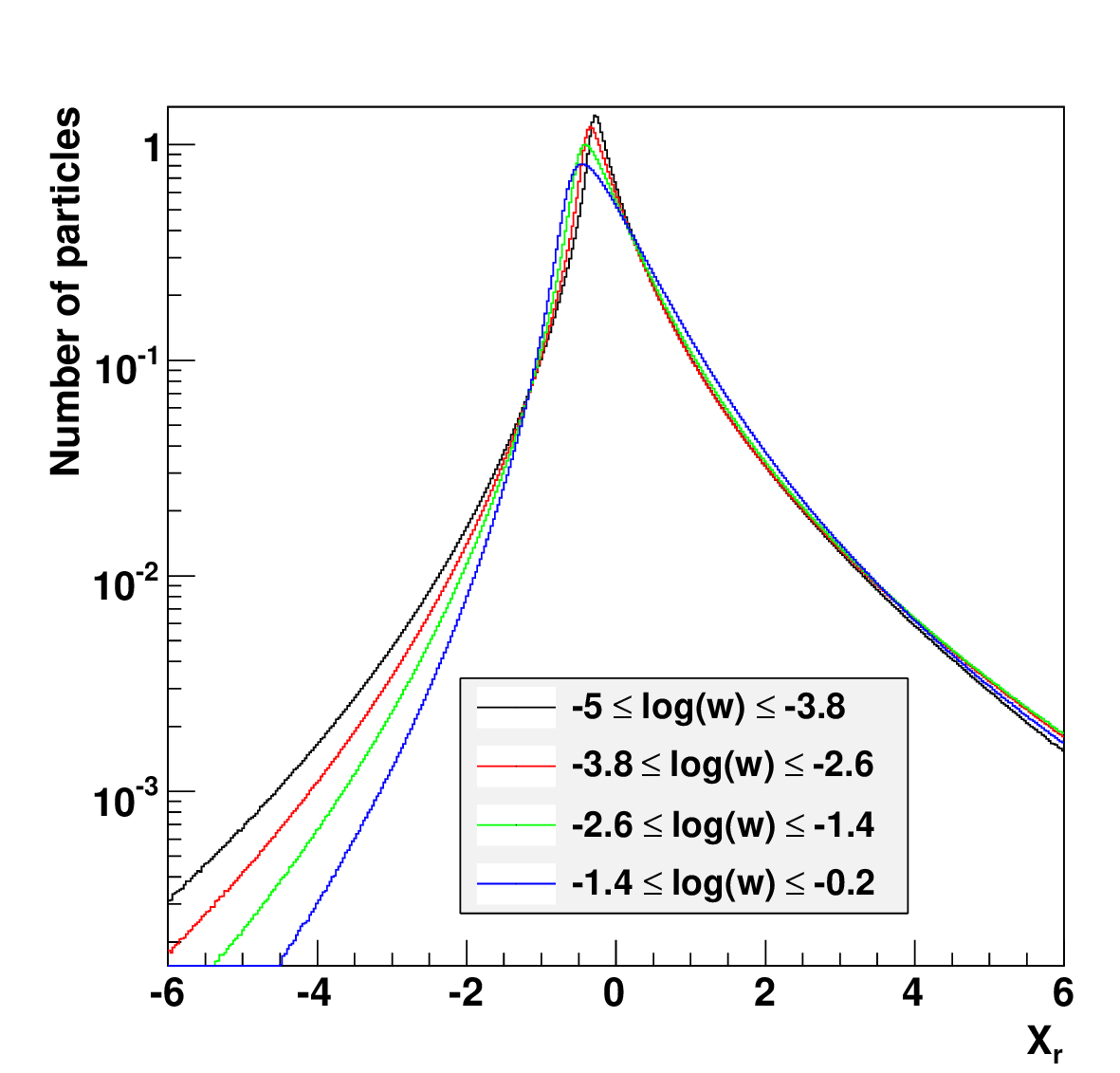,width=0.48\textwidth} & \epsfig{file=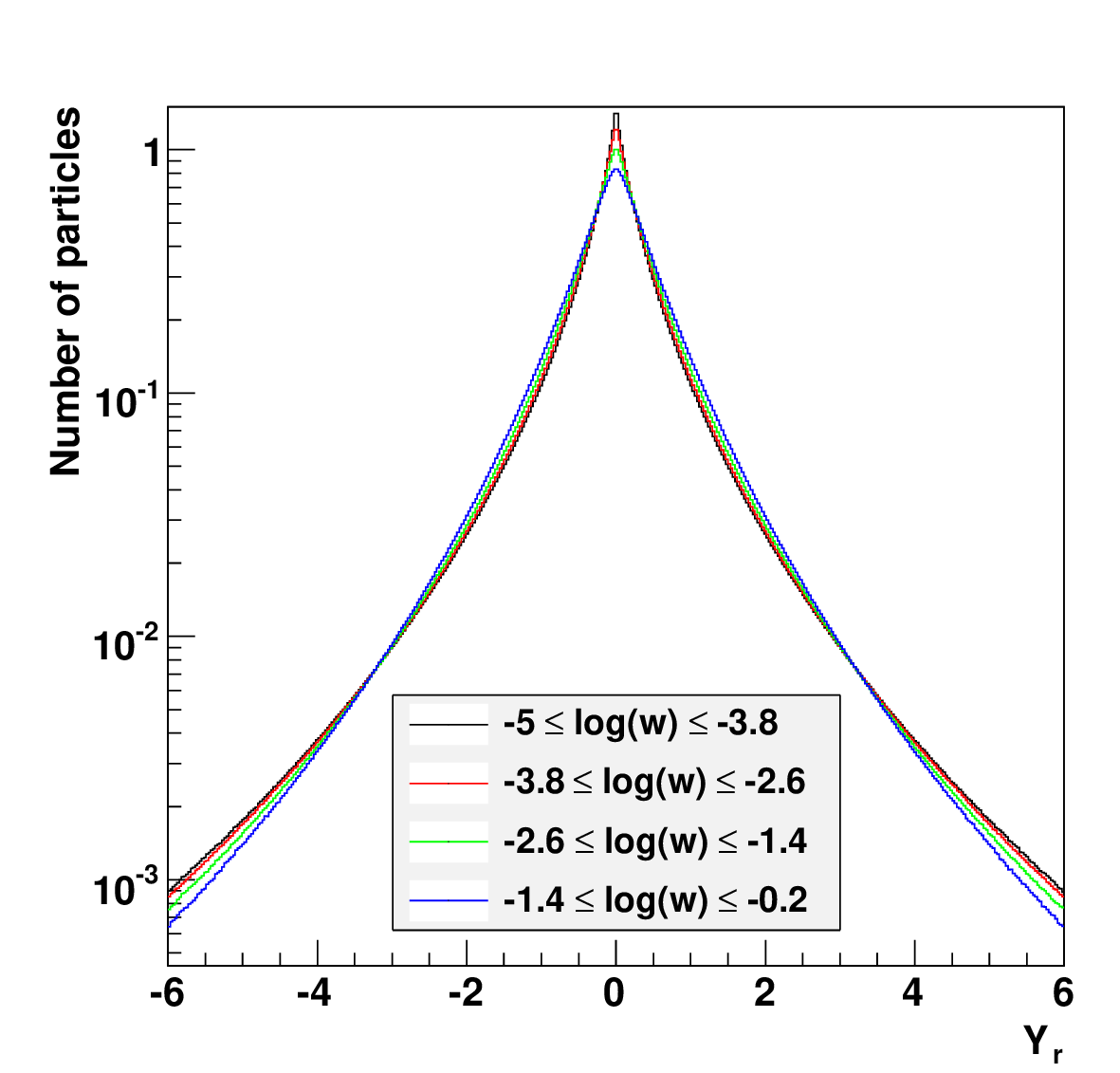,width=0.48\textwidth}\\
\end{tabular}
\end{center}
\caption{\label{fig:ReducedLateralDistri_vs_ESW}Evolution of the reduced lateral distributions
($X_r$: left, $Y_r$:right) with shower age (top) and reduced particle angle (bottom).
The strongest dependency is seen in the dependency of $X_r$ on the reduced angle $w$ and remains acceptable.}
\end{figure}

The dependency of the reduced $X_r$ and $Y_r$ lateral distributions with
shower age and particle angle is shown in
fig. \ref{fig:ReducedLateralDistri_vs_ESW}. The $X_r$ and $Y_r$ distributions remain
stable enough to be considered as independent of these parameters in the shower modelling.
The small variation seen in the distribution of $X_r$ with particle angle affects
mainly particles with a very large angle which are anyhow not likely to reach the telescopes.

\begin{figure}[ht]
\begin{center}
\epsfig{file=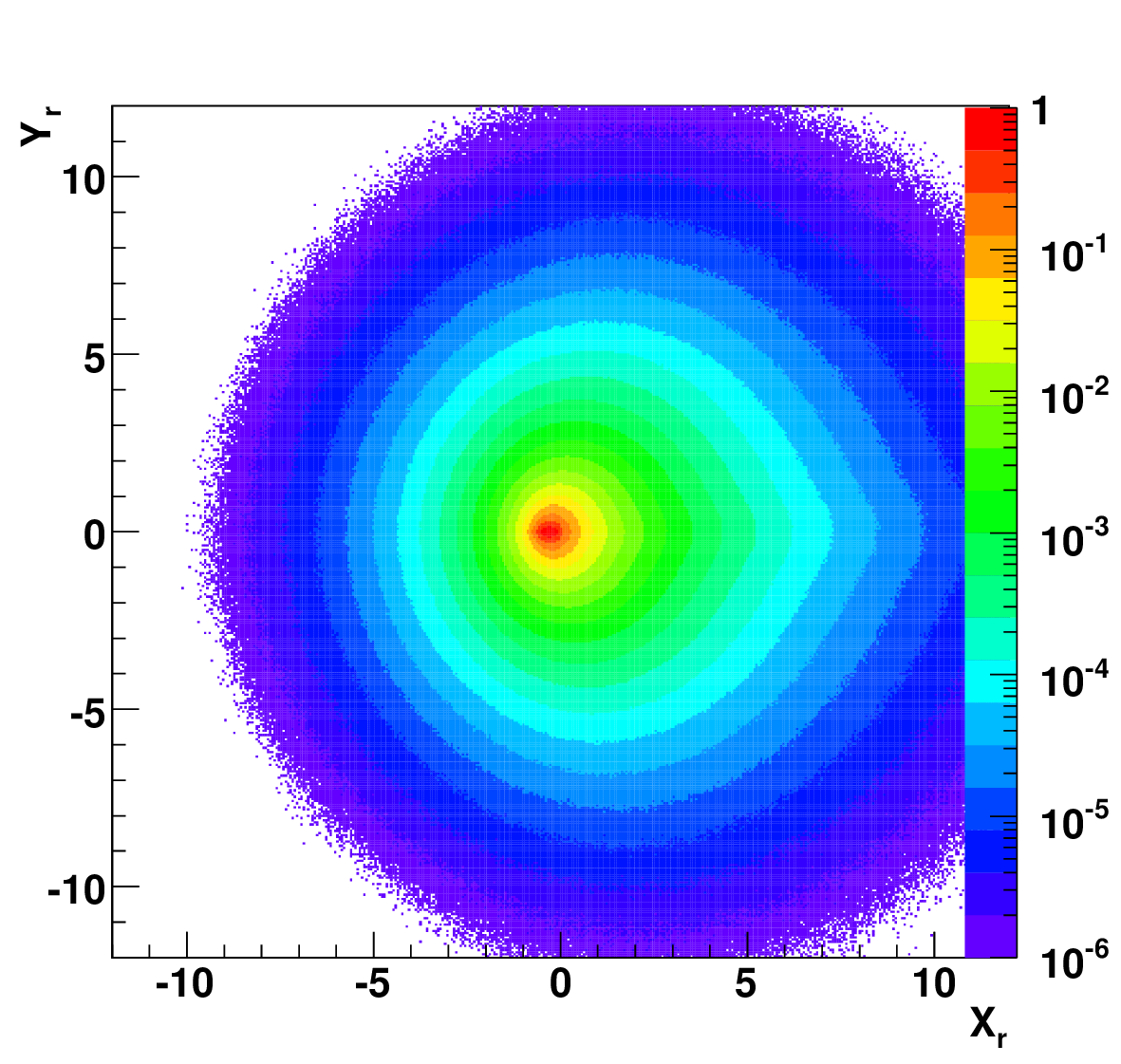,width=0.9\textwidth}
\end{center}
\caption{\label{fig:XYDistri}Bi-dimensional reduces lateral position distribution.
There is a little correlation between the $X_{r}$ and the $Y_{r}$ distribution.
}
\end{figure}

Figure \ref{fig:XYDistri} shows that the $X_r$ and  $Y_r$ distributions
are mostly uncorrelated and can therefore be considered as independent.

\section{Model Generation}\label{sec:generation}

The semi-analytical description of shower development described in the previous section
can be used to construct a {\it shower model}, i.e. a prediction of the light distribution
on the ground for a given primary particle energy, direction, impact parameter and development
depth. This section describes how the shower image model is constructed from the distributions derived in 
the previous section.

\subsection{Principles}

The light density due to a shower in the camera can be calculated by an eight-fold integral:

\begin{itemize}
\item integral over altitude $z$ or depth $t$ (longitudinal development of shower).
\item integral over energy of the electron/positron in shower.
\item integral over electron direction with respect to the telescope ($u$ and $\phi$).
\item integral over electron position with respect to its direction ($X_r$ and $Y_r$).
\item integral over Cherenkov photon wavelengths.
\item integral over Cherenkov photon azimuthal angle around the electron (the angle between the electron
and the Cherenkov photon being fixed for a given electron energy).
\end{itemize}

\begin{eqnarray}
I(x,y) &=& \int \D z \int \D E \times \frac{\D{\mathcal N}_e}{\D E}(t,E) \times \frac{\D t}{\D z}(y) \nonumber \\
&& \int \D u \times F_u(u(E,s)) \int \frac{\D \phi}{2\pi}  \\
&& \int \D X_r\int \D Y_r F_{XY}(X_r,Y_r,E,s,u) \nonumber\\
&& \int \D \phi_{ph} \int \frac{\D \lambda }{\lambda^2} \frac{\D^2 n_\gamma}{\cos\theta\, \D z\,\D \lambda} \times \exp(-\tau(z,\lambda)) 
\times Q_{eff}(\lambda) \\
&& \times Col(z,X_r,Y_r,u,\phi,\phi_{ph})\nonumber 
\end{eqnarray}

Where:

\begin{itemize}
\item $\D{\mathcal N}_e/\D E(t,E)$ is the energy dependant longitudinal distribution of charged particles in the shower
 (from eq. \ref{eq:NeeDeconvolved} and \ref{eq:NeeEDeconvolvedParameters}) as function of atmospheric depth from first
interaction
\item $F_u(u(E,s))$ is the normalised angular distribution of particles (from eq. \ref{fig:MeanWS} and \ref{eq:dNdlogu})
\item $F_{XY}(E,s,u)$ is the normalised lateral distribution of particles, (from eq. \ref{eq:MeanX} to \ref{eq:SigmaY})
\item $1/\lambda^2 \times \D ^2 n_\gamma/(\cos\theta\, \D z\,\D \lambda)$ is the Cherenkov photon production rate
(per unit of vertical track length and emitted photon wavelength) for an electron angle $\theta$ with respect to the shower
axis
\item $\exp(-\tau(z,\lambda))$ is the atmospheric absorption
\item $Q_{eff}(\lambda)$ is the detector quantum efficiency (multiplied by mirror reflectivity and other
wavelength-dependent transmission coefficients in the detector)
\item $Col(z,X,Y,u,\phi,\phi_{ph})$ is the average geometrical collection efficiency for photons emitted by a electron 
at position $(X,Y,z)$ with direction defined by $(u,\phi)$, and with an azimuthal photon angle $\phi_{ph}$ around
the electron.
\end{itemize}

In addition to the aforementioned ingredients,  instrumental effects such as
instrument point spread function and electronic response of the camera, including in particular trigger response and integration duration,
have to be taken into account in the calculation.

These effects, as well as the geometric light collection efficiency are obtained from a detailed simulation of 
the telescope response and parametrised in look-up tables.
Atmospheric absorption of light, wavelength-dependent quantum efficiency and reflectivity used in the model generation are also
implemented as look-up tables.

\begin{figure}[ht]
\begin{center}
\begin{tabular}{cc}
\epsfig{file=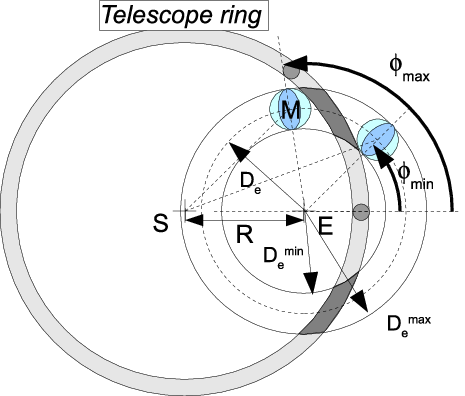,width=0.48\textwidth} &
\end{tabular}
\end{center}
\caption{\label{fig:PhiRange}Definition of integration range for electron azimuthal angle (see text).}
\end{figure}

In practice, for a given set of parameters, the relevant altitude range (in which emitted photons can reach the telescope) is first determined.
Within this range, the shower development is divided into altitude slices (typically 20 slices).
For each slice, the total number of charged particles in the shower is computed from eq. \ref{eq:NeeDeconvolved},
and the particles are distributed in energy bands using eq. \ref{eq:NeeEDeconvolvedParameters}.
For each energy band, the average energy, Cherenkov photon emission angle and emission yield
(integrated over atmospheric transmission, quantum efficiency and dish reflectivity) is derived.
The average shower is then further divided into flying particle angular bands and
lateral displacement (eq. \ref{eq:MeanW} to \ref{eq:LateralDistriY}). 
For each considered position and direction, the geometric collection efficiency is then taken into account to 
compute the contribution of this small shower subset. At each calculation step, a precise estimation of
the required integration range is performed to avoid spending too much time in sampling parts of the shower whose Cherenkov light do not reach the telescope mirror.

An example for such a determination (using a geometrical construction) is shown in fig.~\ref{fig:PhiRange}:
The projection plane ({\it ground}) is taken perpendicular to the shower axis, which is assumed to intersect the ground at point $S$. 
In order to avoid correlations with electron position and direction, 
the azimuthal range is determined in the shower direction frame ({\it i.e.} for a fixed shower and a telescope moving around the shower).
The electron, at a fixed altitude $z$, with an energy $E_e$, is assumed to be at a distance $R$ with an angle $\theta$ from the shower axis. 
Point $E$ in the figure is the projection on the ground of the electron position.
Its impact on the ground lies on the circle of radius $D_e = z \tan \theta$ around the point $E$. Any Cherenkov photon emitted
by this electron (with an angle $\theta_{c}$ with respect to the electron) will fall within the ring defined by the radii $D_e^{min} = z \tan (\theta - \theta_c)$
and $D_e^{max} = z \tan (\theta + \theta_c)$. Intersections of this ring with the possible telescope position (circle
centred on $S$) given the allowed azimuthal range $[\phi_{\mathrm min},\phi_{\mathrm max}]$. A similar approach is used to define the integration range
for the other variables.

\subsection{Parameter space}

Models are generated for:

\begin{itemize}
\item 40 different zenith angles $\theta_z$
\item 40 impact distances ranging from $0$ to $(400~\mathrm{m})/\cos(\theta_z)$ from telescope
\item 65 different energies from $(50~\mathrm{GeV})/\cos(\theta_z)$ to $(20~\mathrm{TeV})/\cos(\theta_z)$
\item 6 first interaction depths from $0~X_0$ to $5~X_0$.
\end{itemize}

The shower images are always generated on-axis ({\it i.e.} for a $\gamma$-ray at the centre of the camera). 
For a perfect telescope, a change of direction will result in a simple translation in the camera frame, that 
can be applied later, in the fit procedure. In a more realistic telescope, the broadening of the point
spread function at large off-axis angles needs to be taken into account. This is currently not needed
for the H.E.S.S. telescopes, which use a Davis-Cotton optical design 
offering a degradation of the PSF from 0.25 mrad (RMS) at the centre to $\sim$1 mrad at the edge of the field of view, 
however always smaller than the pixel size\cite{HESSPSF}.
In total, $624 000$ templates are generated. The model generation procedure requires about 50 to 100 day-machine computing time on recent desktop 
computers and can be easily parallelled. The resulting shower images are stored in a ROOT binary file\cite{ROOT}
for later use.

\subsection{Example}

\begin{figure}[ht]
\begin{center}
\begin{tabular}{cc}
\epsfig{file=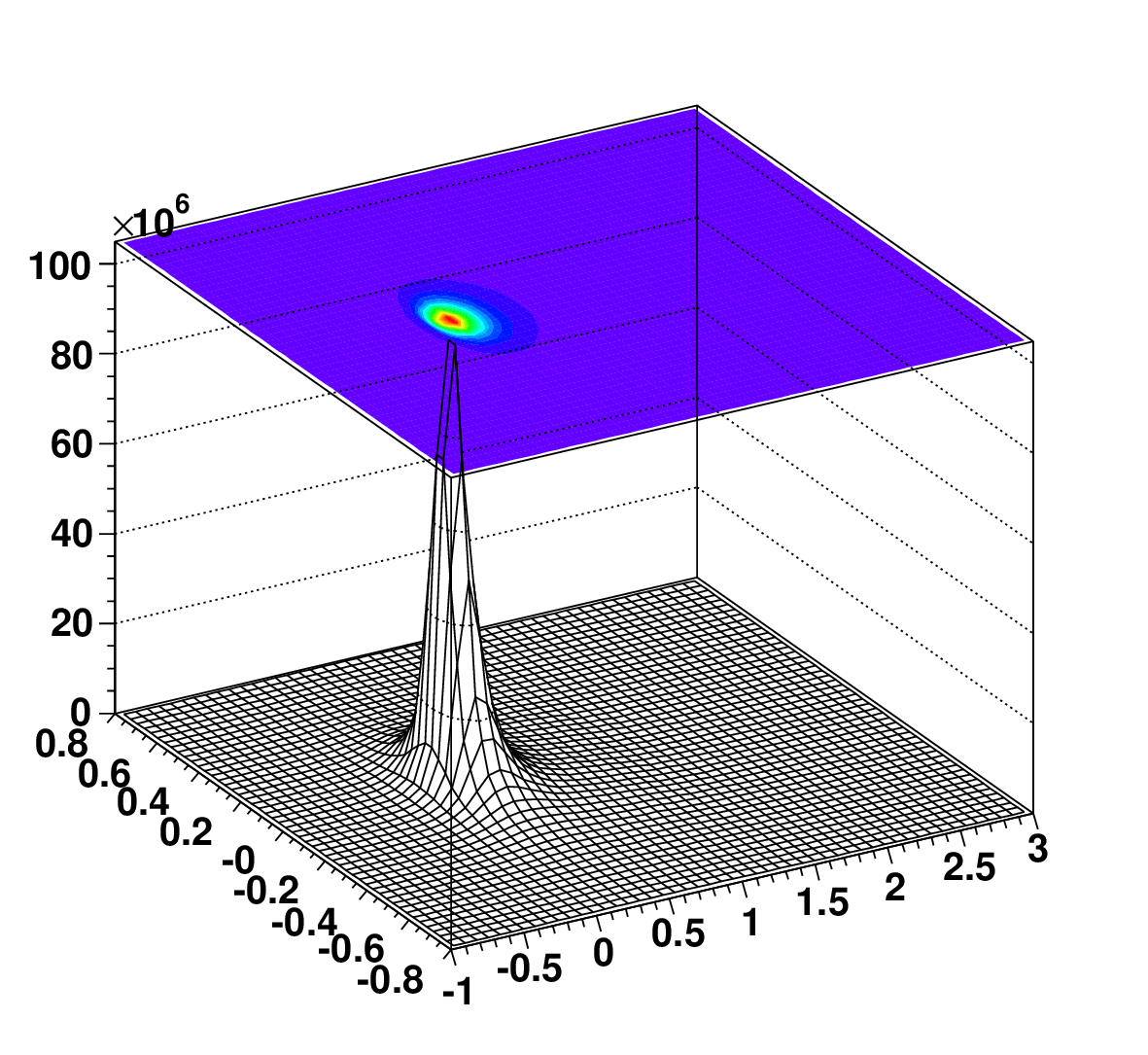,width=0.48\textwidth}&\epsfig{file=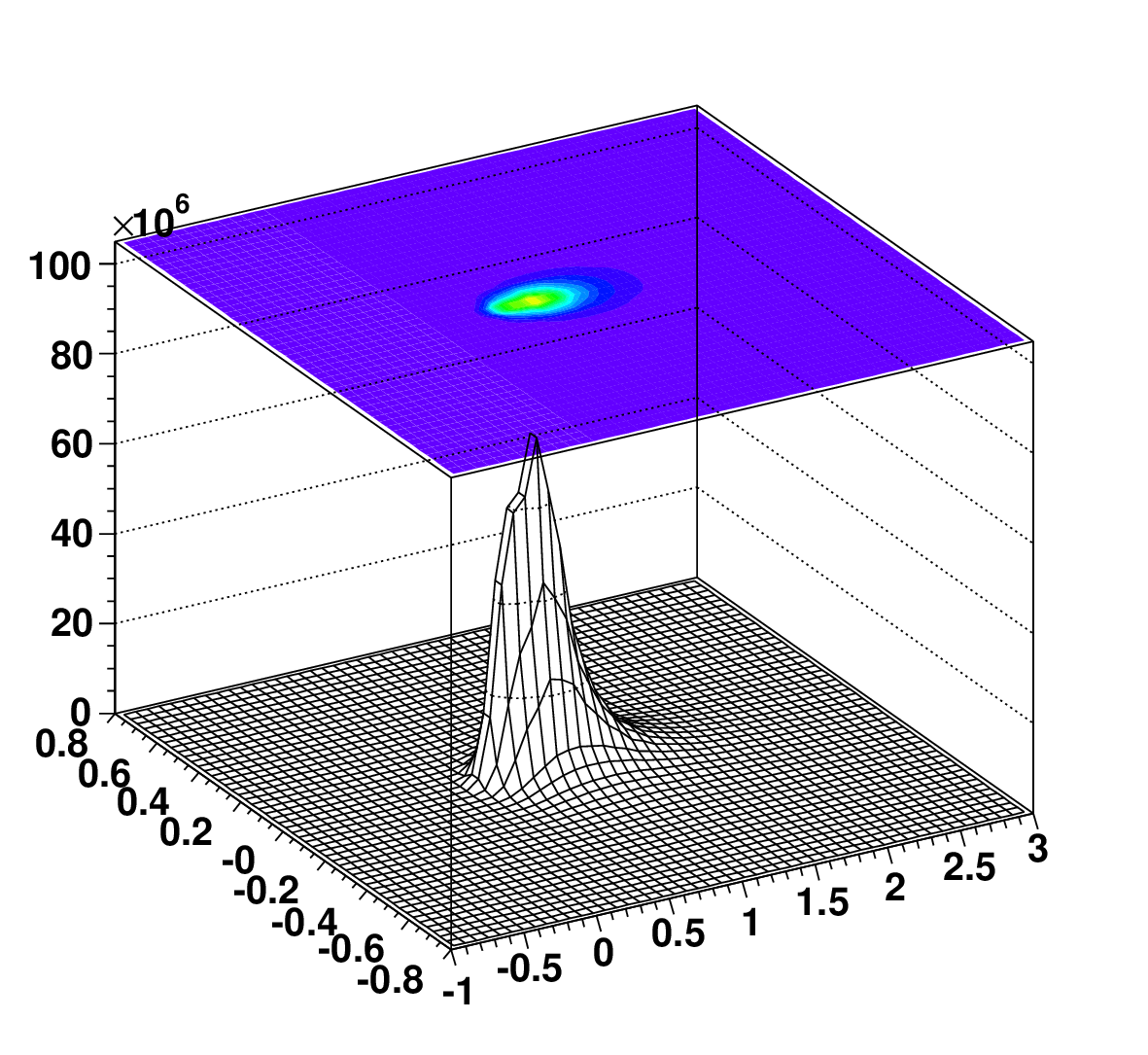,width=0.48\textwidth} \\
\epsfig{file=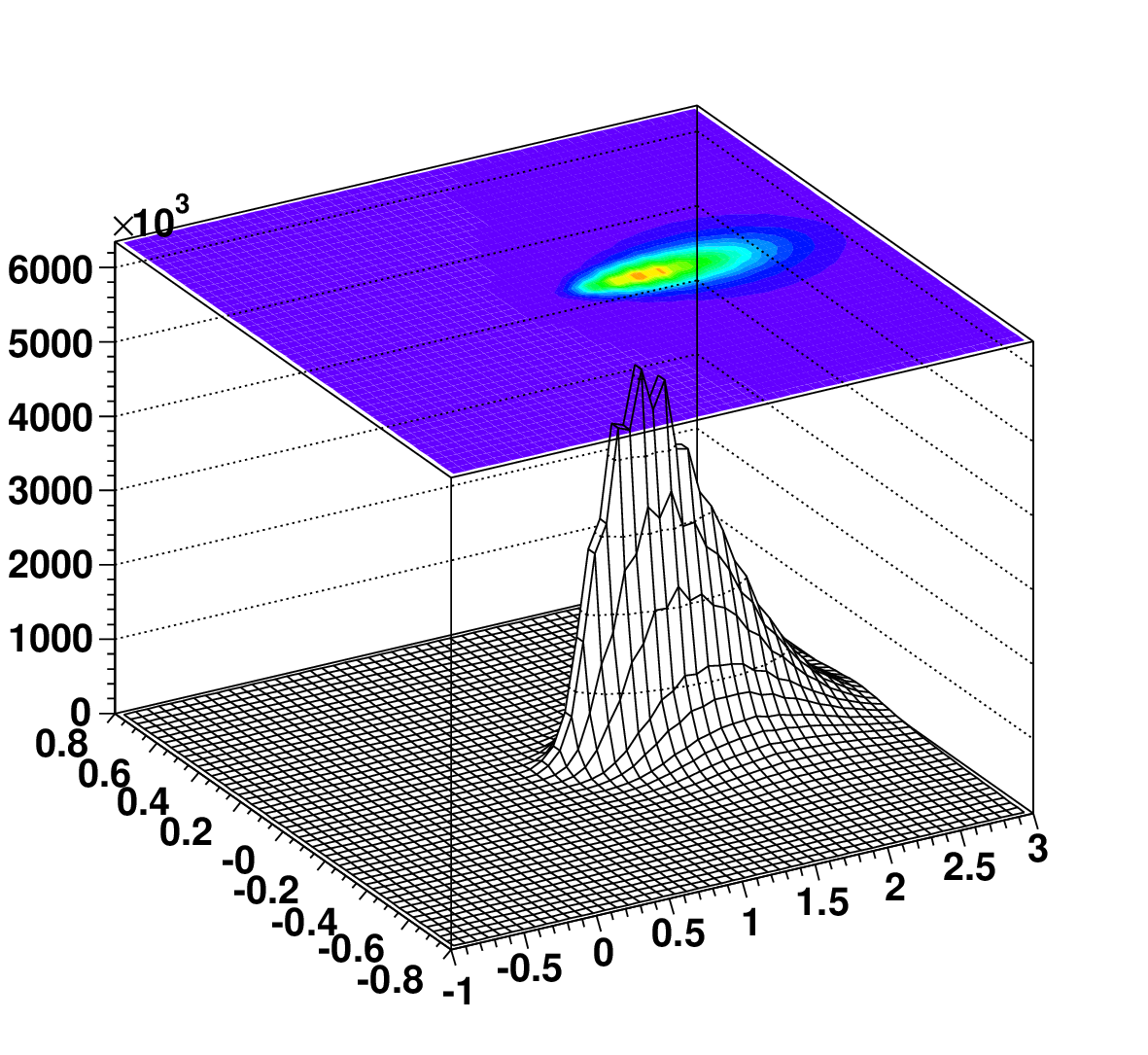,width=0.48\textwidth}&\epsfig{file=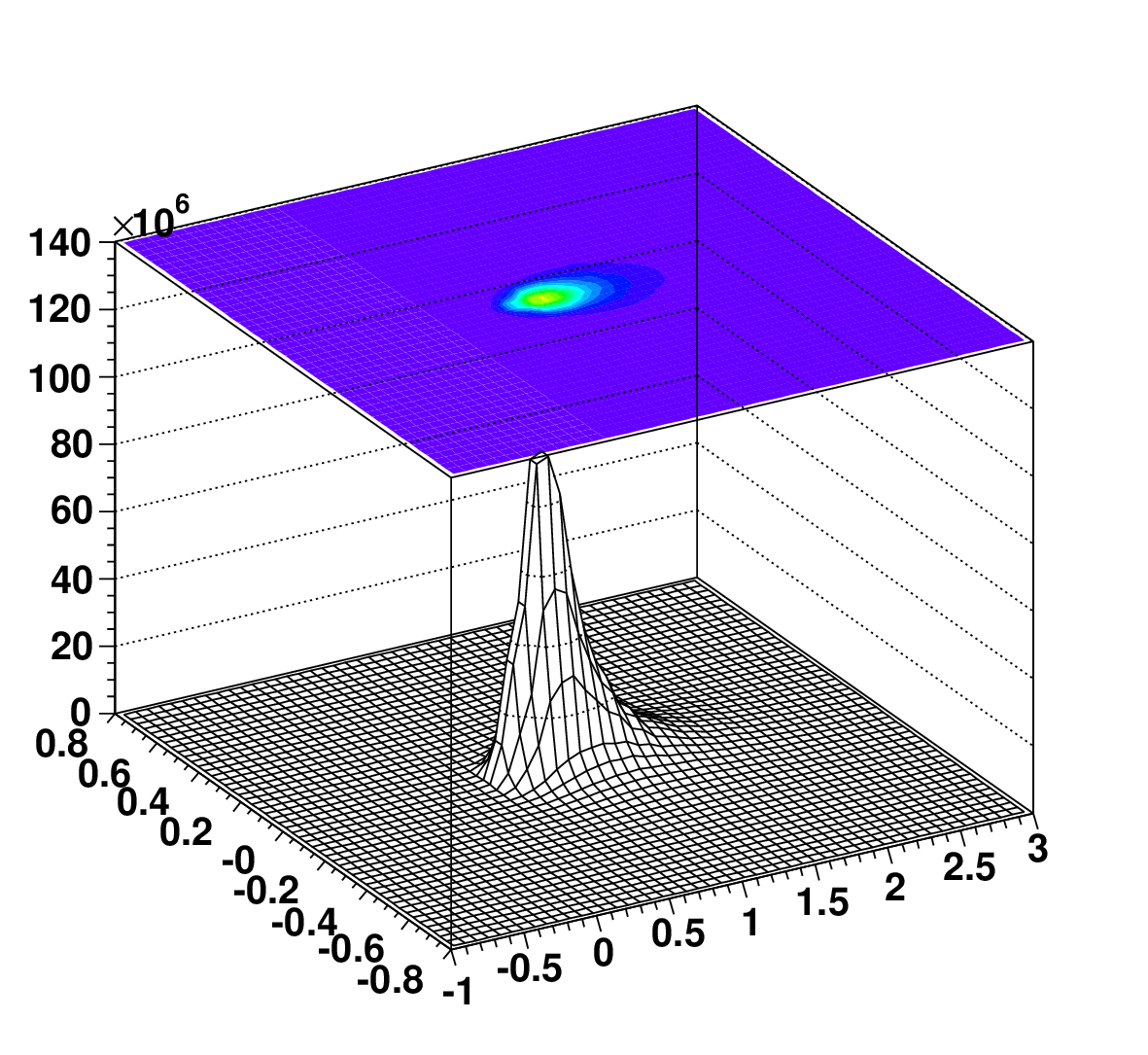,width=0.48\textwidth} \\
\end{tabular}
\end{center}
\caption{\label{fig:ModelExample}Model of a $1~\mathrm{TeV}$ shower started at one radiation length and
falling 20~m (top-left), 100~m (top-right) and 250~m (bottom-left) away from the telescope.
X and Y axes are in units of degrees in the camera frame.
Bottom-right: same as top-right but with a first interaction point located deeper in
the atmosphere (at 3 $X_0$). Note that the vertical scale (image amplitude) differs.}
\end{figure}

The output of the model generation procedure is a bank of two dimensional shower images
in the frame of a perfect camera, with very small ($0.01^\circ$) pixel size. For a given primary particle
set of parameters, a predicted image is computed with an interpolation procedure on the generated
images in a 4 dimensional space (energy, impact distance, primary interaction depth and zenith angle). 
Shower direction and azimuthal angle are then taken into account as a rotation and a translation 
in the camera frame to compute the final predicted image.
Examples of such two dimensional shower images are shown in fig.~\ref{fig:ModelExample} 

\subsection{Comparison with simulation}

\begin{figure}[ht]
\begin{center}
\begin{tabular}{cc}
\epsfig{file=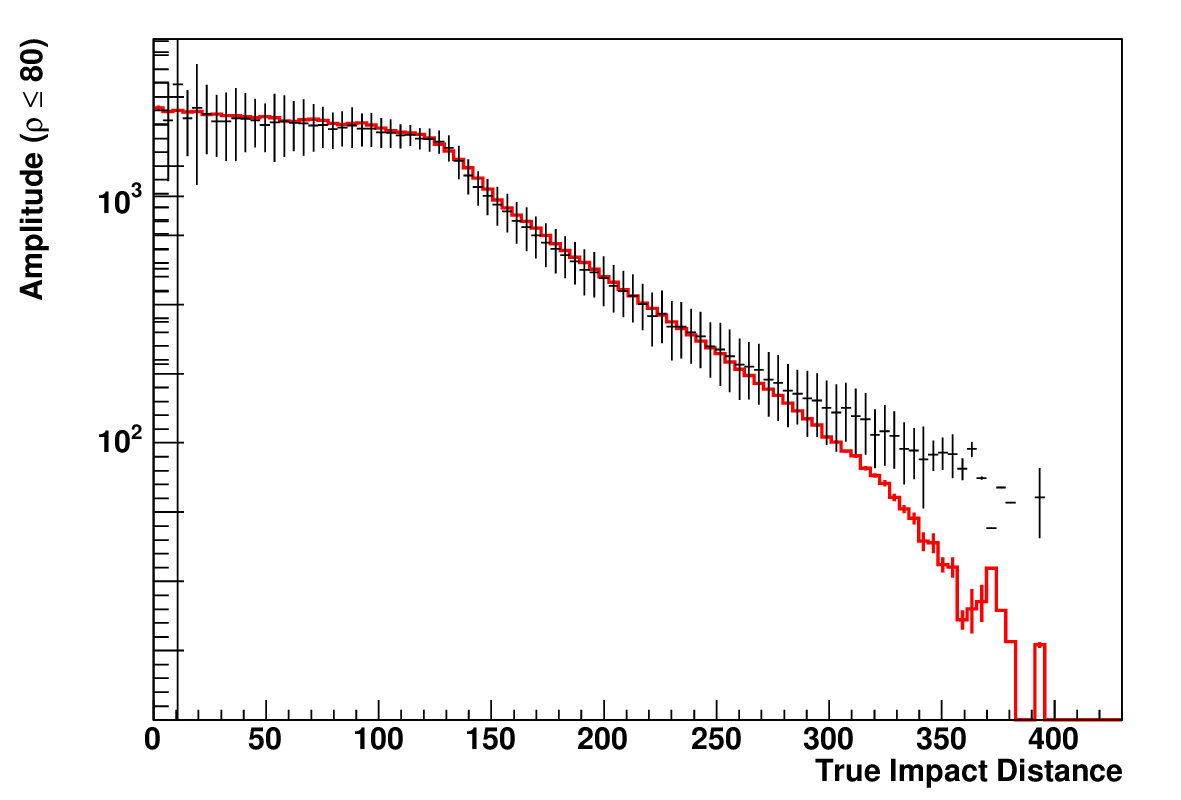,width=0.48\textwidth} & \epsfig{file=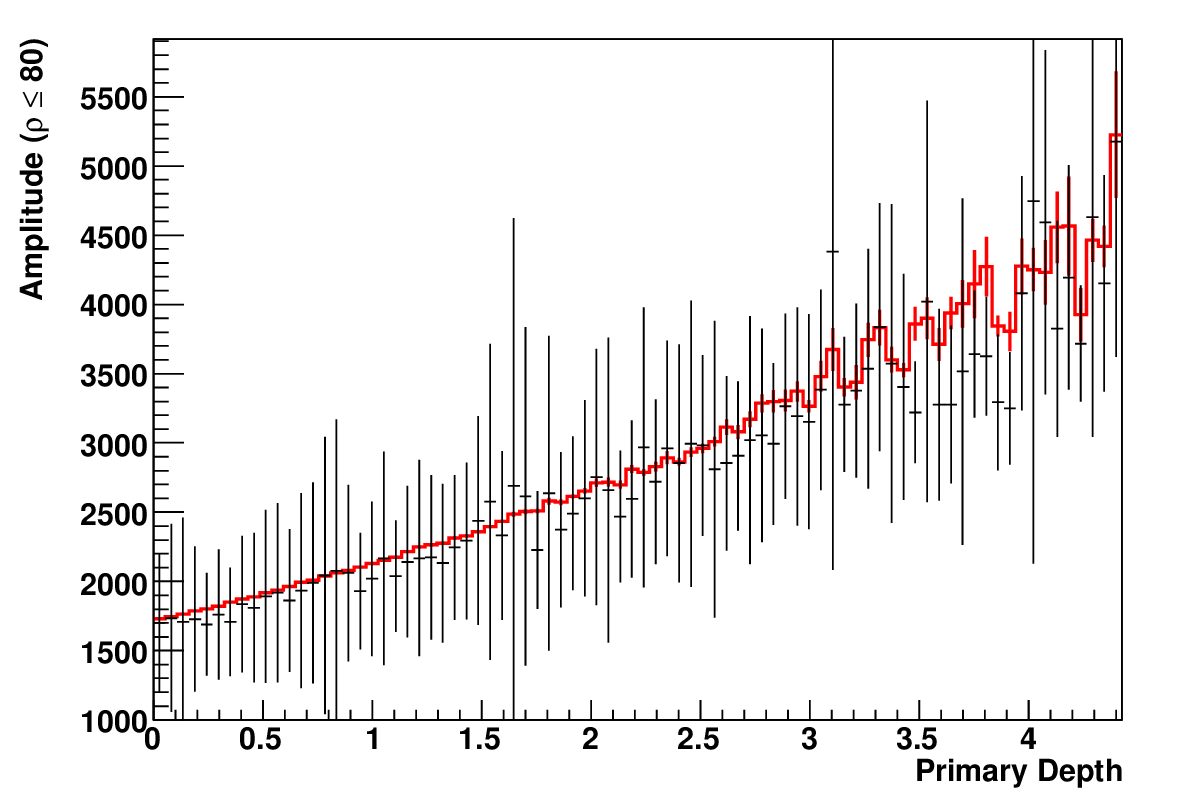,width=0.48\textwidth} \\
\epsfig{file=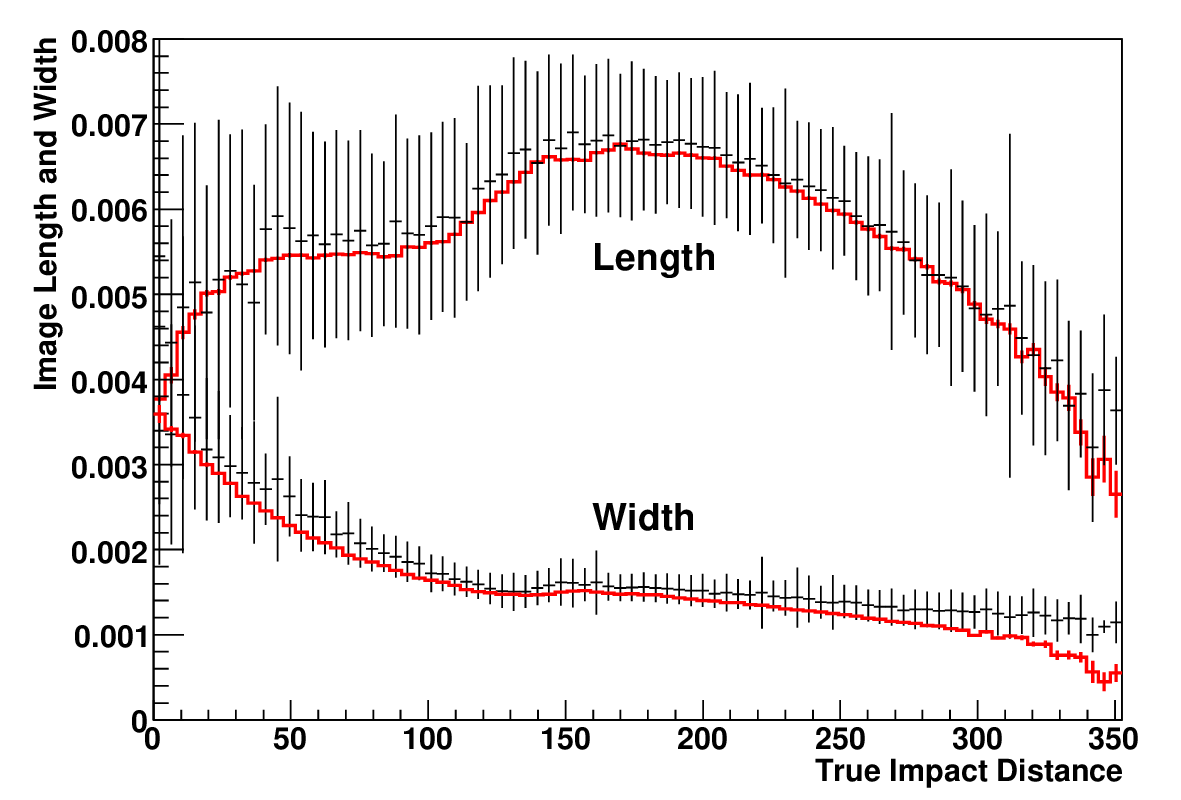,width=0.48\textwidth} & \epsfig{file=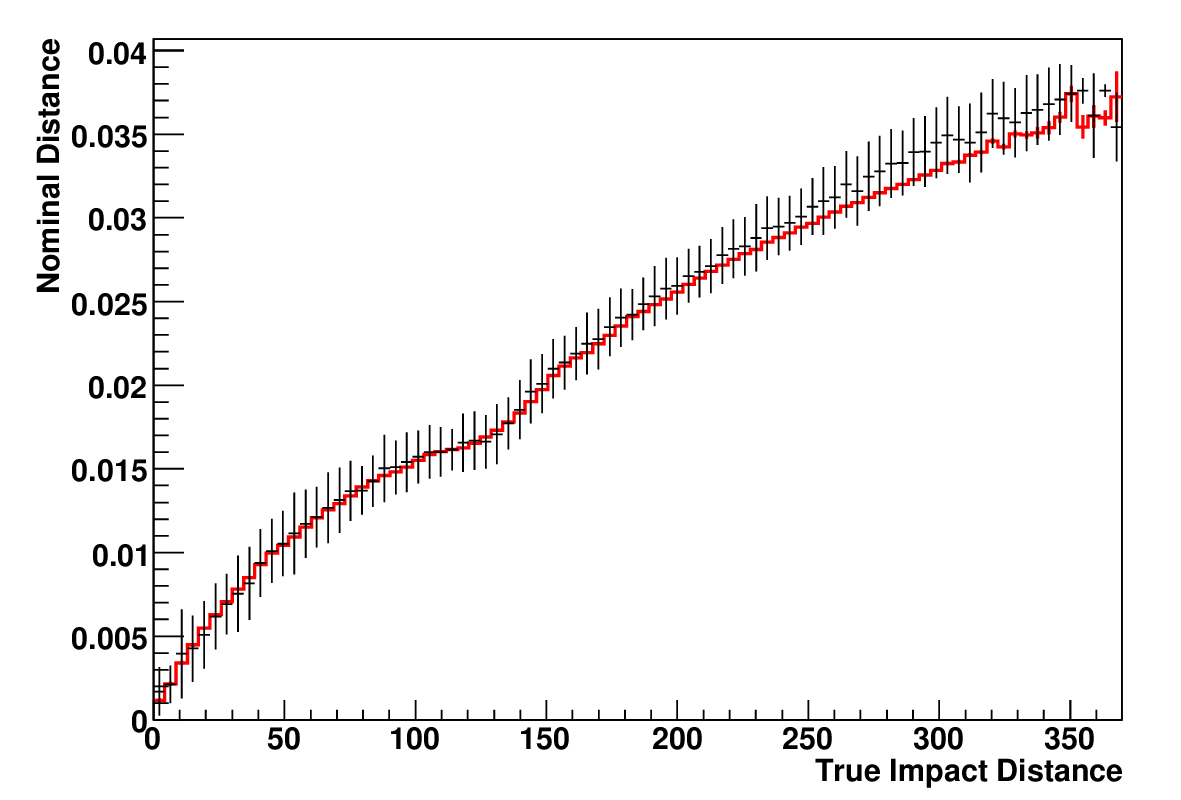,width=0.48\textwidth} \\
\end{tabular}
\end{center}
\caption{\label{fig:CompareSimu1TeV}Comparison between 1~TeV simulated shower images at zenith (black) and model prediction (red).
Top left: Image amplitude as function of impact distance for primary interaction point of one radiation length.
Top right: Image amplitude as function of primary interaction depth for impact distance $\rho \leq 80~\mathrm{m}$ 
(core of light pool). Bottom left: Image length (top) and width (bottom), in units of milliradians, estimated with the standard Hillas
parametrisation technique, and as function of impact distance. Bottom right: Angular distance, in units of milliradians, between the
image centre of gravity and primary direction as function of impact distance. In each plot, the simulation
is drawn as a black histogram, with error bars indicating the shower-to-shower fluctuations. The model
prediction is represented by a solid thick red line.}
\end{figure}

A comparison of the image shapes between simulation and model prediction
is shown in fig.~\ref{fig:CompareSimu1TeV} for 1~TeV gamma-ray showers.
The images were calculated for a H.E.S.S. camera, with pixels of
$0.16^\circ$ diameter. The image length and width were estimated using the 
standard Hillas parametrisation applied to the predicted images.
In each plot, the average value of the simulation is drawn as a black histogram, with error bars indicating 
the shower-to-shower fluctuations. The model prediction (image of average shower) is represented by a solid thick red line.

The agreement between the model and the average values of the simulation is excellent up to core
distance of about 300~m, where some trigger effects can explain the differences:
at this distance, the total image amplitude does not exceed 100 photoelectrons, 
distributed over many pixels. Showers fluctuating up to higher intensities
are more likely to trigger the telescope, thus resulting in a higher 
average image amplitude in the simulation compared to the model.

The  bars in the simulation histograms (in black) are due to shower-to-shower
fluctuations, which are not taken into account in the model generation. 
At small core distance, and when taking into account the evolution
with primary interaction depth, the shower intensity fluctuates by about 20\% from 
shower to shower in addition to the fluctuation related to the depth of first interaction.

\subsection{Conclusions}

The work presented in this section results in the generation of a {\it shower image model}, which is 
nothing more than an accurate prediction of the expected Cherenkov light distribution in the camera
for a given set of primary particle parameters. The key ingredients in the model generation are
the inclusion of depth of first interaction as shower parameter (as the main source
of shower-to-shower fluctuations), and the precise description of longitudinal, lateral, and angular 
distributions of charged particles in the shower. The corresponding shower model is constructed
once for all and can be applied to various zenith angles, telescope impact distance or off-axis angle.

An alternate {\it brute force} approach would be the generation of the shower image model 
from massive simulations performed at every energy, zenith angle, impact distance, off-axis angle 
and primary interaction depth. The size of the parameter space and the need for a smooth model (in order
to avoid local minima that would hamper the quality of the fit procedure) makes this approach less
effective that the generation of an intrinsically smooth shower model.
 
\section{Fit procedure}\label{sec:reconst}

Once a shower model has been obtained, actual images on the camera can be compared to the ones predicted by 
the shower image model for a given set of primary parameters. 
A minimisation procedure is then used to obtain
the most likely parameters of the incoming particle (energy, direction, impact, depth of the first interaction)
under the hypothesis that the particle is a $\gamma$-ray. The minimisation procedure involves
a precise comparison of the intensity in each pixel of the camera with the prediction from the model (interpolated between
grid points to the actual set of parameters).
In order to take into account the Poisson nature of the detected number of photons in each pixel, a
log-likelihood approach has been chosen.

\subsection{Pedestal and calibration}

In the absence of Cherenkov light, each pixel of the H.E.S.S. camera is illuminated by a significant amount 
of Night Sky Background (NSB) light, which largely dominates over the electronic noise and represents a 
single photoelectron rate ranging from $40~\mathrm{MHz}$ 
off the galactic plane to $\sim 300~\mathrm{MHz}$ in the most luminous parts of the galactic plane.
The {\it pedestal} is defined for each pixel as the charge distribution collected in the absence of Cherenkov
signal. It is determined\cite{HessCalib} for each pixel and for each observation run, using a cleaning
procedure that identifies the position of the shower image in the camera and rejects the corresponding
pixels. The pedestal width is the combination of the electronic noise and the night sky background,
the later being largely dominant. Due to varying atmospheric conditions, rotation of the sky on 
the camera\footnote{The H.E.S.S. telescope use a Alt-Az mount} and instrumental effects that depend
on particular on temperature, the pedestal position and width can vary with time, with timescales
of the order of a few minutes. The evolution of pedestal position and width with time for each pixel is 
recorded for every observation run and used during the reconstruction describe below.

\subsection{Pixel log-likelihood}

The shower image model provides a density of Cherenkov light in the focal plane. A convolution by the pixel size is done when
loading the model to compute the expected signal $\mu$ in each pixel. For the sake of simplicity, the dependence of the PSF across the field of view is
not taken into account in this calculation.
The probability density  ({\it likelihood}) to observe a signal of  $s$ (in units of photoelectron) in a pixel
for an expectation value $\mu$ is given by the convolution of the 
Poisson distribution of the photoelectron number $n$ with the photomultiplier
resolution. The latter is well represented by a Gaussian of width $\sqrt {\sigma_p^2 + n \sigma_\gamma^2}$
where $\sigma_p$ is the width of the pedestal (charge distribution under pure noise, including night sky background)
and $\sigma_\gamma$ the width of the single photoelectron peak (photomultiplier resolution):

\begin{equation}
P(s|\mu,\sigma_p,\sigma_\gamma) = 
\sum_n \frac{\mu^n e^{-\mu}}{n!\sqrt{2\pi (\sigma_p^2 + n \sigma_\gamma^2)}} \exp\left(  - \frac{(s - n)^2}{2 (\sigma_p^2 + n \sigma_\gamma^2)} \right) \label{eq:LikProb}
\end{equation}

In eq. \ref{eq:LikProb}, the actual pedestal width $\sigma_p$ is different for each pixel, and depends in particular
on the level of night sky background (NSB) seen by the pixel. The photoelectron resolution $\sigma_\gamma$, also 
specific to each pixel, is measured using calibration runs where the camera is illuminated with a low intensity flashing LED.
The use of measured values of $\sigma_p$ and $\sigma_\gamma$ is an  important aspect that will explain the stability of
the Model Analysis for varying level of NSB (Section \ref{sec:NSB}).

\begin{figure}[ht]
\begin{center}
\epsfig{file=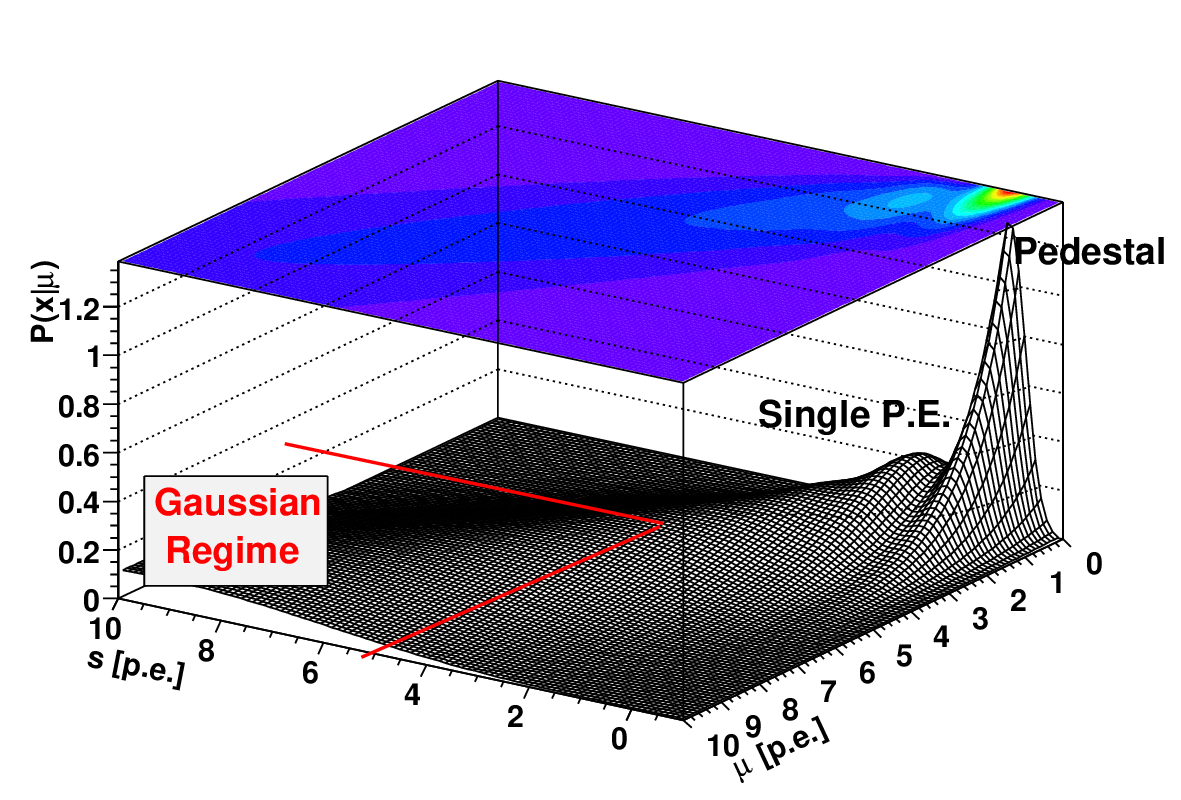,width=0.9\textwidth}
\end{center}
\caption{\label{fig:Likelihood2D}Probability to observe a signal of $s$ (in units of photoelectron) in a pixel
for an expected average value $\mu$, as function of $s$ and $\mu$, and for a pedestal width of $0.3~\mathrm{p.e.}$ and
a single photoelectron resolution of $40\%$, from equation \ref{eq:LikProb}. For low predicted or detected intensity,
the simple Gaussian assumption (as used in a $\chi^2$ test) is not valid anymore.}
\end{figure}

The shape of function \ref{eq:LikProb} is shown in fig.~\ref{fig:Likelihood2D}.
In order to obtain a variable that behaves asymptotically like a $\chi^2$, we define the {\it pixel log-likelihood}

\begin{equation}
\ln {\it L} = -2 \times \ln P(s|\mu,\sigma_p,\sigma_\gamma)
\end{equation}



\subsection{Pixel log-likelihood expectation values}

In order to compare the observed signal $s$ with a model prediction $\mu$, the expectation
value of the log-likelihood under different realisations of the same shower image ({\it i.e.} $\mu$ being fixed and
$s$ fluctuating due to Poisson noise, electronic and NSB fluctuations)
needs to be computed.
If the observed signal is only due to noise (night sky background), the probability density functions
simplifies to a Gaussian of width $\sigma_p$. In this case, the average value of the pixel log-likelihood reads

\begin{eqnarray}
\left<\ln {\it L}\right>|_{\mu = 0}  &=& \int ds \, \ln L(s|\mu= 0,\sigma_p) \times P(s|\mu= 0,\sigma_p)  \nonumber \\
&=&\int ds \left(\ln (2\pi) + \ln \sigma_p^2 +  \frac{s^2}{\sigma_p^2}\right) \frac{1}{\sqrt{2\pi \sigma_p^2}} \exp \left( -\frac{s^2}{2\sigma_p^2} \right)  \nonumber \\
&=& 1 + \ln (2\pi) + \ln \sigma_p^2 \label{eq:LikNull}
\end{eqnarray}

In a similar manner, one obtains

\begin{equation}
\left<\ln^2 {\it L}\right> = \left(1 + \ln (2\pi) + \ln \sigma_p^2\right)^2 + 2 \quad \Longrightarrow \quad \sigma^2(\ln L) = 2 
\end{equation}

At high $\mu$, the Poisson distribution can be replaced by a Gaussian of width
$\sqrt{\mu}$, and  the probability density function can be well approximated by the convolution of two Gaussian:

\begin{eqnarray}
P(s|\mu \gg 0,\sigma_p,\sigma_\gamma) &\approx&  \frac{1}{\sqrt{2\pi \left(\sigma_p^2 + \mu(1+\sigma_\gamma^2)\right)}}\exp\left(-\frac{(s-\mu)^2}{2\left(\sigma_p^2 + \mu(1+\sigma_\gamma^2)\right)}\right) \nonumber
\end{eqnarray}

In this limit, the log-likelihood behaves like a $\chi^2$:

\begin{eqnarray}
\left<\ln {\it L}\right>|_{\mu} &=& 1 + \ln (2\pi) + \ln \left(\sigma_p^2 +\mu(1+\sigma_\gamma^2)\right), \quad \sigma^2(\ln L) = 2 \label{eq:LikAverage}
\end{eqnarray}

\begin{figure}[ht]
\begin{center}
\epsfig{file=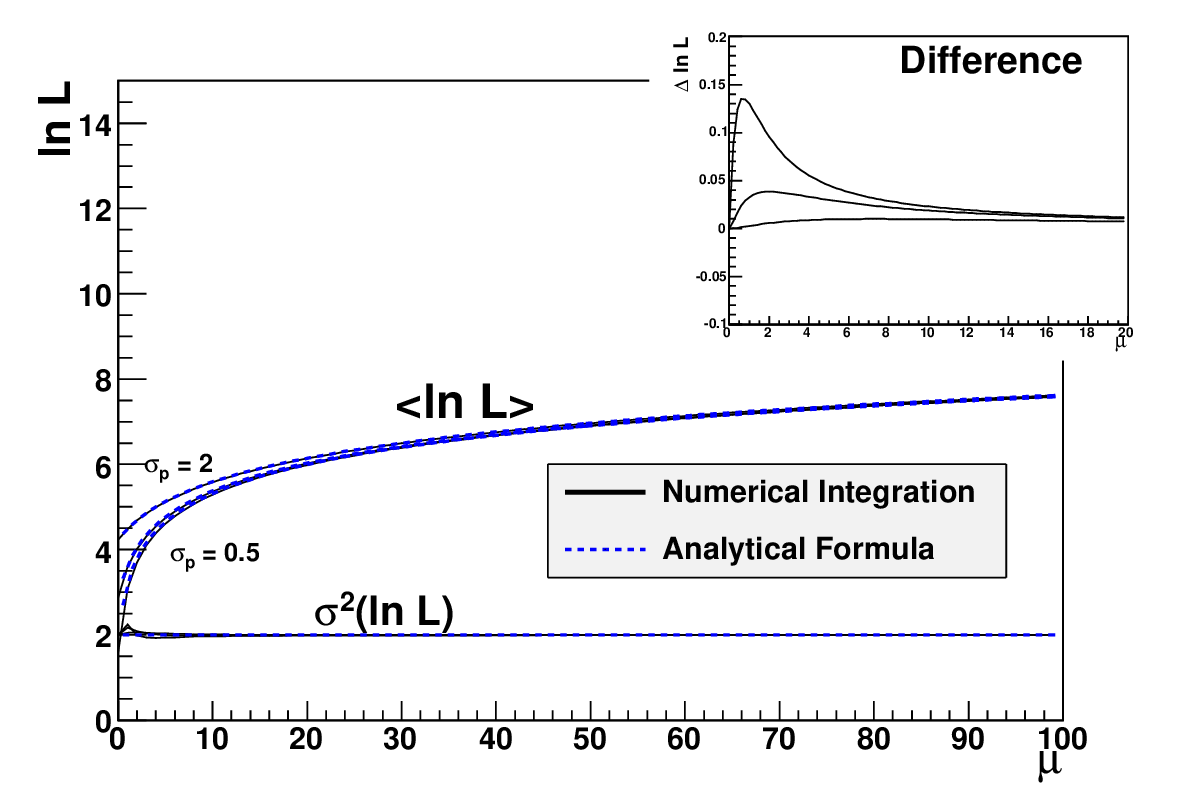,width=0.95\textwidth}
\end{center}
\caption{\label{fig:LikAverage} Average value and sigma of pixel log-likelihood as function of expected amplitude $\mu$
(see text) for different pedestal width ($\sigma_p = 0.5,~1,\mathrm{and}~ 2~\mathrm{p.e.}$). The solid line are the results of a numerical integration, whereas the dashed
line are the analytical expression \ref{eq:LikAverage}. Inset: difference between analytical expression and numerical integration.}
\end{figure}

This expression reduces to eq. \ref{eq:LikNull} when $\mu=0$.
Figure \ref{fig:LikAverage} shows the comparison of this analytical expression with a numerical
integration of the average log-likelihood. The analytical expression is valid for $\mu=0$ and, as expected,
is also asymptotically valid for $\mu \gg 1$. In the transition regime, the analytical expression slightly 
overestimates the likelihood value. 

In order to properly calibrate the average log-likelihood, the discrepancy between the analytical expression and
the numerical integration is computed for every expected amplitude $\mu=0$ and pedestal width $\sigma_p$ (using a Monte Carlo
simulation) and stored in a look-up table. This look-up table will be used at the end of the fit procedure (see below) to
produce calibrated discriminating variables.

\subsection{Telescope log-likelihood}

Pixels belonging to a camera are assumed to be independent. We define the {\it telescope log-likelihood} 
as the sum over all pixels of the {\it pixel log-likelihood}:

\begin{equation}
\ln {\it L}_{tel} = \sum_{\mathrm{pixel}\ i}  \ln {\it L}_i = \sum_{\mathrm{pixel}\ i} -2 \times \ln P(s_i|\mu_i,\sigma_p,\sigma_\gamma) 
\label{eq:TelLik}.
\end{equation}

\subsection{Reconstruction - Fit algorithm}

The model photon reconstruction relies on the pixel-per-pixel comparison of the actual shower images with the ones that are predicted for
a given set of parameters. A minimisation procedure is used to find the best parameters (direction, impact, depth of the first
interaction and energy). In contrast to Hillas parameters based reconstruction techniques, the raw image amplitudes are directly used, without any
image cleaning procedure. All pixels are used in the fit, not only those close to the actual image.

A Levenberg-Marquardt fit algorithm \cite{Levenberg,Marquardt} is used
to minimise the telescope log-likelihood (eq. \ref{eq:TelLik}). 
This algorithm is very efficient in the case the first and second derivative
of the minimised function (log-likelihood or $\chi^2$) can be expressed analytically and depend mostly on the first derivative of the model
(the second derivative being negligible). This is in general valid when the minimised function is a quadratic form ($\chi^2$ or similar) and
when the model function exhibits a smooth behaviour when varying the model parameters.
The algorithm assumes a quadratic form and uses the inverse matrix of second derivatives to estimate the position of the minimum. 
In order to improve convergence stability, a combination of
pure steepest descent and parabolic assumption is used with weights that vary during the minimisation, depending on whether the quadratic assumption
is valid or not. The Levenberg-Marquardt fit algorithm always converges - sometimes to a local minimum - and provides reliable estimates of the 
model parameters uncertainties, taken from the correlation matrix.

An important issue in the fit algorithm is the starting point. The standard Hillas reconstruction technique\cite{HessCrab}
with different sets of image cleaning parameters is used to derive a handful of possible estimates. The estimate that provides 
the best initial log-likelihood is chosen as starting point of the minimisation.

In our case, about 40 iterations are needed for the algorithm to converge, whereas simpler algorithms such as Minuit\cite{Minuit}
needed in average about ten times more. On recent desktop computers, the reconstruction of a single event takes about $50$ to $100~\mathrm{ms}$.

The actual telescope reflectivity, as measured from ring-shape images of muons passing through the telescope\cite{HessCalib}, is 
used to scale the model prediction to the observation conditions, on a run-by-run basis. Since the optical efficiency is taken into
account directly at the reconstruction level, no further energy correction is needed (which is completely different to
the way the optical efficiency is handled in Hillas parameters based analyses).

The output of the minimisation procedure are:

\begin{itemize}
\item Best guess of the 6 shower parameters: direction (2 parameters), impact (2 parameters), depth of the first interaction and energy
\item Correlation matrix and therefore uncertainty on these best fit parameters
\item Final log-likelihood value
\end{itemize}

\subsection{Goodness of fit}

Since Atmospheric Cherenkov Telescopes are background dominated systems, the performance of any analysis  is
mainly driven by its ability to discriminate between gamma-ray induced showers and the much more numerous hadronic background.
In the Model Analysis, a {\it goodness-of-fit} approach is chosen to compare the model prediction and the actual shower images,
in order to check the compatibility of the recorded events with a pure $\gamma$-ray hypothesis. 
The {\it goodness-of-fit} is defined as a normalised sum over all pixels of the difference between the actual pixel log-likelihood and 
its expectation value, properly normalised:

\begin{eqnarray}
G = \frac{\displaystyle \sum_{\mathrm{pixel}\ i} \Big[ \ln L(s_i|\mu_i) - \left<\ln {\it L}\right>|_{\mu_i} \Big] }{\sqrt{2 \times \mathrm{NdF}}} \label{eq:Goodness}
\end{eqnarray}

where $\mathrm{NdF}$ is the number of degrees of freedom (number of pixels - 6). The goodness of fit behaves asymptotically like
a $\chi^2$ and provides therefore a measure of the fit quality. This will be used later on the hadronic discrimination. 
By construction, if the pixels likelihood behave like independent random variables, $G$ is expected to behave like a normal variable:

\begin{equation}
\left<G\right> = 0 \quad \hbox{and} \quad \sigma^2(G) = 1 \label{eq:AvgGoodness}
\end{equation}

\subsection{Goodness of fit calibration}

The goodness of fit average value relies on the assumption of a Gaussian pedestal (of width $\sigma_p$).
In the absence of night sky background, the pedestal reduces to the electronic noise and the assumption is
always valid\cite{HessCalib}. The H.E.S.S. camera use two different gains: a {\it high gain}, with dynamical range from 0 to
200 photoelectrons with single photoelectron resolution and an electronic noise representing about $0.2$
p.e., and a {\it low gain}, with dynamical range up to 2000 photoelectrons but worse resolution. In the low
gain, where the electronic noise represents about $2.5$ p.e., the pedestal is also Gaussian to a good approximation.

\begin{figure}[htb]
\begin{center}
\epsfig{file=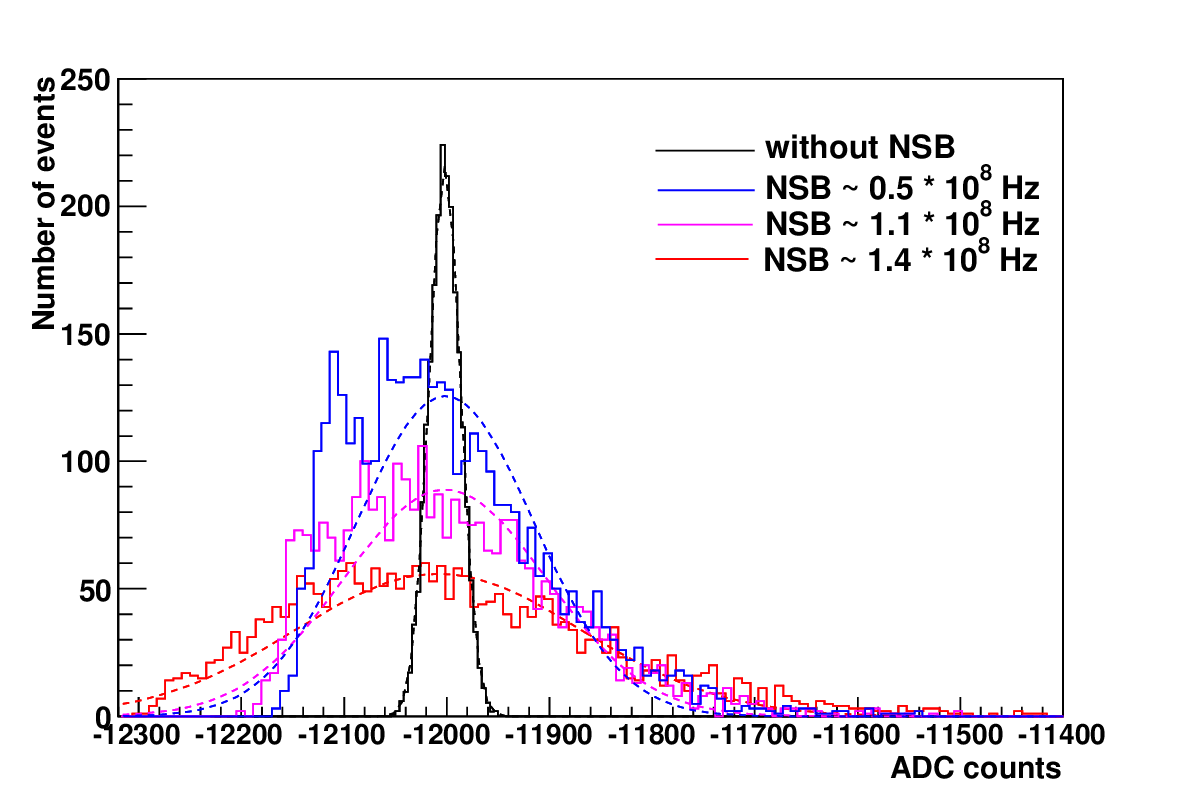,width=0.98\textwidth}
\end{center}
\caption{\label{fig:PedestalShift} Evolution of pedestal shape with increasing night sky background level (solid line) 
and Gaussian fits to the distributions (dashed lines), using only the high gain channel. From \cite{HessCalib}. }
\end{figure}

In the high gain channel, figure \ref{fig:PedestalShift}, reproduced from \cite{HessCalib}, shows that the Gaussian assumption
is also almost valid for high night sky background (above $\sim 150~\mathrm{MHz}$), but not in the intermediate regime
where the pedestal shape exhibits two peaks, similar to a single photoelectron spectrum,
and produced by fractions of single photoelectron pulses falling by chance within the acquisition window.

\begin{figure}[htb]
\begin{center}
\begin{tabular}{cc}
\epsfig{file=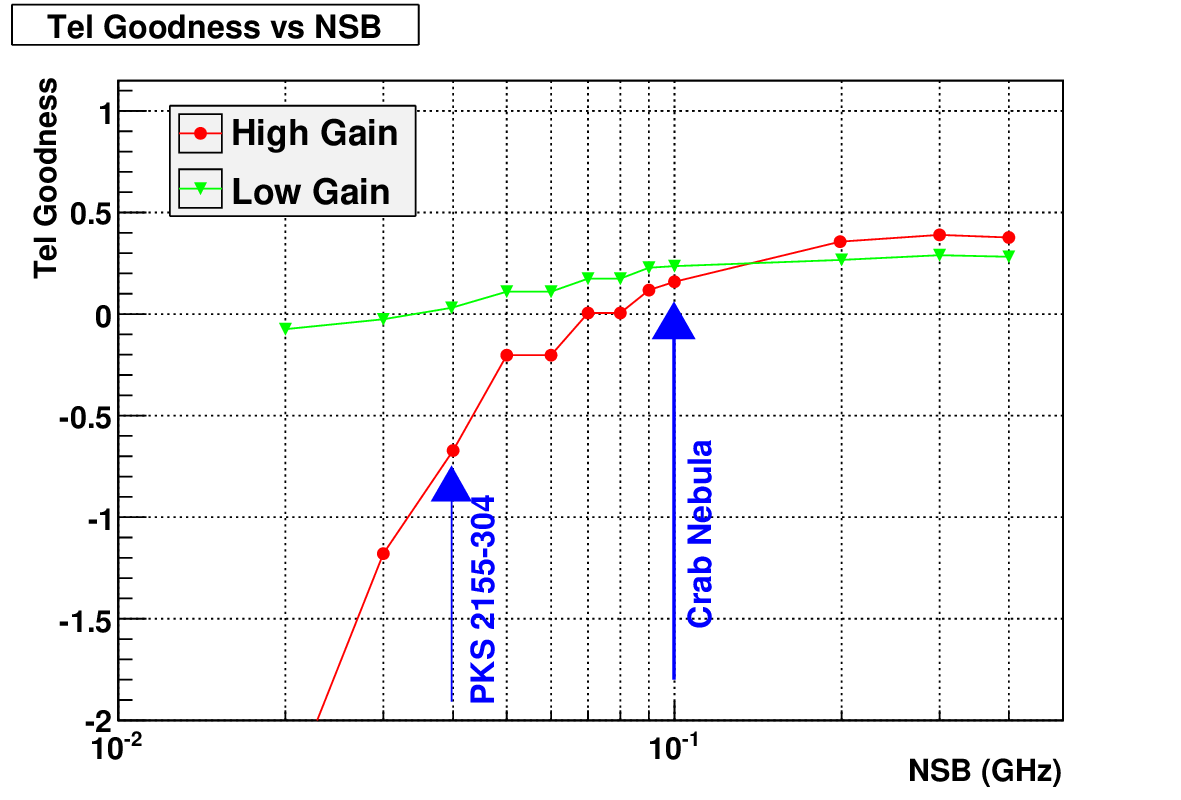,width=0.48\textwidth} & \epsfig{file=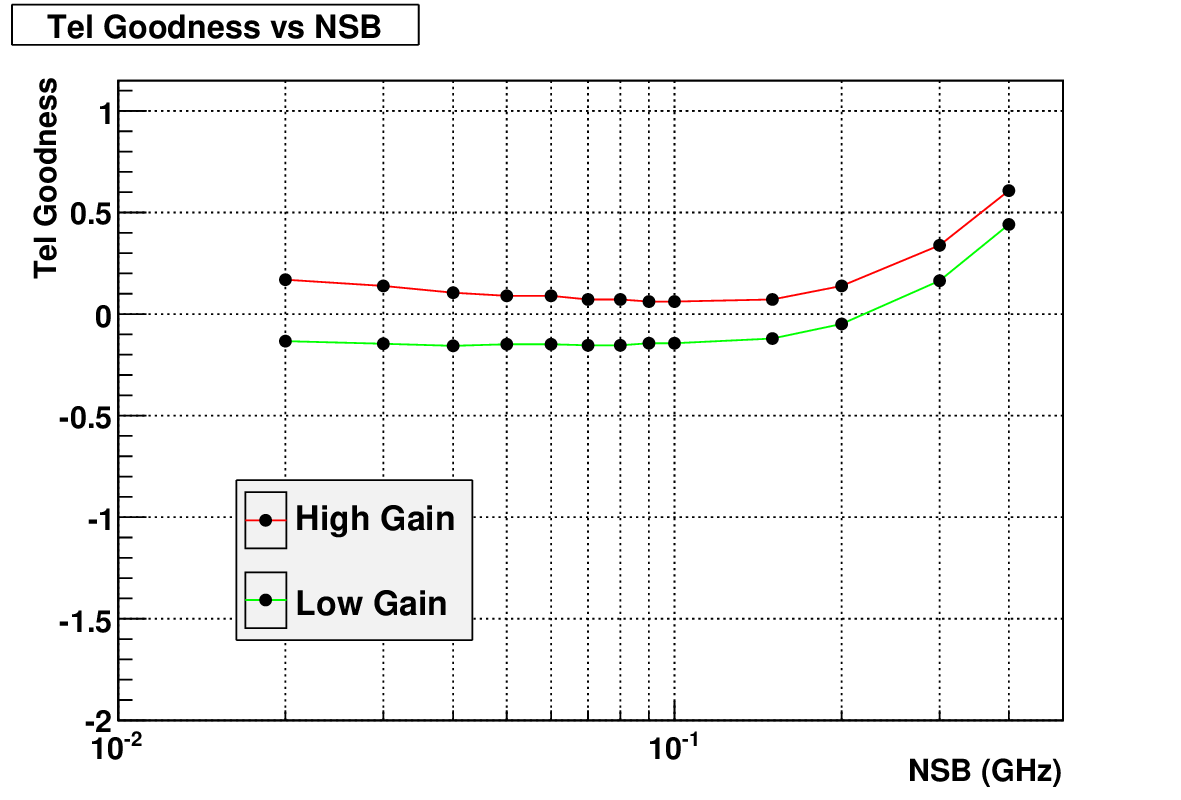,width=0.48\textwidth}  \\
\end{tabular}
\end{center}
\caption{\label{fig:TelGoodnessNsb} Average value of goodness-of-fit per telescope as function of night sky background level,
and as obtained from simulations, before (left) and after (right) goodness calibration.}
\end{figure}

Simulation of $\gamma$-ray spectra were conducted with different levels of night sky background (from $10$ to $800~\mathrm{MHz}$).
Figure \ref{fig:TelGoodnessNsb} shows the average goodness of fit, computed on a single telescope basis (to keep a constant number of
pixels), and as function of the night sky background level for the two gains of the H.E.S.S. camera.

The goodness of fit calculated with low gain channels on each pixel appears to be rather stable with night sky background, thus confirming
the validity of the Gaussian assumption for this channel. In contrast, the goodness of fit calculated with high gain channels exhibit a
strong deviation at low night sky background level, where the pedestal is not Gaussian anymore. Simulations were used to derive a correction
table as function of night sky background level in each pixel independently. This table is not used during the reconstruction, 
and is only applied to the final goodness of fit estimator. The resulting goodness-of-fit is shown in fig.~\ref{fig:TelGoodnessNsb},
before correction (left) and after correction (right). After calibration, the goodness-of-fit is stable for a night sky background
variation from  $10$ to $~200~\mathrm{MHz}$, which corresponds the bulk of the H.E.S.S. observations.

\subsection{ShowerGoodness and BackgroundGoodness}

\begin{figure}[htb]
\begin{center}
\epsfig{file=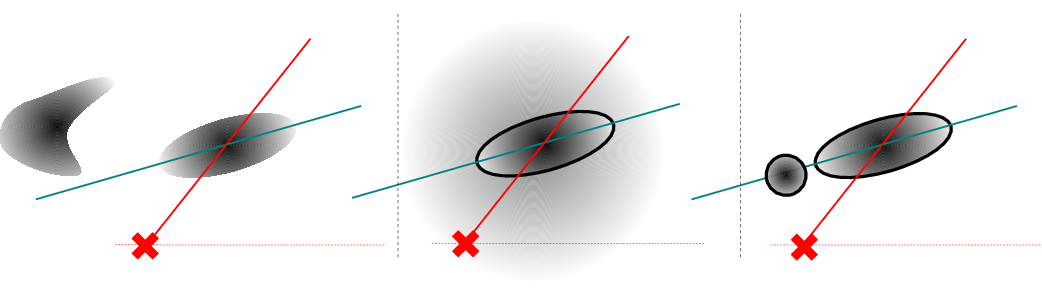,width=0.98\textwidth}
\end{center}
\caption{\label{fig:HadronicShowers}Examples of image topologies that can
discriminate between a $\gamma$-ray induced showers and an hadronic one. Left: image
subdivided in several clusters corresponding to electromagnetic sub-showers. Middle: {\it hadronic rain}
emitted by the hadronic component of the shower. Right: Cherenkov emission from the primary particle (nucleus)
entering the atmosphere, before the actual shower development. The red cross denotes the centre of the camera.}
\end{figure}

The discrimination between $\gamma$-ray induced showers and hadron induced ones
can use several distinct features (fig. \ref{fig:HadronicShowers}):
\begin{itemize}
\item Hadron induced showers are more irregular, and contain several electromagnetic sub-showers
initiated in particular by disintegration of neutral pions. As a consequence, the image in a
Cherenkov camera often exhibits several clusters separated apart (fig. \ref{fig:HadronicShowers}, left)
\item In addition, the hadronic component of the shower itself emits a low intensity Cherenkov light
spread over a large fraction of the camera. This emission, denoted as {\it Hadronic rain}, is in
general eliminated by cleaning procedures used in standard reconstruction techniques. (fig. \ref{fig:HadronicShowers}, middle).
\item Finally, the charged nucleus entering the atmosphere can emit a Cherenkov light before
interacting. This emission is produced very high in the atmosphere and therefore it generates, in the camera,
a faint spot close to the shower direction. 
\end{itemize}

In order to fully exploit the differences between the $\gamma$ and hadron induced showers, 
individual pixels contributing to the goodness-of-fit (eq. \ref{eq:Goodness}) are classified into two different groups
at the end of the fit:

\begin{itemize}
\item Pixels belonging to the {\it shower core}, defined as pixels whose predicted amplitude is above $0.01~\mathrm{p.e.}$,
are grouped together with three rows of neighbours around them to construct a variable named {\it ShowerGoodness} (SG)
in a similar way to eq. \ref{eq:Goodness}. 
These pixels are selected at the end of the fit procedure to avoid changes of the number of degrees of freedom  during the fit.
Due to the large reduction of the number of degrees freedom, this 
variable is more sensitive than the Goodness to discrepancies between the model prediction and the actual shower
images. 
\item Remaining pixels, denoted as {\it background pixels}, are grouped together to construct a variable 
named {\it BackgroundGoodness} (BG), which is sensitive to hadronic clusters outside the main image, hadronic
rains and other irregularities.
\end{itemize}

\section{Results}\label{sec:results}

\subsection{Data sets}

For a detailed comparison of the Model Analysis with previously existing reconstruction techniques, two data sets are used:

\begin{itemize}
\item A data set ({\it PKS Flare}) of 7.7 hours (live time) of data taken on the blazar PKS~2155-304 during flaring period in July 2006,
with 4 telescopes, yielding a test sample of more than 10~000 $\gamma$-rays with very small background contamination,
at relatively low zenith angles (typically $20^\circ$). The average night sky background for this data set corresponds to a
single pixel triggering rate of $40~\mathrm{MHz}$, representative for extragalactic space. Since the data has been taken within 3 days,
the system state can be considered as stable.
\item A large set ({\it Crab Full}) of $\sim$ 34 hours (live time) of data taken on the Crab Nebula, from 2004 to 2008,
with 4 telescopes and larger zenith angles (from $40^\circ$ to $~60^\circ$). This data set yields a
sample of $\sim 10~000$ excess events. The average NSB for this data set, representative for galactic plane, is more than twice higher ($\sim 100~\mathrm{MHz}$)
than for extragalactic space.
\end{itemize}

The model analysis will be compared to the standard Hillas parameters based reconstruction in use in the H.E.S.S. collaboration\cite{HessCrab}.
Two sets of cuts will be defined: {\it Hillas 60} will denote a minimum image amplitude of 60~photoelectrons (p.e.), same as for the Model Analysis,
and {\it Hillas 200} will use a larger minimum image amplitude of 200~p.e.. When needed,  {\it Hillas Std} and {\it Hillas Hard} will denote
configurations used in the Crab publication\cite{HessCrab} with respectively 80 and 200~p.e. minimal image amplitude.

\subsection{Shape cuts\label{sec:ShapeCuts}}

In order to have well reconstructed showers, the following {\it shape cuts} are used throughout the current section
(unless specified differently):

\begin{itemize}
\item A minimum image amplitude of 60 photoelectrons per telescope.
\item A maximal nominal distance (distance of the shower image centre to the centre of the camera) of $2^\circ$
for a camera radius of $2.5^\circ$. The nominal distance cut removes truncated images, close to the camera edge, that often
lead to misreconstruction of shower direction.
\item At least images from two different telescopes passing the previous cuts to ensure stereoscopic reconstruction.
\end{itemize}

These cuts will be used both for the Model Analysis and for the Hillas parameters based reconstruction techniques. The later needs
an additional cleaning procedure. We use a two-level filter cleaning\cite{HessCrab} with pixel
thresholds of respectively $5$ and $10~\mathrm{p.e.}$. The image amplitude is computed after cleaning.
The Model Analysis doesn't require any cleaning procedure, so the image amplitude for the model photon reconstruction
is slightly larger than for the Hillas analysis (as pixels belonging to the shower image tail are taken into
account for the model reconstruction and not for the Hillas reconstruction).

\subsection{Hadron discrimination}

\begin{figure}[htb]
\begin{center}
\begin{tabular}{cc}
\epsfig{file=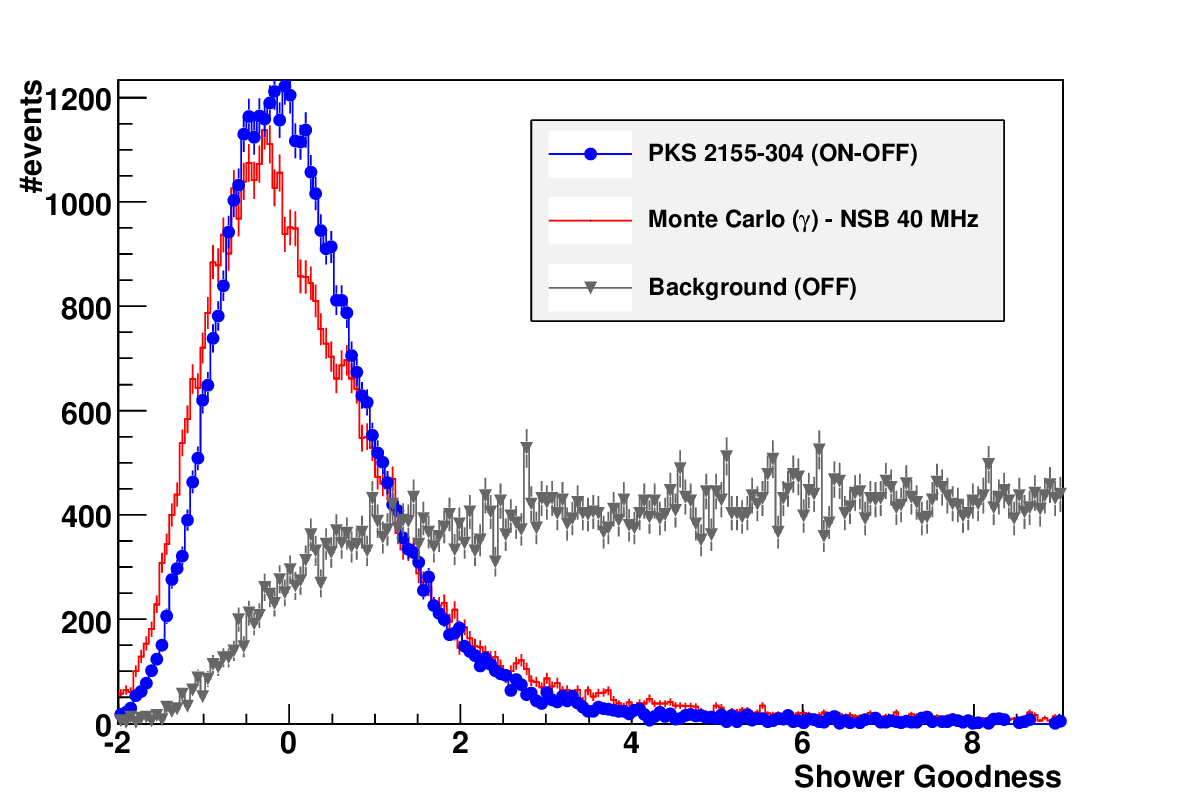,width=0.48\textwidth} & \epsfig{file=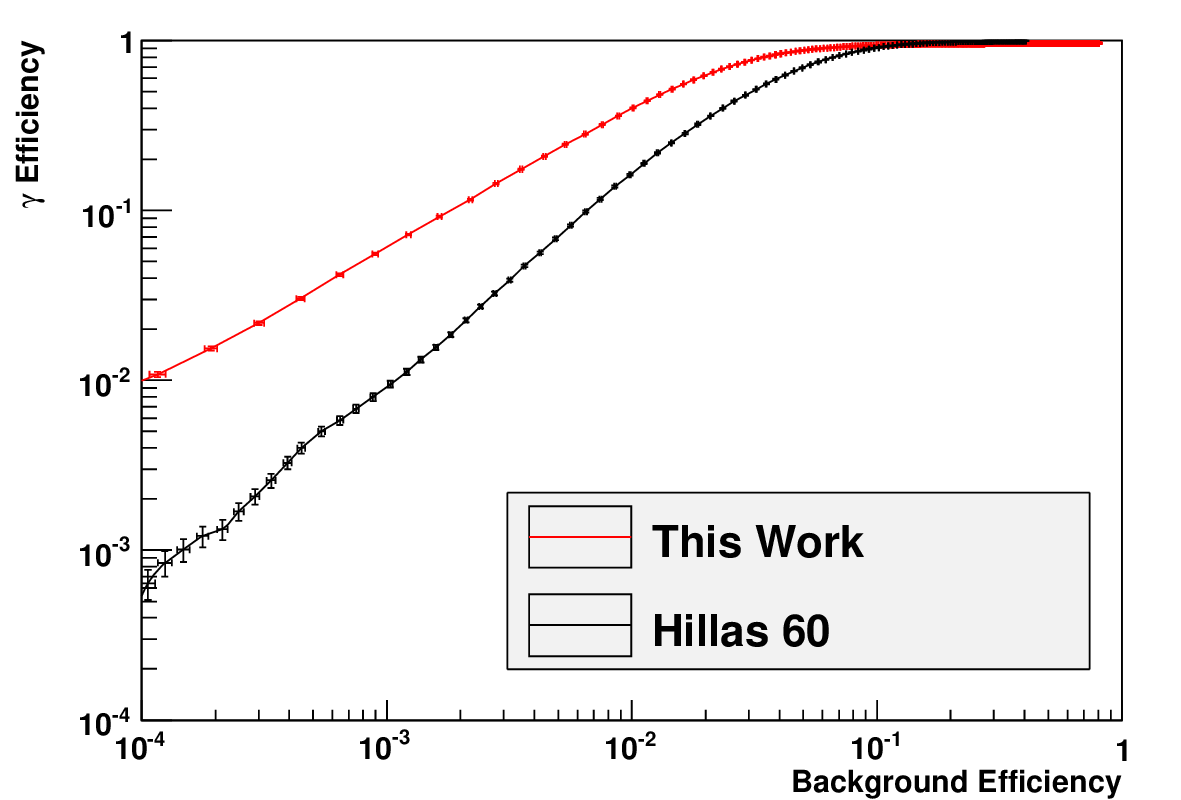,width=0.48\textwidth}  \\
\end{tabular}
\end{center}
\caption{\label{fig:MeanScaledShowerGoodness}Left: Distribution of Shower Goodness for real data taken on the blazar PKS~2155-304 (blue point for 
excess events, grey triangles for background events), compared with a simulation (red histogram) with a similar night sky background level. Right: $\gamma$-ray efficiency 
as function of background rejection for a selection based on Shower Goodness, compared with selection achieved in a standard Hillas parameters reconstruction at the same
minimum image amplitude (60 p.e.) and with the Mean Scaled Width and Mean Scaled Length variables.
}
\end{figure}

The distribution of ShowerGoodness (SG) in the shape cuts of section \ref{sec:ShapeCuts} is shown in fig.~\ref{fig:MeanScaledShowerGoodness} 
for the data taken on the blazar PKS~2155-304, and for a simulation with 40~MHz of night sky background noise, corresponding to the average  
night sky luminosity around PKS~2155-304. 
The Monte Carlo distribution (red line) and the distribution for the  PKS~2155-304 excess events (blue filled circles) are in overall good agreement, 
and are very different from the distribution  obtained for background events (grey triangles), 
thus confirming the discrimination capabilities of the ShowerGoodness variable. 
The small shift between the two distribution is responsible for a systematic error in acceptance determination
of 4\% for a cut at $SG \leq 0.6$.

Fig. \ref{fig:MeanScaledShowerGoodness}, right, shows the $\gamma$-ray efficiency as function of background rejection for a selection 
based on ShowerGoodness only\footnote{several other variables, described later in this paper, can provide additional rejection}, 
after {\it shape cuts}, compared to what is achieved in the standard Hillas parameters based reconstruction (at the same minimal image amplitude, and using the
{\it Mean Scaled Width} as varying selection variable).
A large squared angular distance cut of $\theta^2 \leq 0.04~\mathrm{deg}^2$ was used to ensure that all events are accepted
(extended source assumption) and to decorrelate angular resolution performance(sec. \ref{sec:AngularResolution}) from hadronic discrimination. 
The model reconstruction provides, for a given $\gamma$-ray efficiency, a much better background rejection than Hillas parameters based reconstruction.
Moreover, as will be shown in the following sections, the Model Analysis provides additional discriminating variables that improve further
the sensitivity.

In the standard configuration of the Model Analysis, a cut ShowerGoodness $SG \leq 0.6$ will be used
as main discriminating parameter. This cut retains $70\%$ of $\gamma$-rays and rejects more than
$95\%$ of background events, yielding a quality factor

\begin{equation}
Q = \frac{\epsilon_\gamma }{\sqrt \epsilon_\mathrm{hadrons}} \approx 4 
\end{equation}

\subsection{Primary interaction reconstruction}

\begin{figure}[htb]
\begin{center}
\epsfig{file=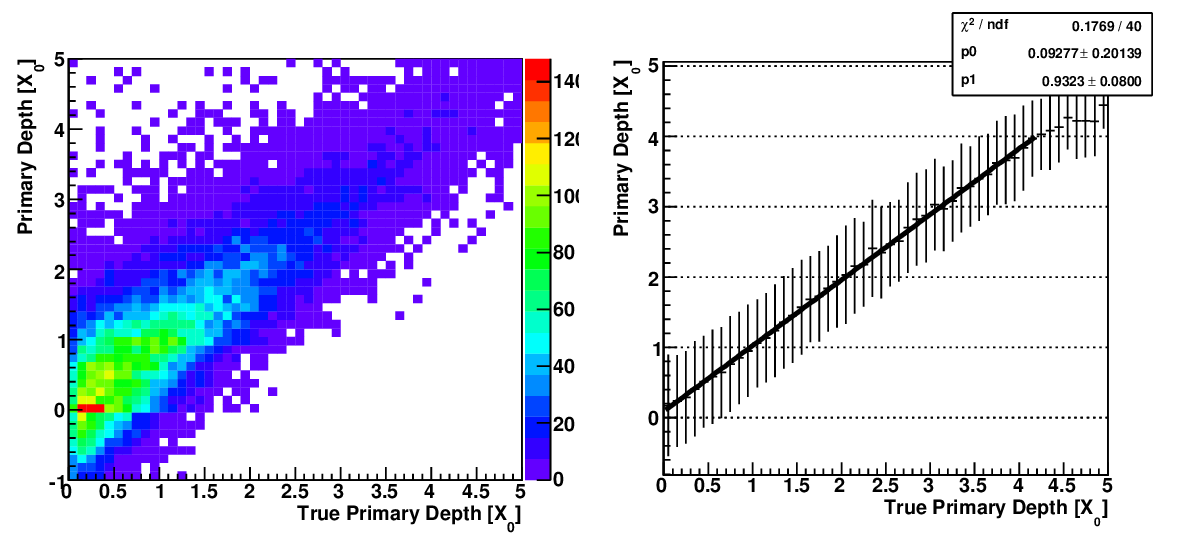,width=0.98\textwidth}
\end{center}
\caption{\label{fig:PrimaryDepth} Reconstruction of the depth of the first interaction for a $\gamma$-ray spectrum
of differential index 2.0 at zenith. Left: Reconstructed depth of the first interaction (in radiation length) versus
true one. Right: profile showing the linear relation between true and reconstructed depth of the first interaction.}
\end{figure}

The depth of the first interaction being a parameter of the model, is also a direct product
of the reconstruction procedure (instead of being calculated, as in the Hillas parameters based reconstruction,
from the shower maximum). Fig. \ref{fig:PrimaryDepth} shows the ability of the Model Analysis
to reconstruct the depth of the first interaction with almost no bias and a resolution of $0.7~ X_0$
(slightly worse at large zenith angles). This is better than Hillas parameters based analyses,
which obtain resolution not better than $1~X_0$ in the best cases. 

\begin{figure}[hbt]
\begin{center}
\begin{tabular}{cc}
\epsfig{file=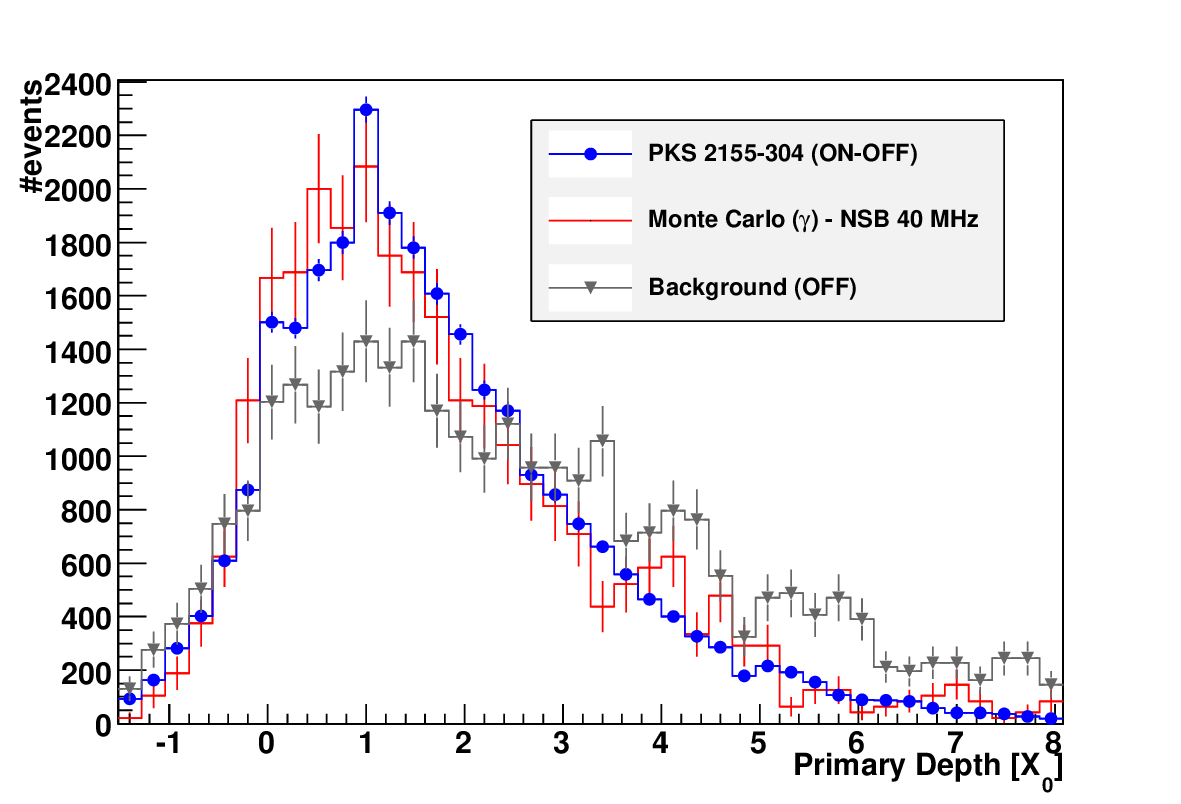,width=0.48\textwidth} & \epsfig{file=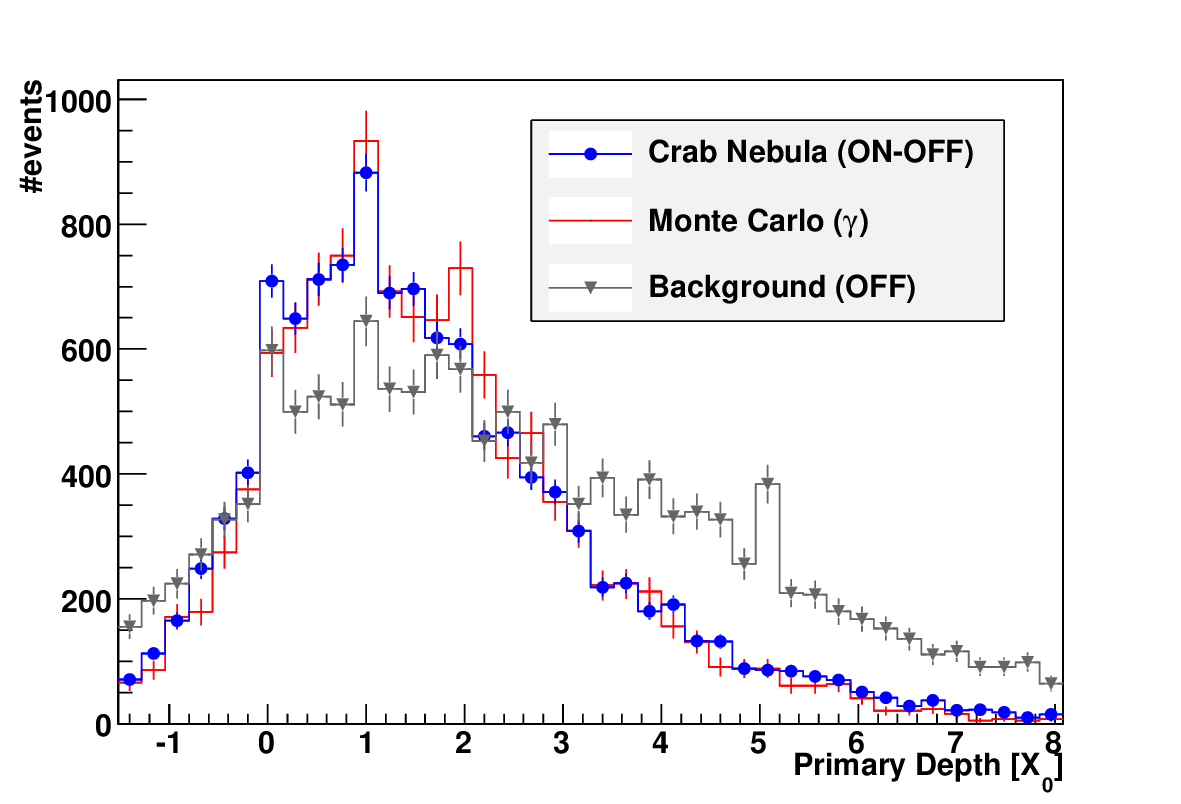,width=0.48\textwidth}  \\
\end{tabular}
\end{center}
\caption{\label{fig:PrimaryDepthData} Distribution of depth of the first interaction for real $\gamma$-ray (from observations
on PKS~2155-304 for the left figure, and on the Crab nebula on the right) compared to Monte-Carlo simulations, respectively 
at zenith with differential index $-3.4$ and at a zenith angle of $46^\circ$ with differential index of $-2.6$. The agreement
between the simulation and the actual data is very good.}
\end{figure}

A comparison of the actual depth of the first interaction obtained with real data and with simulation is shown in fig.~\ref{fig:PrimaryDepthData},
in {\it shape cuts} and $SG \leq 0.6$, under a point-like source assumption (squared angular distance cut of $\theta^2 \leq 0.01~\mathrm{deg}^2$).
The agreement between data and simulations is excellent. The PKS~2155-304 data are compatible with a $0.7~X_0$ resolution, and the Crab data with 
a slightly worse ($0.9~X_0$) resolution due to a larger average zenith angle.
Moreover, since the distribution for OFF data is significantly different, the depth of the first interaction can be used to improve the $\gamma$-hadron separation. 
The irregularities in the OFF distributions are artifacts of the reconstruction procedure: models are only available for depths of $0,1,2,3,4$ and $5~X_0$ and are 
linearly interpolated between these values
(and extrapolated above $5~X_0$). Poorly constrained showers tend to accumulate at the grid model points and avoid interpolated values. 

In order to improve $\gamma$-background separation, a cut in reconstructed primary interaction depth 
$-1 X_0 \leq t_0 \leq 4 X_0 $ is used.
This cut retains $90\%$ of $\gamma$-rays (resp.$91\%$) and rejects $30\%$ of hadrons (resp. $25\%$) , yielding a quality factor $Q=1.07$
(resp. $1.06$) respectively for the Crab and PKS~2155-304 data sets, leading to an improvement of the signal over
background ratio by about $20\%$.

\begin{figure}[htb]
\begin{center}
\epsfig{file=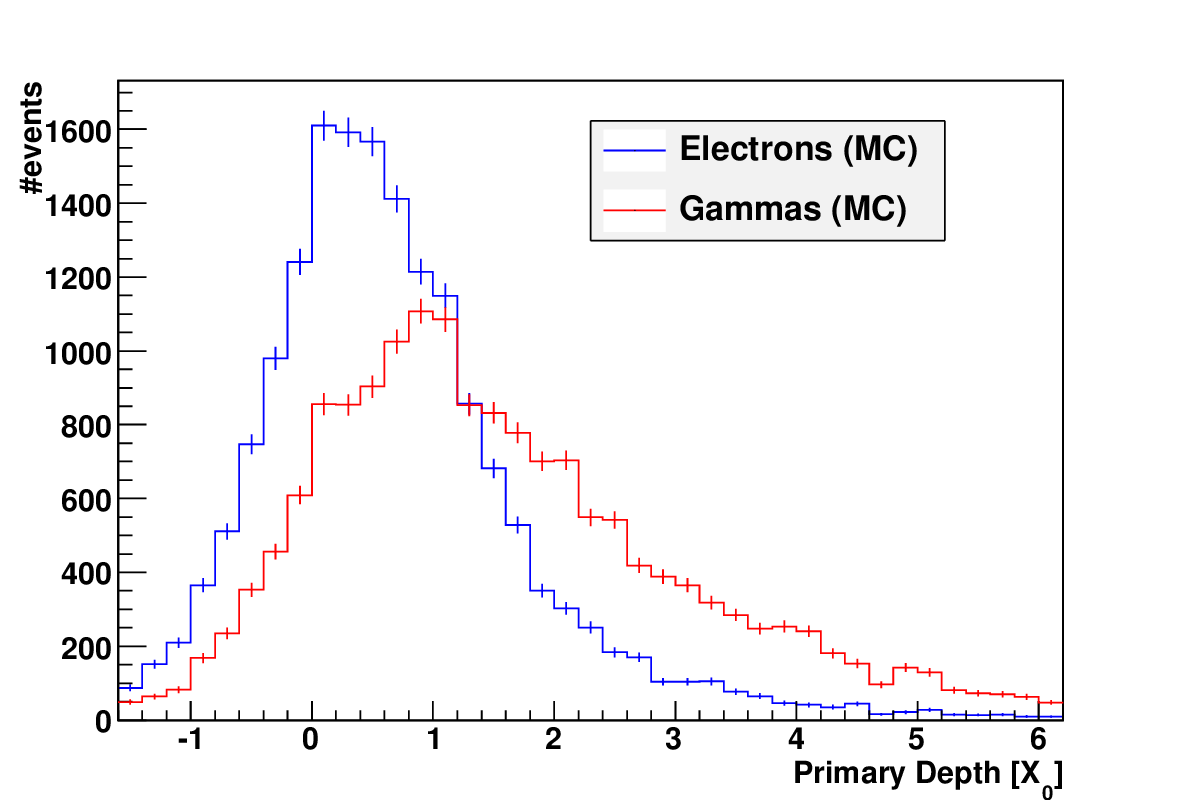,width=0.55\textwidth}
\end{center}
\caption{\label{fig:PrimaryDepthElectrons} Distribution  of the depth of the first interaction for a $\gamma$-ray spectrum
of differential index 2.6 at zenith (in red) compared to electrons with same spectrum (in blue): electron induced showers
start to emit Cherenkov light a little bit earlier in the atmosphere.}
\end{figure}

At low energies, the electron background becomes dominant for atmospheric Cherenkov telescopes. It is usually considered as
irreducible, since electron induced showers are almost identical to $\gamma$-ray induced one. However, electron induced showers
start to produce Cherenkov light a little bit earlier in the atmosphere (about a radiation length before $\gamma$).
This is confirmed by simulation shown in fig.~\ref{fig:PrimaryDepthElectrons}: the distribution of reconstructed primary depth 
for a simulated electrons spectrum shows a peaks centred around $0~X_0$, earlier than for $\gamma$'s.
The difference is not big enough to distinguish between electrons and gammas on an event by event basis, but a statistical 
discrimination seems to be possible.

\subsection{Analysis Configurations}

The $\gamma$-hadron separation is essentially based on the {\it ShowerGoodness} and primary interaction depth
variables. For completeness, two additional cuts are used: a cut on {\it BackgroundGoodness} ($BG \leq 2$), with 
$\gamma$-ray efficiency close to $100\%$, is designed to remove a small number of shower misreconstructed
at very large distance from the camera centre ($4$ degrees or more). A second cut on the Goodness ($G$) variable 
(eq. \ref{eq:Goodness}), in which the model is replaced by a null ($\mu = 0$) assumption, removes images that are
consistent with pure night sky background noise, and reduces the effect of night sky background inhomogeneities
across the field of view.

The {\it standard cuts} for the Model Analysis are defined as:

\begin{itemize}
\item A minimum image amplitude of 60 photoelectrons per telescope
\item A maximal nominal distance (distance of the shower image centre to the centre of the camera) not more than $2^\circ$
\item At least two telescopes passing the previous {\it shape} cuts
\item A maximum ShowerGoodness (SG) of $0.6$
\item A reconstructed primary interaction depth $t_0$ between -1 and 4 $X_0$.
\item For a point-like source, a squared angular distance cut of $\theta^2 \leq 0.01~\mathrm{deg}^2$,
independent of zenith angle\footnote{Since the angular resolution degrades at large zenith angle,
a varying cut could perform better. This is not specific to the Model Analysis and applies
to other reconstruction techniques as well.}
\end{itemize}

Two additional sets of cuts are defined as reference point for future H.E.S.S. publications: 
the {\it Faint source} configuration is optimised for
source fainter than a few percent of the Crab flux. The {\it Loose cuts} configuration
is designed to maximise the $\gamma$-ray efficiency for strong sources at the expense
of a poorer background rejection.
For analysis of extended source, the $\theta^2$ cut is usually replaced by a selection
that encompasses the whole source. Cuts for the three aforementioned configurations
are summarised in Tab. \ref{tab:AnalysisConfigs}.
of cuts 

\begin{table}[htb]
\begin{center}
\begin{tabular}{|c|c|c|c|c|c|c|}
\hline
Name & Min.  & Max. Nom. & \#Tels & $SG_{\mathrm{max}}$ & $t_0$ & $\theta^2_{\mathrm{max}}$ \\ 
& Charge & Distance & & & & \\
& (p.e.) & (deg.) & & & ($X_0$) & ($\mathrm{deg}^2$) \\
\hline
Standard & 60 & 2 & 2 & 0.6 & $[-1,4]$ & 0.01 \\
Faint Source & 120 & 2 & 2 &  0.4 & $[-1,4]$ & 0.005 \\
Loose Cuts & 40 & 2 & 2 & 0.9 & N/A & 0.0125 \\ \hline
\end{tabular}
\vspace{1em}
\end{center}
\caption{\label{tab:AnalysisConfigs} Reconstruction configurations: Minimum charge, maximum nominal distance, maximum ShowerGoodness,
primary interaction depth range and maximum squared angular distance from the reconstructed shower position to the source. 
A minimum of two telescopes passing the per-telescope cuts, on image
amplitude and distance from the centre of the field of view, are also required.}
\end{table}

\subsection{Effective Areas}

\begin{figure}[htb]
\begin{center}
\epsfig{file=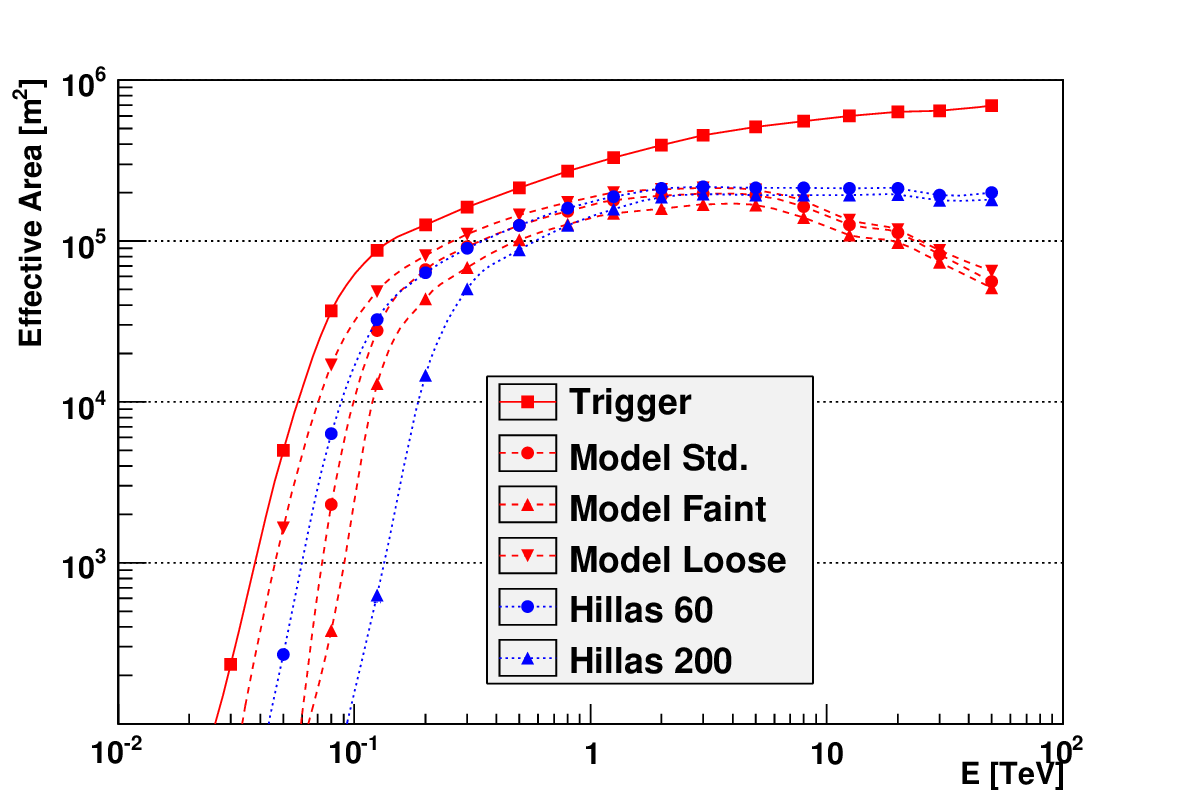,width=0.98\textwidth}
\end{center}
\caption{\label{fig:EffectiveArea} Effective area as function of energy, at zenith, compared
with the values obtained for the standard and hard cuts Hillas parameters based analyses. }
\end{figure}

The effective area as function of energy for the Model Analysis at zenith is shown in fig.~\ref{fig:EffectiveArea},
compared to areas obtained with Hillas parameters based reconstructions. The reconstruction efficiency for the standard
configuration is similar to the values obtained with Hillas reconstruction with a minimum image amplitude of 60 p.e..
As expected, the loose configuration has a larger effective area and a lower threshold.
In most of the H.E.S.S. publications so far, a minimum image amplitude of 200 p.e. was used, yielding a much 
larger threshold of $\sim 300~\mathrm{GeV}$ at zenith.
Models are currently generated only up to $20~\mathrm{TeV}$, thus leading to a loss of acceptance
at high energy. In addition, showers above $10~\mathrm{TeV}$ can reach the ground, resulting into 
large fluctuations that are not fully reproduced by the model.

Performances obtained on the Crab Nebula and on the Blazar PKS~2155-304 are shown in tab. \ref{tab:Significance}.
The Model Analysis yields  a $\gamma$-ray efficiency similar to the 60 p.e. Hillas reconstruction, but with a background
rejection improved by a factor of $6.2$ (PKS 2155-304) to $6.8$ (Crab Nebular), yielding a signal over background 
ratio similar or better to the one obtained with hard cuts Hillas reconstruction. 
As a result, the significance obtained with the Model reconstruction is larger, and 
the sensitivity improved by a factor of more than $2$. The faint source configuration provides
an additional factor of $\sim 2$ in S/B compared to standard configuration.

\begin{table}[htb]
\begin{center}
\begin{tabular}{c|c|c|c|c|c|c|c}
Data Set &  Analysis & ON & OFF & $1/\alpha$ & $\gamma$ & $\sigma$ & S/B  \\ \hline 
Crab Full & Hillas 60   & 12768 & 26154 & 16.2 & 11148.6 & 162.6 & 6.9 \\ 
Crab Full & Hillas 200  & 3742  & 1435  & 16.9 & 3657.1  & 125.0 & 43.1   \\ 
Crab Full & Model Std   & 10249 & 3848  & 18.2 & 10037  & 210.7 & 47.3   \\  
Crab Full & Model Faint & 5920  & 1605  & 25.8 & 5857.7 & 176.8 & 94.0   \\  
Crab Full & Model Loose & 20107 & 22137 & 16.7 & 18782.3  & 244.3  & 14.2 \\  \hline
PKS Flare & Hillas 60   & 24964 & 7025 & 10.9 & 24320.4 & 302.1 & 37.8 \\ 
PKS Flare & Hillas 200  & 5148 & 490 & 12.7 & 5109.3 & 153.9  & 132.1  \\ 
PKS Flare & Model Std   & 24388 & 1303 & 12.7 & 24285.4 & 342.9 & 236.7 \\ 
PKS Flare & Model Faint & 11047 & 427 & 18.1 & 11023.4 & 248.1 & 466.5 \\ 
PKS Flare & Model Loose & 38308 & 3676 & 11.0 & 37972.7 & 407.2 & 113 \\ \hline
\end{tabular}
\vspace{1em}
\end{center}
\caption{\label{tab:Significance} Number of excess events and significances obtained with the Model Analysis
on some H.E.S.S. sources compared to standard reconstruction techniques. The published results on the Crab Nebula\cite{HessCrab}
where obtained with a minimum image amplitude of 80 photoelectrons, yielding a better S/B ratio than results obtained
with a 60 pe cut, although with similar significance.}
\end{table}

\subsection{Energy resolution\label{sec:EnergyResolution}}

\begin{figure}[htb]
\begin{center}
\epsfig{file=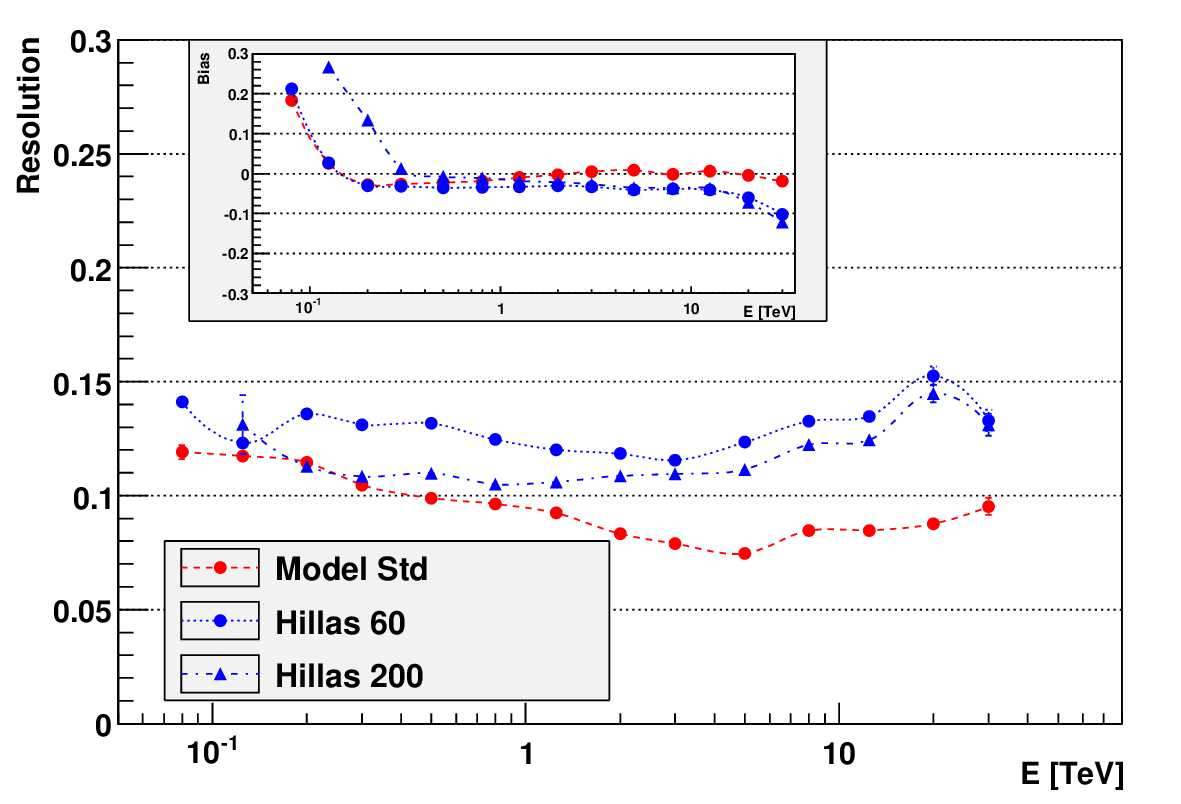,width=0.98\textwidth} 
\end{center}
\caption{\label{fig:EnergyResolution} Energy resolution (main plot) and bias (inset) as function of energy, at zenith, compared
with the values obtained for the standard and hard cuts Hillas parameters based analyses (resp. blue circles and triangles). }
\end{figure}

The energy resolution (in cuts) as function of energy is shown in fig.~\ref{fig:EnergyResolution} for zenith,
where the energy resolution is defined as the RMS of the $\Delta E/E$ distribution. The energy resolution is better than $15\%$ for the whole
energy range (from $80~\mathrm{GeV}$ up to more than $20~\mathrm{TeV}$), with biases not exceeding $5\%$ in the central range.
The very central energy range ($500~\mathrm{GeV}$ to more than $10~\mathrm{TeV}$), the energy resolution is better than $10\%$ and reaches
values as low as $7$ to $8\%$. 
Larger energy biases appear at very low energy (up to $20\%$ at $80~\mathrm{GeV}$), due to trigger selection effects. These biases are however smaller than those obtained with an
Hillas parameters based reconstruction. In a similar manner, negative energy biases appear at very high energies, due to very distant high energy 
showers reconstructed closer to the telescopes\footnote{Very distant shower produce almost parallel images in the camera, which introduces a degeneracy in the reconstruction.}
and with an underestimated energy. 
In the medium energy range ($500~\mathrm{GeV}$ to a few 10 $\mathrm{TeV}$), the Model Analysis largely outperforms
standard reconstruction techniques.

\subsection{Angular resolution\label{sec:AngularResolution}}

\begin{figure}[htb]
\begin{center}
\begin{tabular}{cc}
\epsfig{file=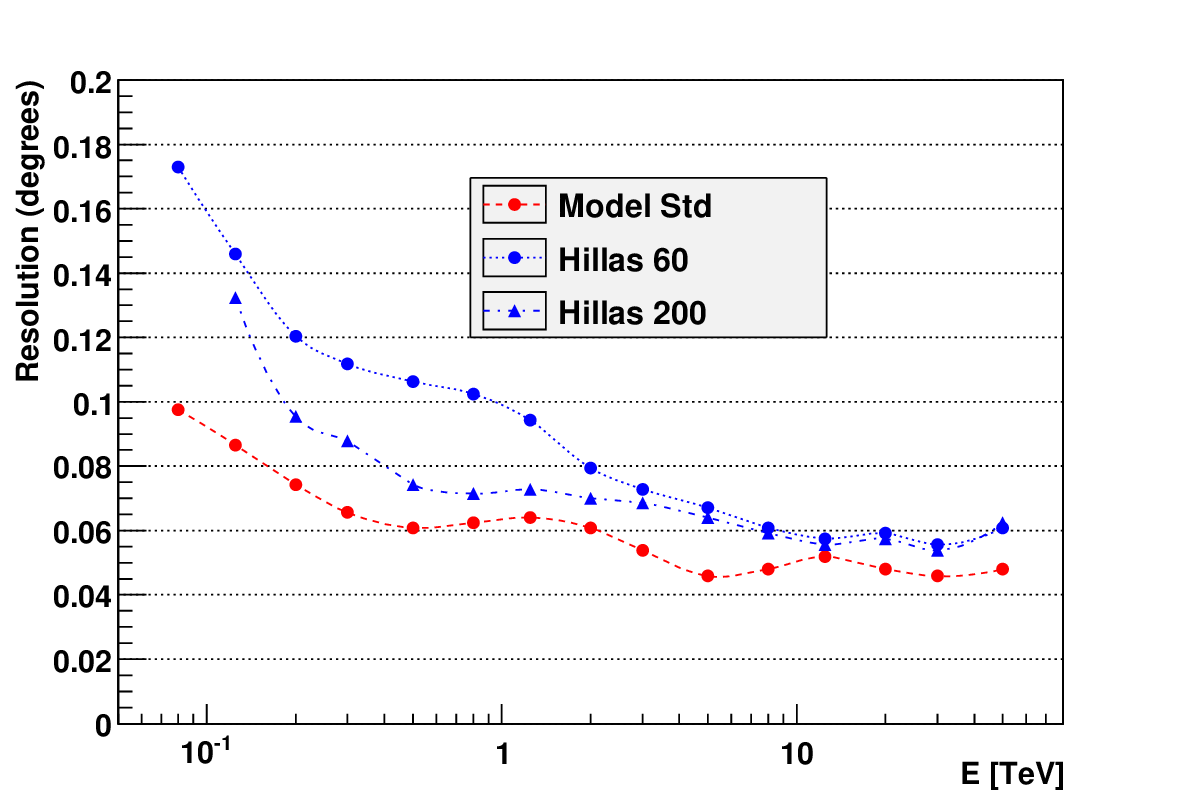,width=0.48\textwidth} & \epsfig{file=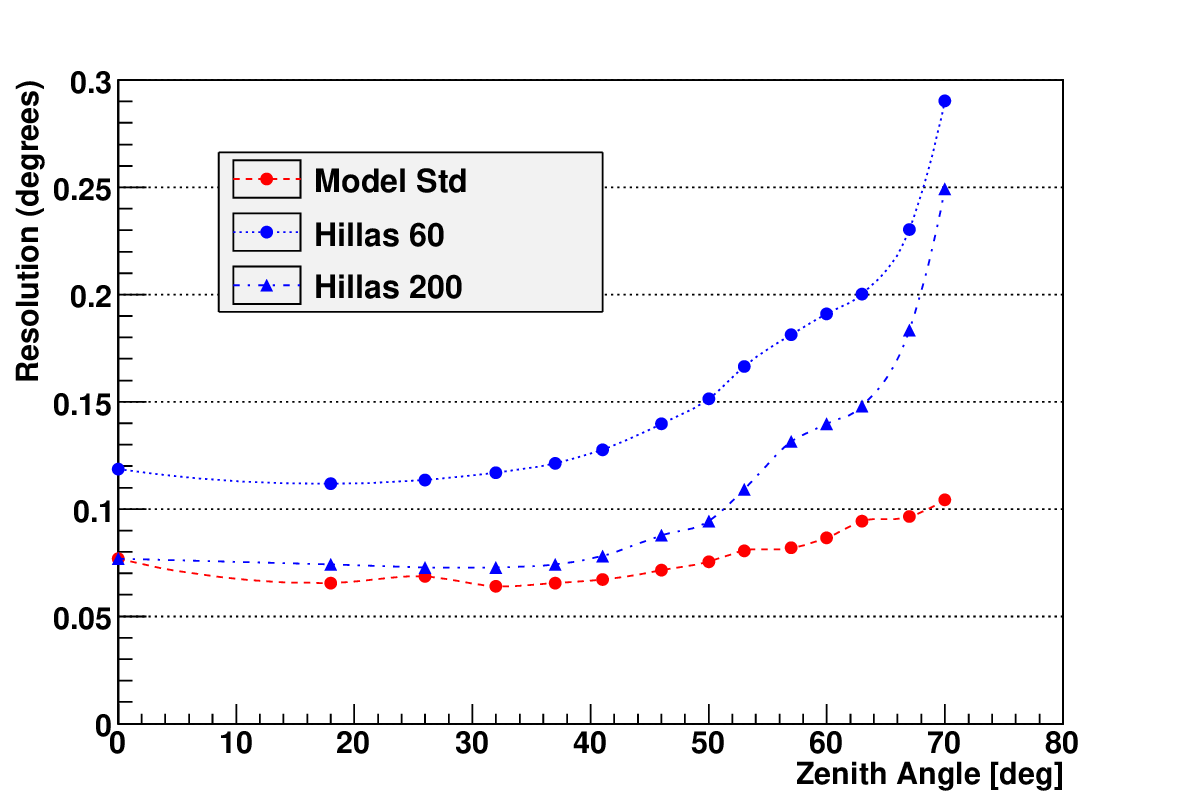,width=0.48\textwidth} \\
\end{tabular}
\end{center}
\caption{\label{fig:AngularResolution} Angular resolution (defined as the $68\%$ containment radius) as function of energy, at zenith, compared
with the values obtained for the standard and hard cuts Hillas parameters based reconstruction techniques. Right: Average angular resolution for a $E^{-2}$ spectrum, as function of zenith angle. }
\end{figure}

The angular resolution, defined as the $68\%$ containment radius, is shown in fig.~\ref{fig:AngularResolution} for showers at zenith (left),
and is of the order of $0.06^\circ$ in most of the energy range, which is much better than values obtained with Hillas parameters based analyses
(using algorithm 1 from \cite{Hofmann99}) for similar minimal image amplitude. An immediate consequence is an improved sensitivity for point like sources
and improved morphological studies of extended source. The angular resolution is stable for zenith angles up to $50^\circ$ (fig. \ref{fig:AngularResolution}, right),
ans rises slowly up to very large zenith angles. The degradation observed for very large zenith angles is much larger for Hillas parameters based reconstruction techniques.

\begin{figure}[htb]
\begin{center}
\begin{tabular}{cc}
\epsfig{file=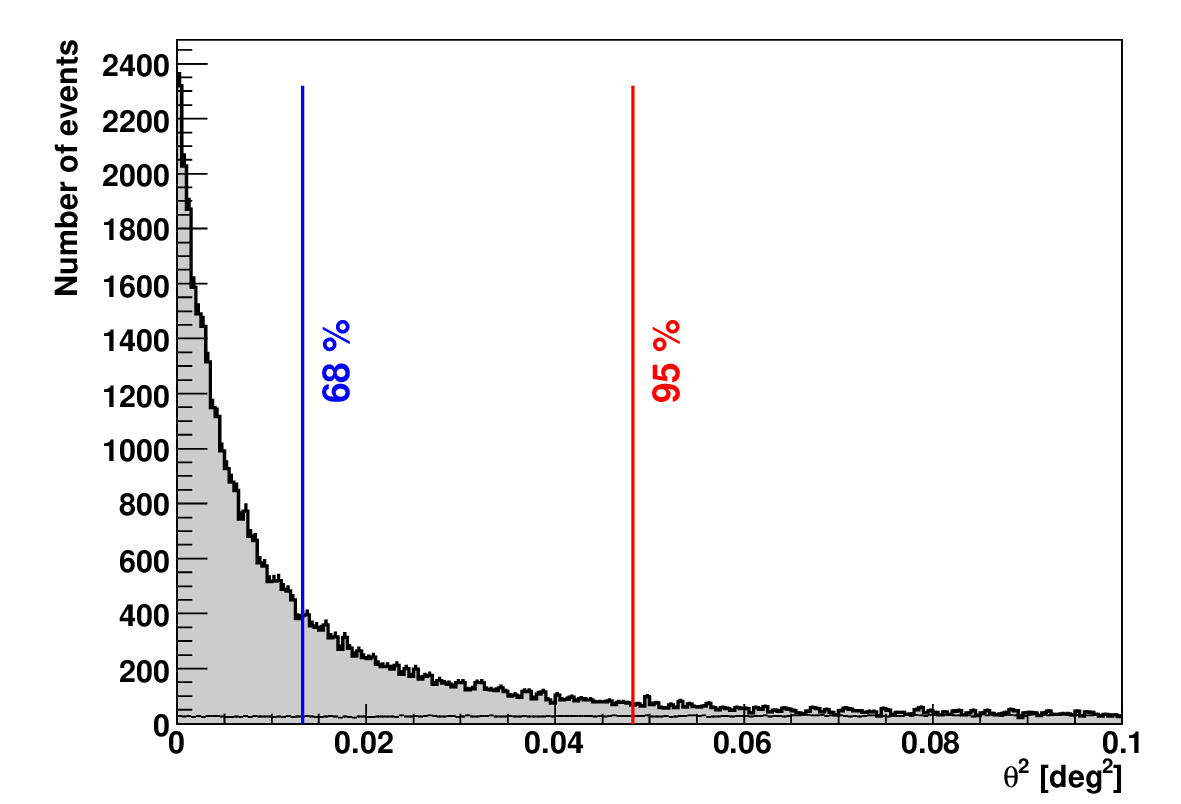,width=0.48\textwidth} & \epsfig{file=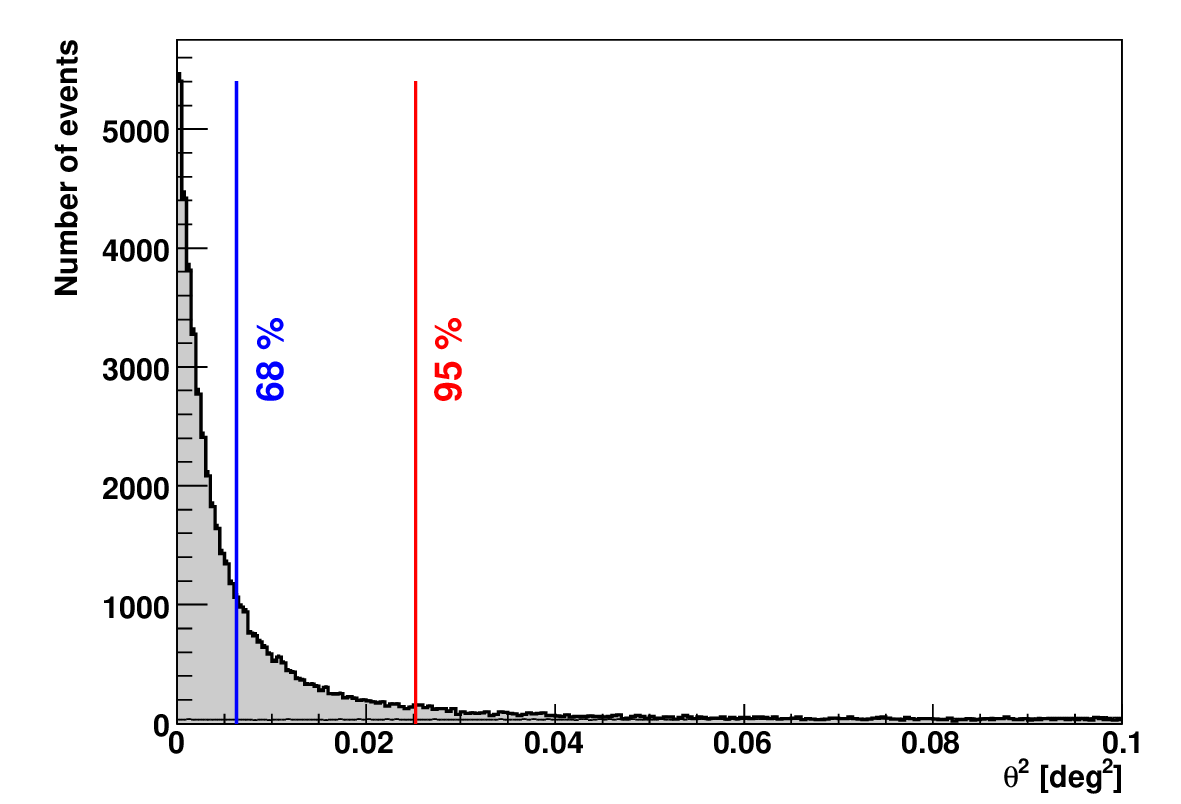,width=0.48\textwidth} \\
\end{tabular}
\end{center}
\caption{\label{fig:Theta2_2155} Squared angular distribution ($\theta^2$) obtained on the blazar PKS~2155-304 with the standard 
Hillas parameters based analysis with a minimal image amplitude of 60 p.e. (left) and with the Model Analysis (right).}
\end{figure}

\begin{figure}[htb]
\begin{center}
\begin{tabular}{cc}
\epsfig{file=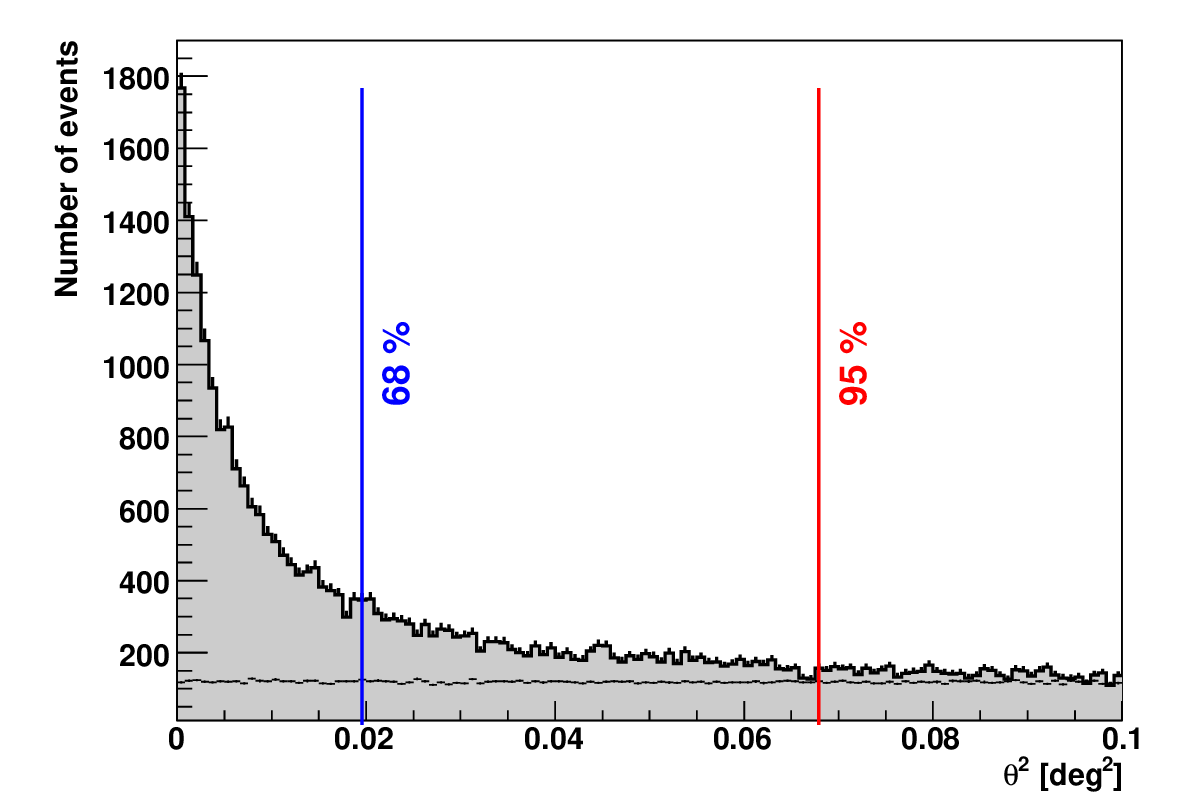,width=0.48\textwidth} & \epsfig{file=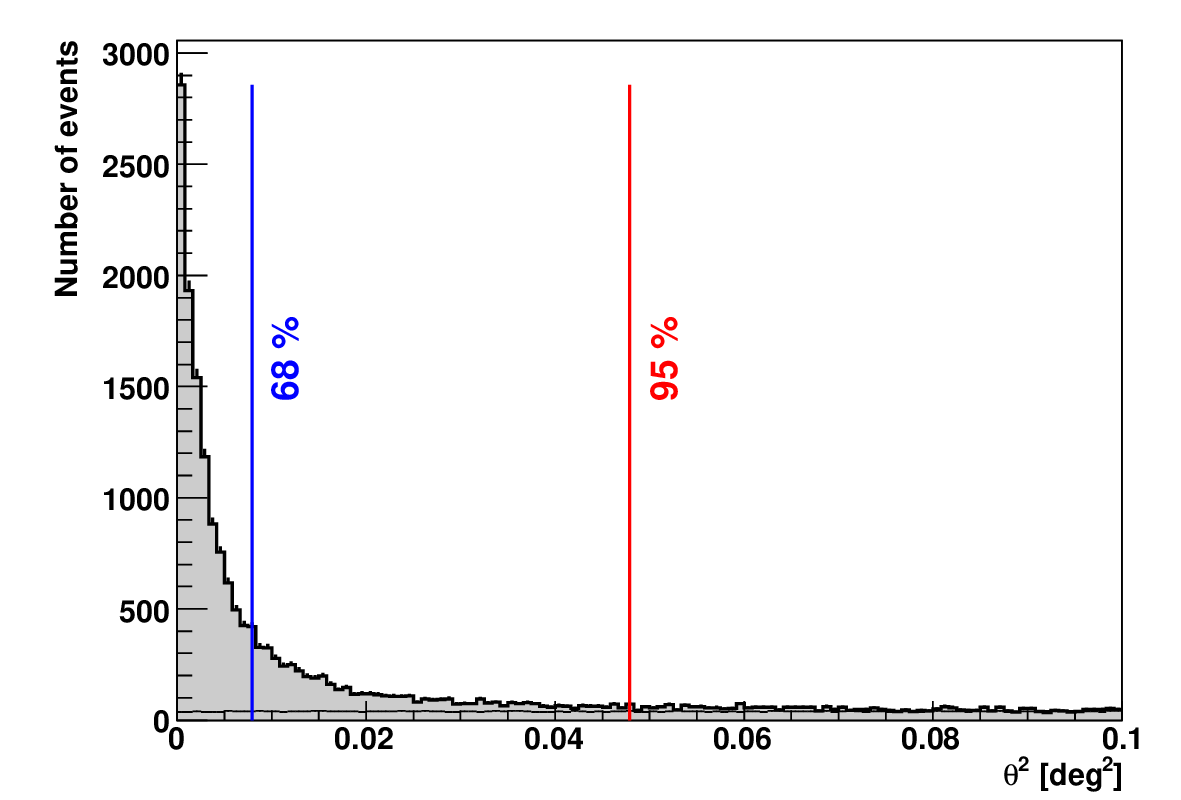,width=0.48\textwidth} \\
\end{tabular}
\end{center}
\caption{\label{fig:Theta2_Crab} Squared angular distribution ($\theta^2$) obtained on the Crab Nebula with the standard 
Hillas parameters based analysis with a minimal image amplitude of 60 p.e. (left) and with the Model Analysis (right).}
\end{figure}

The superior angular resolution of the Model Analysis is demonstrated in fig.~\ref{fig:Theta2_2155} on data taken on PKS~2155-304.
The $\theta^2$ distribution is twice as more peaked for the Model Analysis (right), compared to the Hillas reconstruction (left).
The same behaviour is observed at larger zenith angles for the Crab Nebula (fig. \ref{fig:Theta2_Crab}).

Detailed numerical comparison shown in tab. \ref{tab:AngularResolution} demonstrates that, on average, the Model Analysis
angular resolution performances with a minimal image amplitude of 60 p.e. are  equally good as the Hillas Analysis
with a much higher minimal image amplitude of 200 p.e..

\begin{table}[htb!]
\begin{center}
\begin{tabular}{c|c|c|c}
Data Set &  Analysis & $\sigma_{68} [\mathrm{deg}]$ &  $\sigma_{95} [\mathrm{deg}]$    \\ \hline 
Crab Full & Hillas 60 & 0.14 & 0.26  \\ 
Crab Full & Hillas 200 & 0.09 & 0.21  \\ 
Crab Full & Model Std & 0.09 & 0.22 \\  \hline
PKS Full & Hillas 60 & 0.11 & 0.22  \\ 
PKS Full & Hillas 200 & 0.07 & 0.14 \\ 
PKS Full & Model Std &  0.08 & 0.16 \\ \hline
\end{tabular}
\vspace{1em}
\end{center}
\caption{\label{tab:AngularResolution}Angular resolutions obtained on real data (Crab Nebula and PKS~2155-304 for the Model Analysis
and Hillas parameters based analyses.}
\end{table}

\subsection{Uncertainty on parameters}

The Model Analysis provides uncertainty estimation for each reconstructed parameter, 
as byproducts of the correlation matrix. This can be used to select in a simple and natural way
a sub-sample of the events with improved angular or energy resolution (for more precise
morphological or spectral analysis).

\begin{figure}[htb]
\begin{center}
\begin{tabular}{cc}
\epsfig{file=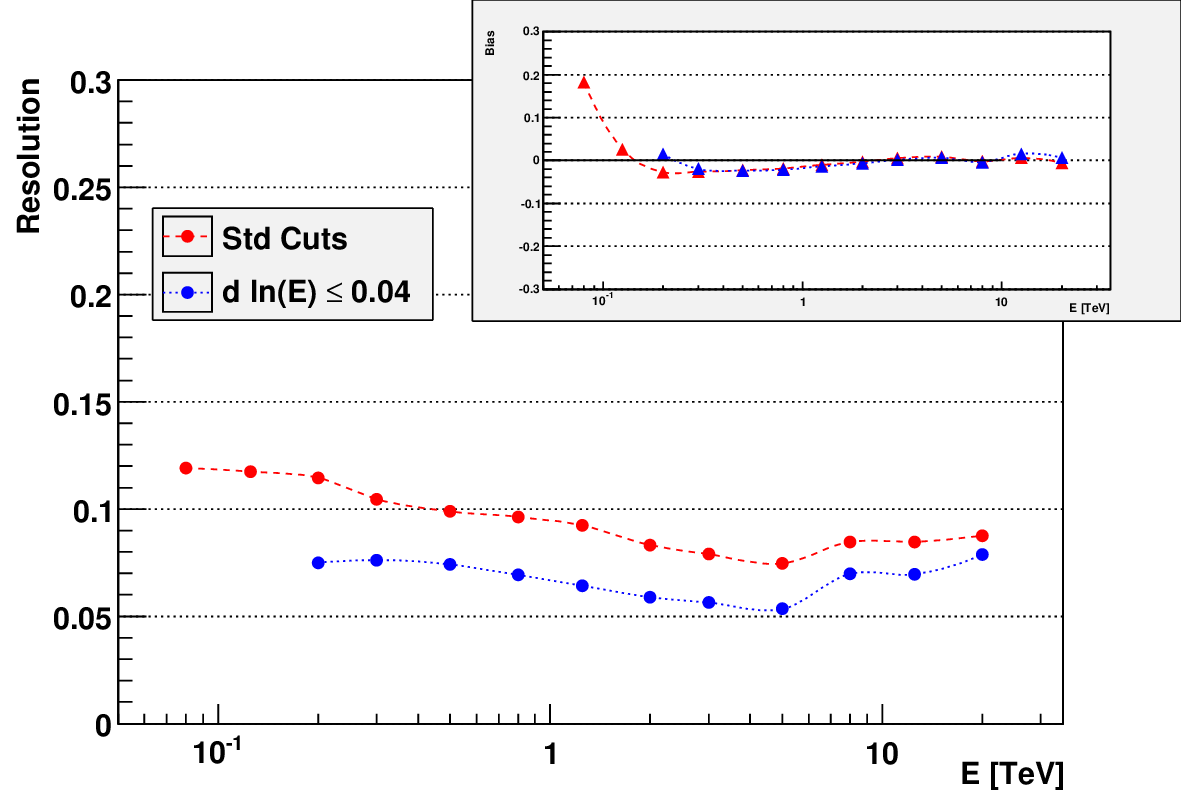,width=0.48\textwidth} & \epsfig{file=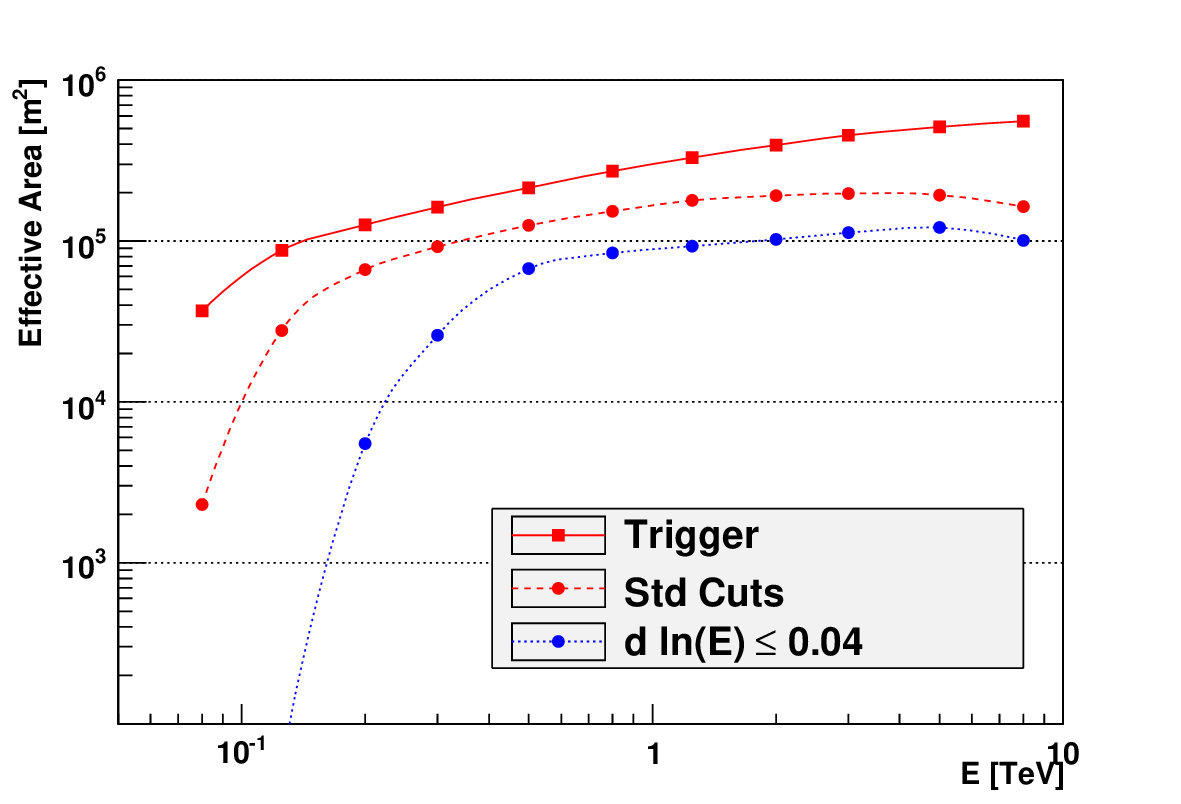,width=0.48\textwidth}  \\
\end{tabular}
\end{center}
\caption{\label{fig:EnergyResolution_Improved} Effect of an additional selection on the energy uncertainty $d\ln E\leq 0.04$ 
on the energy resolution (left) and effective area (right).}
\end{figure}

An example of such a subset selection is shown in fig.~\ref{fig:EnergyResolution_Improved}, where 
an additional selection on the energy uncertainty $d\ln E\leq 0.04$ is applied to the data. The energy
resolution becomes better than $8\%$ from $100~\mathrm{GeV}$ up to $10~\mathrm{TeV}$ with values
as low as $\sim 6~\%$ at a few $\mathrm{TeV}$. The prices to pay for such a {\it golden events} selection is
an increase of energy threshold and a reduction of statistics (fig. \ref{fig:EnergyResolution_Improved}, right) by a factor of $\sim 2$ above 
$1~\mathrm{TeV}$.

\begin{figure}[htb]
\begin{center}
\begin{tabular}{cc}
\epsfig{file=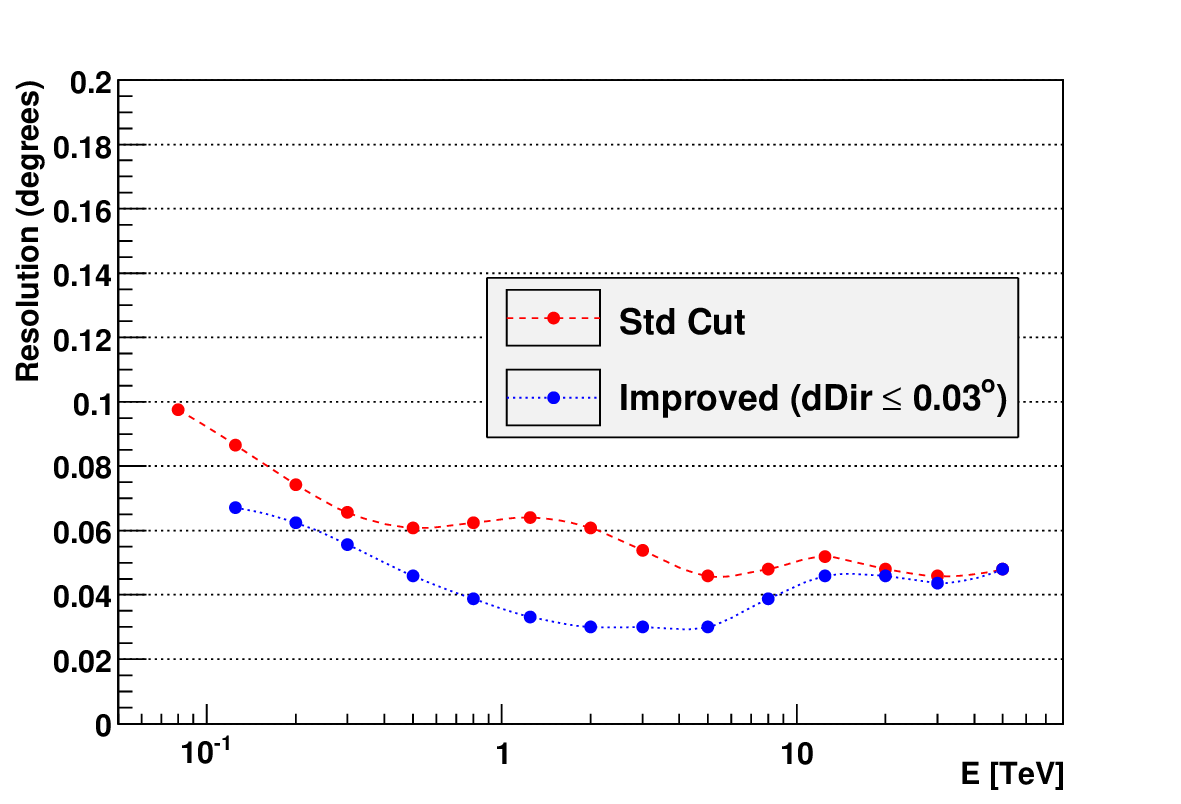,width=0.48\textwidth} & \epsfig{file=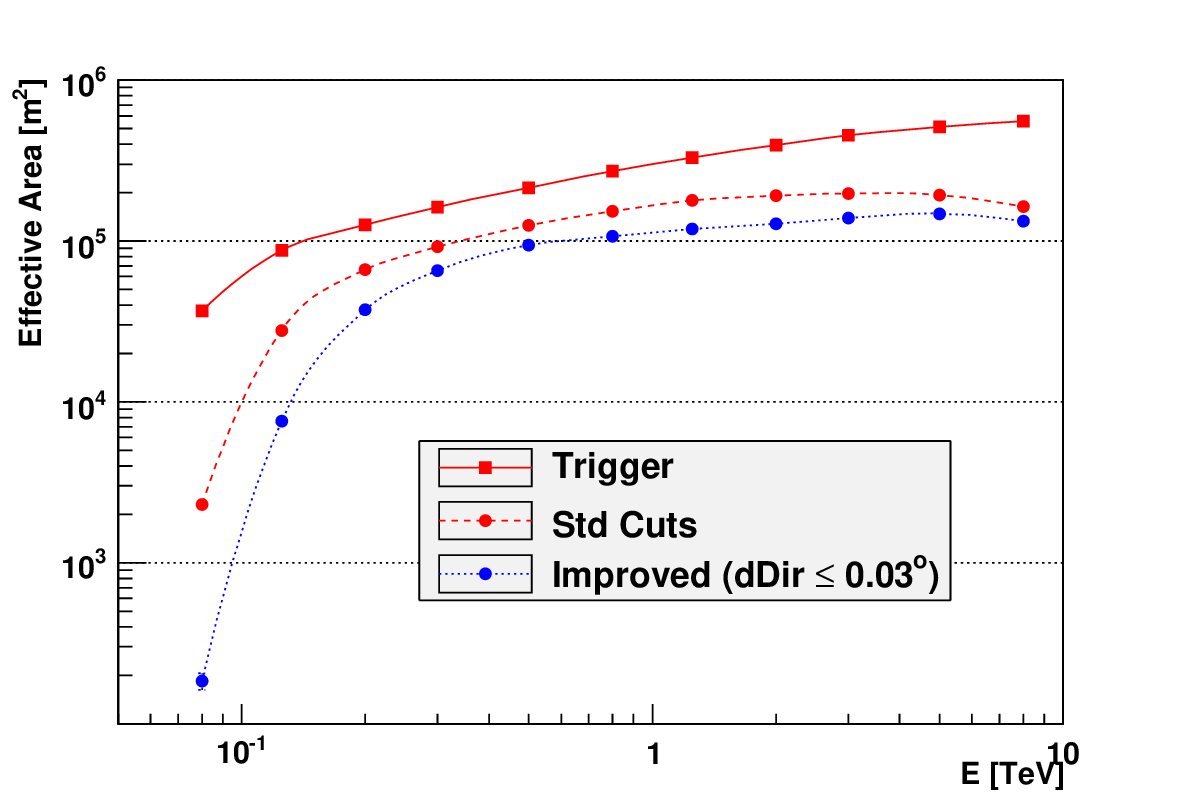,width=0.48\textwidth}  \\
\end{tabular}
\end{center}
\caption{\label{fig:AngularResolution_Improved} Effect of an additional selection on the direction uncertainty $dDir \leq 0.03^\circ$ 
on the angular resolution (left) and effective area (right).}
\end{figure}

In a similar manner, fig. \ref{fig:AngularResolution_Improved} shows the effect of an additional selection
on the direction uncertainty $dDir \leq 0.03^\circ$ on the angular resolution. 
It is possible to achieve a resolution as good as $0.03^\circ$
in the TeV energy range at the expense of a factor less than 2 in statistics. This can greatly improve the ability to
pin-point the source position in challenging regions such as the Galactic Centre.

\begin{figure}[htb]
\begin{center}
\begin{tabular}{cc}
\epsfig{file=Theta2_Model_2155.eps,width=0.48\textwidth} & \epsfig{file=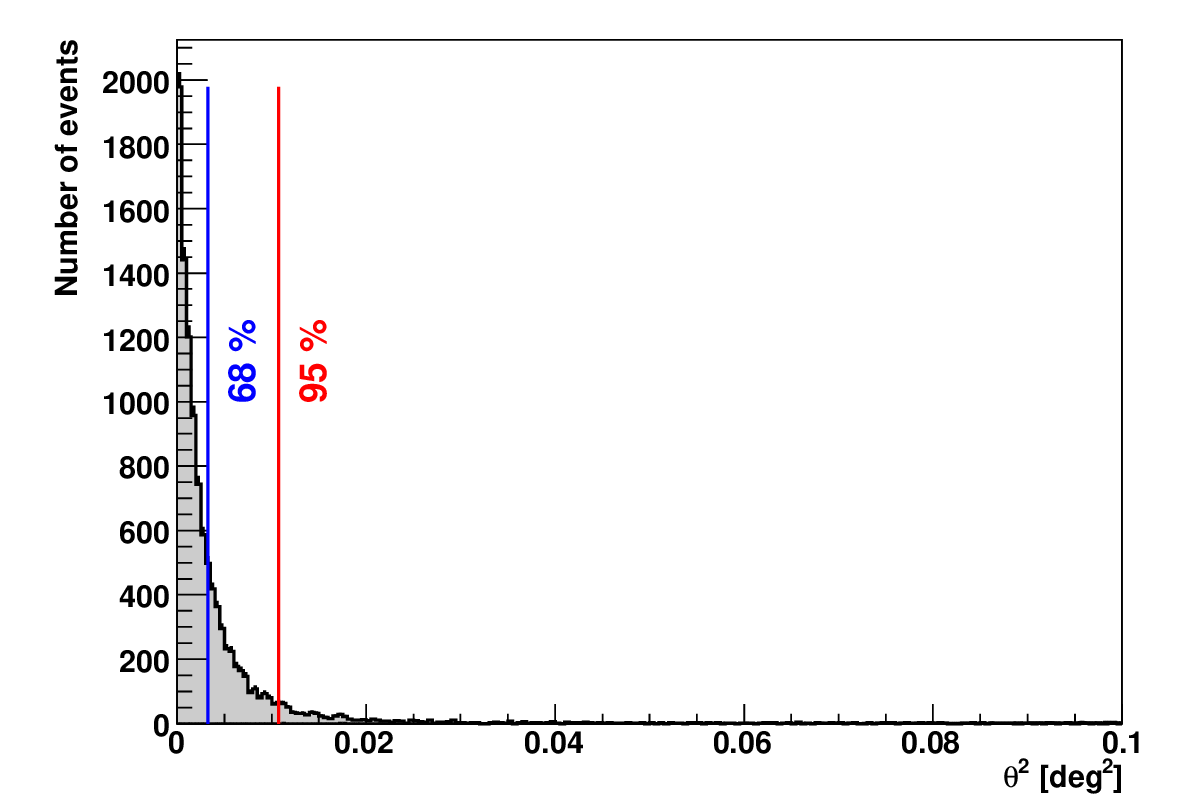,width=0.48\textwidth} \\
\end{tabular}
\end{center}
\caption{\label{fig:Theta2_Improved_2155} Effect of an additional selection on the direction uncertainty $dDir \leq 0.03^\circ$ 
on the squared angular resolution of events from the blazar PKS~2155-304.}
\end{figure}

This is demonstrated in fig.~\ref{fig:Theta2_Improved_2155}, where the same selection $dDir \leq 0.03^\circ$ is applied
to data from PKS~2155-304. The net result of the selection is an unprecedented resolution $\sigma_{68}= 0.055$ and $\sigma_{95}= 0.1$ with 
number of excess event being still around 10000 $\gamma$. The background is also much more suppressed than the signal, 
resulting into a $S/B$ ratio of $454$ compared to $236$ before selection. The direction uncertainty could therefore serve as an additional
discriminating parameter, but its effect would not be independent of energy.

\subsection{Combination of reconstruction methods}

In the early stages of the Model Analysis, it was realised that the Goodness variable was
almost uncorrelated with the Mean Scaled Width and Mean Scaled Length variables obtained with 
standard reconstruction techniques\cite{Cherenkov2005,XEFF}. As a consequence, the combination of several
reconstruction techniques was improving rejection capabilities and thus final sensitivity.

\begin{figure}[htb]
\begin{center}
\begin{tabular}{cc}
\epsfig{file=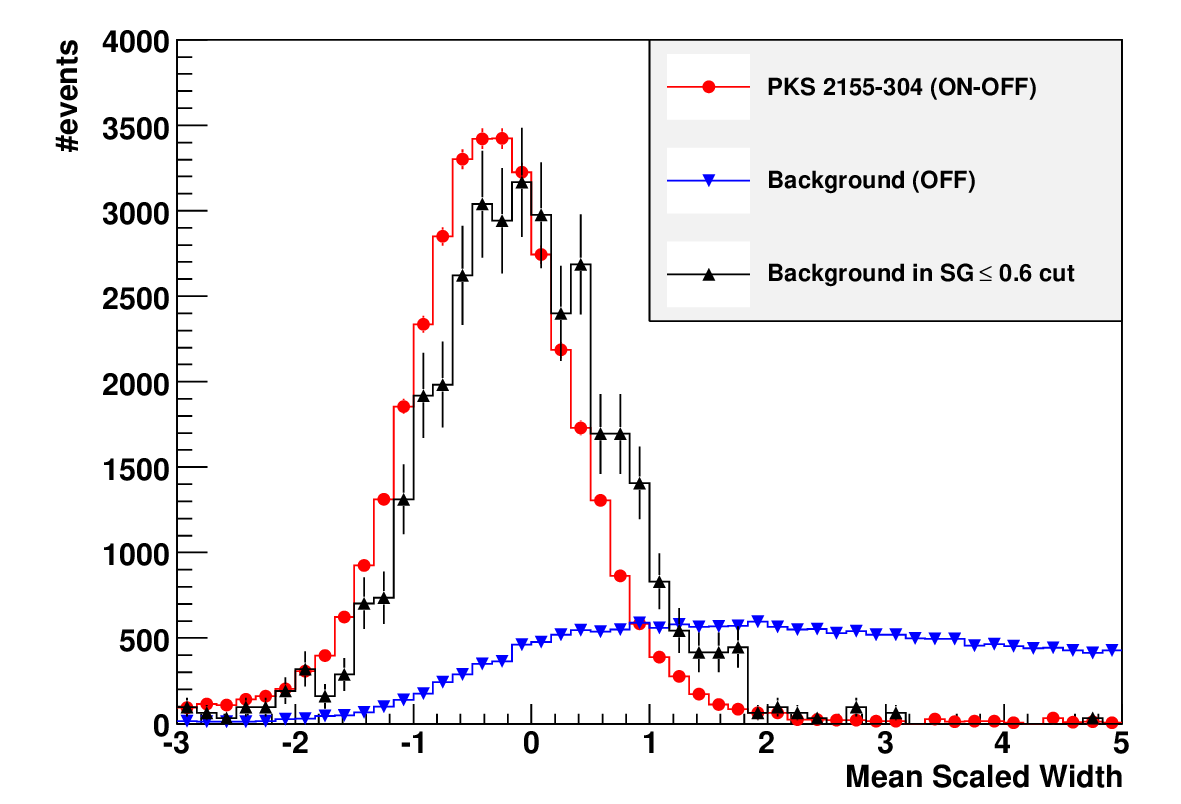,width=0.48\textwidth} & \epsfig{file=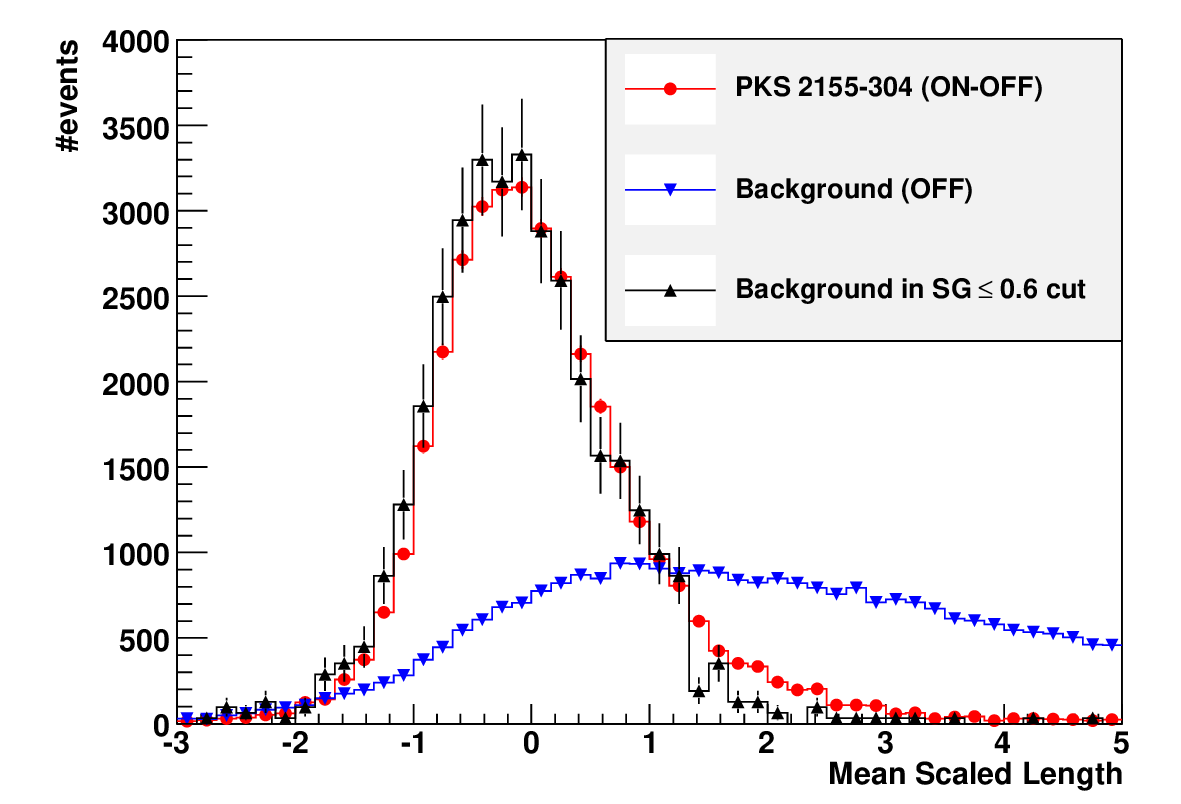,,width=0.48\textwidth} \\
\end{tabular}
\end{center}
\caption{\label{fig:MeanScaledVariables}Distribution of Mean Scaled Width (left) and Mean Scaled Length (right) on real data
taken on PKS~2155-304, for excess events (red), background events before selection on ShowerGoodness (blue) and after (black).
All histograms are normalised to the same number of events.}
\end{figure}

Fig \ref{fig:MeanScaledVariables} shows the behaviour of the Mean Scaled Width and Mean Scaled Length variable (from Hillas reconstruction)
when imposing the ShowerGoodness cut. The background data histograms (in blue) are much more extended than the excess events ($gamma$-rays)
histograms (red), thus confirming the rejection capabilities of the Mean Scaled Width and Mean Scaled Length variables.
However, after imposing a selection on the ShowerGoodness ($SG \leq 0.6$), most of the discrimination capability
vanishes (black histograms): the Mean Scaled Width still provides only a very small  additional discrimination, whereas the Mean Scaled Length does not anymore.
The games of combining several analyses appears therefore not as promising as it was, when using a older version of the Model Analyis with less performance.

\subsection{Sensitivity to stars and broken pixels}

\begin{figure}[htb]
\begin{center}
\epsfig{file=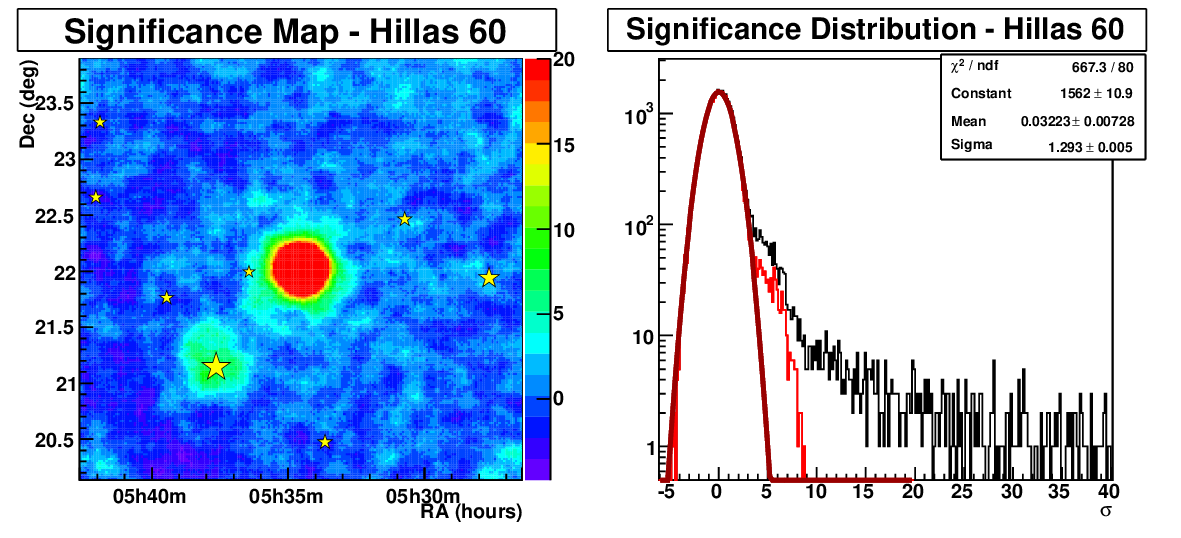,width=0.95\textwidth}
\epsfig{file=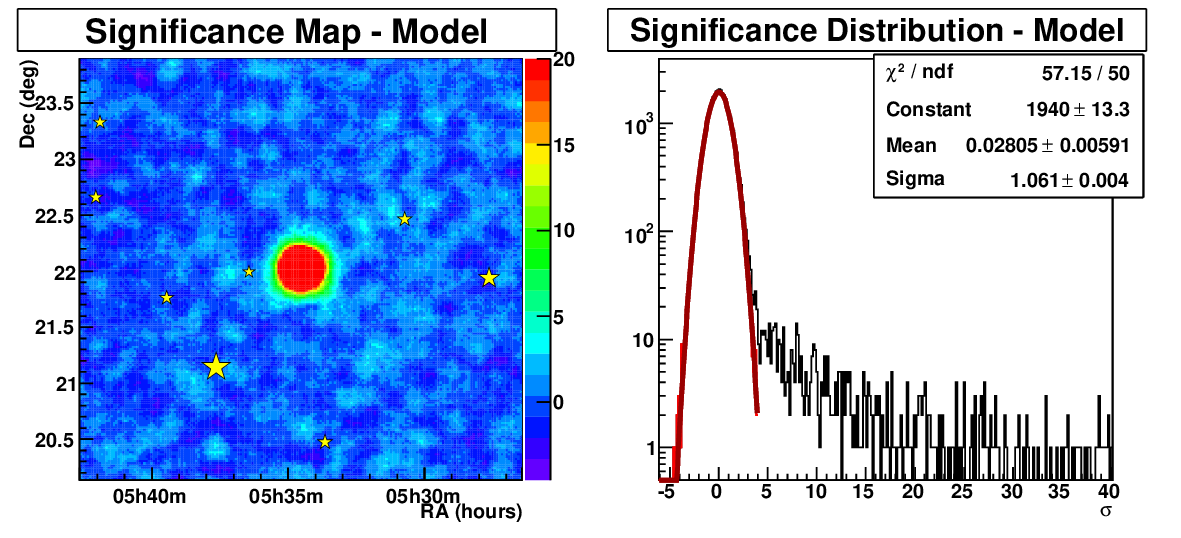,width=0.95\textwidth}
\end{center}
\caption{\label{fig:Crab_Template}Significance Maps (left) and distributions (right) obtained respectively
with the Hillas Analysis (top) and Model Analysis (bottom), using Template Model background subtraction.
The Hillas Analysis is sensitive to a bright star ($\zeta$-Tauri, visual magnitude 2.97) in the field of view
that causes a deficit in hadron acceptance. }
\end{figure}

During H.E.S.S. observations, camera pixels illuminated by a bright star are turned off to avoid damage to the
photocathode. In addition, some (in general less than $5\%$) pixels are removed from the analysis due to 
instrumental malfunction (non working high voltage, noisy pixel, badly behaving analogue memory, ...). 
Missing pixels affect more Hillas parameters based analyses (as they produce truncated images)
than the Model Analysis, which just ignores the missing pixels.

The effect of bright stars is demonstrated in fig.~\ref{fig:Crab_Template}, where the large Crab Nebula data set
(32 live hours) was used to produce a significance map using the template model\cite{Template}. A bright star, $\zeta$-Tauri 
of visual magnitude 2.97, is present in the field of view. On average, about three to four pixels surrounding the 
star are turned off during observations.
An artificial excess of events, up to the level of $8\sigma$, is visible in the Hillas Analysis significance map. This excess is due to a deficit
in background events which translates into an artificial excess of events. The Model Analysis significance map
shows no deviation at the same position, demonstrating that it is less sensitive to missing pixels.
The artifact in the Hillas Analysis significance map weakens at larger minimal image amplitude (200 p.e.) as the
shower images in the camera become larger and are therefore less affected by a few missing pixels.

\subsection{Sensitivity to varying night sky background\label{sec:NSB}}

\begin{figure}[htb]
\begin{center}
\begin{tabular}{cc}
\epsfig{file=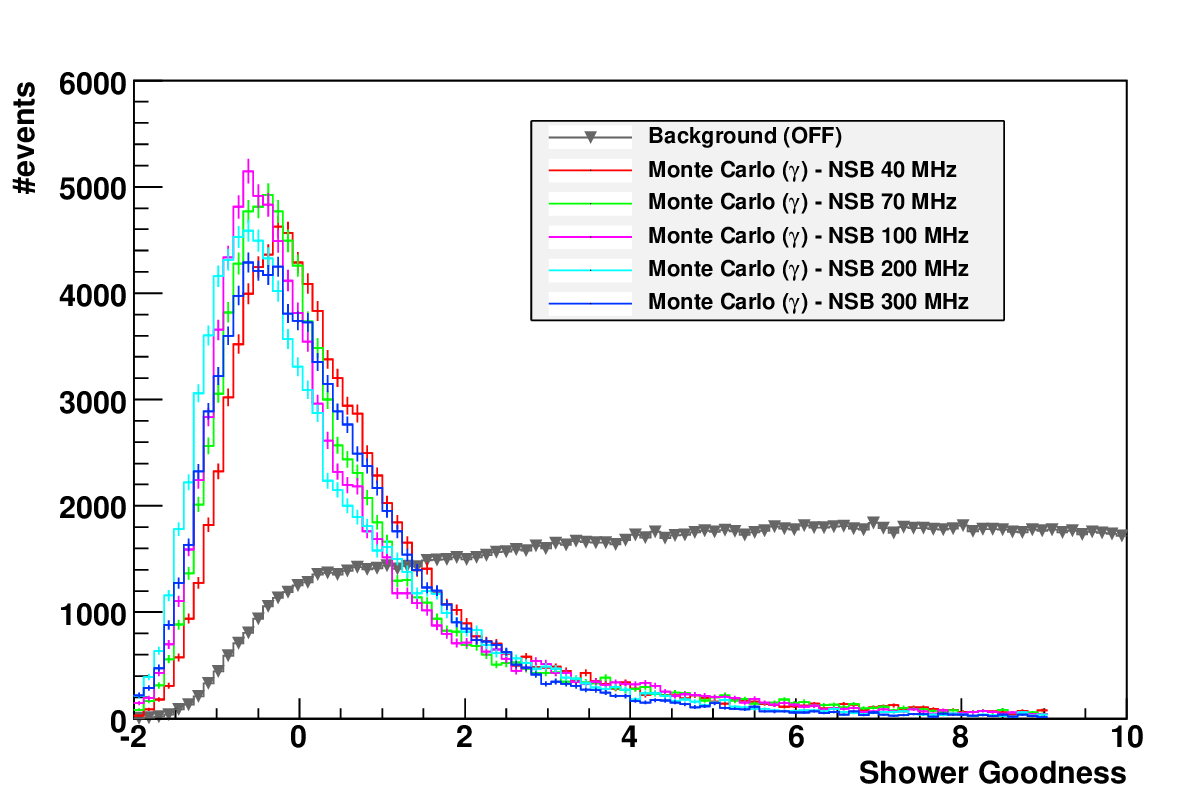,width=0.48\textwidth} & \epsfig{file=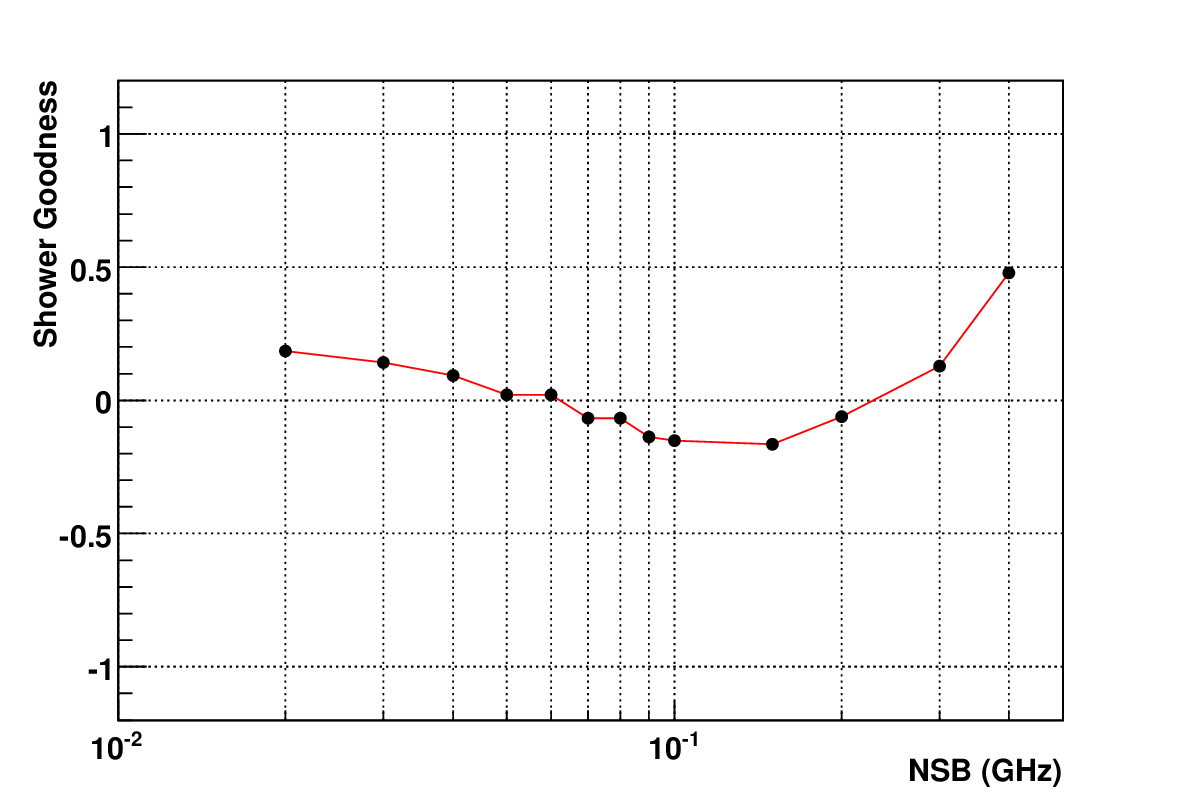,width=0.48\textwidth} \\
\end{tabular}
\end{center}
\caption{\label{fig:ShowerGoodness_NSB}Left: Evolution of Shower Goodness distribution for simulated $\gamma$-rays
with varying night sky background level. Right: Evolution of the mean Shower Goodness value with increasing night sky background level.}
\end{figure}

The evolution of Shower Goodness distribution for  simulated $\gamma$-rays (with photon index 2.0 at zenith)
is shown in fig.~\ref{fig:ShowerGoodness_NSB}. No strong evolution is seen up to $300~\mathrm{MHz}$.
The bulk of the HESS observations correspond to an average NBS level of $100~\mathrm{MHz}$, varying between
$40~\mathrm{MHz}$ to $~300~\mathrm{MHz}$ in the most luminous parts of the Galactic Plane.

\section*{Conclusion}

We have presented a sophisticated $\gamma$-ray reconstruction technique for Atmospheric Cherenkov Telescopes, 
based on an accurate pixel per pixel comparison of observed intensity with a pre-calculated model.
This significant improvement over earlier works leads to an improved angular and energy resolution, and background rejection, 
resulting in an increase of the sensitivity of the H.E.S.S. telescope system by a factor close to 2.
The model analysis is less sensitive than reconstruction techniques based
on Hillas parameters to instrumental or environmental effects (missing pixels, night sky background 
variation across the field of view, \dots),
thus allowing an optimal use of the full dynamical range of the instruments. 
This novel reconstruction technique 
should also benefit to upcoming projects such as CTA and AGIS.

\section*{Acknowledgements}

We thank Prof. W. Hofmann, spokesman of the H.E.S.S. Collaboration and Prof. G. Fontaine,
chairman of the Collaboration board, for allowing us to use H.E.S.S. data in this publication. 
We
are grateful to Dr. B. Degrange and Prof. C. Stegmann for carefully reading the manuscript and for providing us with
very useful suggestions. 
Finally, our thanks go to all the members of the H.E.S.S. Collaboration
for their technical support and for many stimulating discussions.

\end{document}